\documentclass[11pt, preprint]{aastex63}
\usepackage{textcomp}
\usepackage{gensymb}
\usepackage{rotating}
\usepackage{bm}
\usepackage{graphicx}
\usepackage{rotating} 
\usepackage[para,online,flushleft]{threeparttable}
\usepackage{tabularx}
\usepackage{diagbox}
\usepackage{subfigure}

\newcommand{\Plsq}{P_{{\rm LS},q}}
\newcommand{\mPlsq}{\langle P_{{\rm LS},q}\rangle}
\newcommand{\Dls}{\Delta_{\rm LS}}
\newcommand{\rev}{\textcolor{black}}
\received{}
\revised{}
\accepted{}
\submitjournal{ApJ}

\shorttitle{Photometric and asteroseismic
measurements of stellar rotation periods}
\shortauthors{Lu, Benomar, Kamiaka, \& Suto}
\graphicspath{{./}{figures/}}
\begin{document}

\title{Meta-analysis of photometric and asteroseismic
measurements \\ of stellar rotation periods:
the Lomb-Scargle periodogram, autocorrelation \\ function, wavelet and rotational splitting analysis for 92 Kepler asteroseismic targets}

\correspondingauthor{Yuting Lu}
\email{luyuting923@g.ecc.u-tokyo.ac.jp}

\author{Yuting Lu}
\affiliation{Department of Physics, The University of Tokyo,  
Tokyo 113-0033, Japan}

\author[0000-0001-9405-5552]{Othman Benomar}
\affiliation{National Astronomical Observatory of Japan,
  Mitaka, Tokyo 181-0015, Japan}
\affiliation{Research Center for the Early Universe, School of Science,
The University of Tokyo, Tokyo 113-0033, Japan}

\author[0000-0001-6036-3194]{Shoya Kamiaka}
\affiliation{Department of Physics, The University of Tokyo,  
Tokyo 113-0033, Japan}
\affiliation{Daikin Industries, Ltd., Kita-ku, Osaka 530-8323, Japan} 

\author[0000-0002-4858-7598]{Yasushi Suto}
\affiliation{Department of Physics, The University of Tokyo,  
Tokyo 113-0033, Japan}
\affiliation{Research Center for the Early Universe, School of Science,
The University of Tokyo, Tokyo 113-0033, Japan}
\affiliation{\rev{Laboratory of Physics, Kochi University of Technology,
  Tosa Yamada, Kochi 782-8502, Japan}}

\begin{abstract}
  We perform \rev{intensity variability analyses (photometric
    analyses: the Lomb-Scargle periodogram, autocorrelation, and
    wavelet)} and asteroseismic analysis of 92 Kepler solar-like
  main-sequence stars to understand the reliability of the measured
  stellar rotation periods. We focus on the 70 stars without reported
  stellar companions, and classify them into four groups according to
  the quarter-to-quarter variance of the Lomb-Scargle period and the
  precision of the asteroseismic period.  We present detailed
  individual comparison among photometric and asteroseismic
  constraints for these stars.  We find that most of our targets
  exhibit significant quarter-to-quarter variances in the photometric
  periods, suggesting that the photometrically estimated period should
  be regarded as a simplified characterization of the true stellar
  rotation period, especially under the presence of the latitudinal
  differential rotation. On the other hand, there are a fraction of
  stars with a relatively small quarter-to-quarter variance in the
  photometric periods, most of which have consistent values for
  asteroseismically and photometrically estimated rotation periods. We
  also identify over ten stars whose photometric and asteroseismic
  periods significantly disagree, which would be potentially
  interesting targets for further individual investigations.
\end{abstract}


\vspace{2cm}

\section{Introduction} \label{sec:intro}

Observational studies of stellar rotation play a fundamental role in
understanding the physics of stars.  The rotation rate of stars
evolves substantially over their main-sequence stage due to a variety
of angular momentum loss processes.  Various empirical relations
between the period, age and mass have been derived, known as
gyrochronology
\citep[e.g.][]{Skumanich1972,Garcia2014,Saders2016}. The determination
of rotation periods for a large number of stars with different masses
and ages is critical for verifying and understanding the physics
behind these relations. In reality, stars are supposed to exhibit
latitudinal and internal radial differential rotation. Thus the
observed stellar rotation period is not a unique quantity, but should
be interpreted as a weighted average dependent on specific
observational methods.

Indeed, the rotation period of the Sun is known to vary along both
latitudinal \cite[via spectroscopy, e.g.][]{Howard1970} and radial
\cite[via helioseismology, e.g.][]{Schou1998} directions.  The
internal radiative core of the Sun ($r<0.67R_\odot$) is
well-approximated by a solid body rotation of $P_{\rm rot}\approx
27$ days. In contrast, beyond the transition region between the
radiative and outer convective zones at $r\approx 0.7R_\odot$
(so-called tachocline), the rotation period $P_{\rm rot}$ becomes
significantly dependent on the latitude \citep{Thompson2003}, varying from $\approx 25$days
($\ell=0^\circ$) to $\approx 32$days ($\ell=60^\circ$).

Such detailed rotational profiles, however, are poorly understood for
other solar-like stars except the Sun. Furthermore, the Sun shows a
magnetic activity cycle of 11-years, that is thought to be related
to near-surface convection and stellar rotation
\cite[e.g.][]{Thompson2003,Hartmann1987}. Due to this link, the large
range of activity observed among stars suggest caution when
interpreting the measures of rotation periods.

With the high precision photometry from Kepler and TESS, measurements
of stellar rotation periods have been significantly improved in both
quantity and accuracy. Since the starspots over the stellar surface
are mainly responsible for the variation of stellar light intensity,
the surface rotation period may be inferred from the \rev{intensity
  variability analyses} using the Lomb-Scargle periodogram
\cite[e.g.][]{Nielsen2013}, auto-correlation function \cite[ACF,
  e.g.][]{McQuillan2013,McQuillan2014} and wavelet analysis
\cite[e.g.][]{Garcia2014,Ceillier2016}.

Note that those photometric measurements are sensitive to the observed
distribution of spots on the stellar surface.  The number of spots
varies in a time-dependent manner due to the continuous creation and
annihilation processes, and those spots spread over a range of
different latitudes. All of them inevitably complicate the
correspondence between the observed photometric variation and the
stellar rotation in particular due to the stellar differential
rotation.  Furthermore, the photometric variation is produced solely
by those spots located at the range of latitudes on the hemisphere
visible to the observer.  Thus, the measured rotation periods should
depend on both the stellar inclination relative to the observer's
line-of-sight and the non-stationary latitudes of the visible spots as
discussed in detail by \citet{SSNB2022}.

Spectroscopic data of the stellar lines measure the projected
rotation velocity $v\sin i_*$.  If additional information for the
stellar radius $R$ and inclination $i_*$ is available, it also
provides an independent estimate for the stellar rotation period;
further discussion is given in \citet[e.g.][]{Kamiaka2018}.

Yet another complementary estimate for the stellar rotation period can
be derived from asteroseismic analysis, which models the stellar
oscillation pattern in power spectrum
\rev{\cite[e.g.][]{Kjeldsen1995,Gizon2004,
    Appourchaux2008,Kamiaka2018,Garcia2019}}. Unlike the the
photometric variation of the stellar light curves, asteroseismic
analysis estimates the averaged rotation period over a range of
stellar latitude and radius. In this sense, the stellar rotation
periods estimated with the two complementary methods do not have to be
identical, depending on the degree of the differential rotation. In
turn, the comparison of the two results for a sample of stars puts
interesting constraints on the differential rotation, if the
reliability of both measurements is confirmed.

It should be noted that the rotation profile of stars has been
investigated using the high-quality photometry. \cite{Reinhold2015}
attempted to detect surface differential rotation by studying
significant peaks in the Lomb-Scargle periodogram for more than ten
thousand Kepler stars. In addition, \cite{Benomar2018} measured the
latitudinal differential rotation in the convection zones for 40
Sun-like stars by asteroseismology.

In this paper, we compute the rotation periods using three independent
photometric methods \rev{({\it i.e.,} intensity variability analyses
  using the Lomb-Scargle periodogram, auto-correlation function, and
  wavelet)} for 92 Kepler Sun-like stars, which were selected as
targets for asteroseismic analysis by \cite{Kamiaka2018}.  \rev{The
  present paper performs systematic and complementary analyses of the
  intensity variability of the same stars, and \rev{compares the different methods for measuring} 
  the stellar
  rotation periods. The detailed comparison against their
  asteroseimically determined rotation periods by \citet{Kamiaka2018}
  enables us to understand the reliability and possible biases of the
  rotation periods measured from different methods on an
  individual basis.}

\cite{Nielsen2015} compared the rotation periods derived from the
Lomb-Scargle and asteroseismic analyses, $P_{\rm LS}$ and $P_{\rm
  astero}$, for their six Sun-like target stars, and found that they
agree within the measurement uncertainty. We expand the sample size to
92 Kepler solar-type stars, and compute the stellar rotation period
from three photometric methods. The reliability of such photometric
rotation periods can be tested through the agreement among the three
independent methods, while their variation in different quarters over
four years may probe the latitudinal differential rotation as proposed
by \citet{SSNB2022}. In addition, comparison against the asteroseismic
estimates of the rotation periods may constrain the stellar
inclination \citep{Kamiaka2018,Kamiaka2019,Suto2019}.  Thus, the
result enables us to evaluate the reliability of the measurements, as
well as to extract the possible signature of, and/or put constraints
on, their differential rotations.

The rest of the paper is organized as follows. Section 2 presents our
sample of 92 Sun-like stars selected from the Kepler data, and the
flow of our data processing. We identify 22 possible binary/multiple
star systems. In section 3, we classify the remaining 70 stars into
four groups according to the variance of the photometric rotation
periods from the Lomb-Scargle analysis in different quarters, and the
fractional uncertainties in the asteroseismic periods.  Section 4
compares photometric and asteroseismic rotation periods for the 70
stars individually, while their statistical comparison among
photometric analyses and between photometric and asteroseismic results
is given in sections 5 and 6, respectively. Section 7 presents
implications of our results.  Finally, section 8 is devoted to summary
and conclusions of the paper. In appendix, we describe our methods to
estimate the rotation period, various comparison against previous
results, the Lomb-Scargle periodograms and asteroseismic
constraints for potentially interesting stars that are mentioned in
the main text.

\section{Our sample and data processing of light curves \label{sec:sample}}

\subsection{Our sample of stars}

\begin{figure}[t]
\centering
\includegraphics[width=13cm]{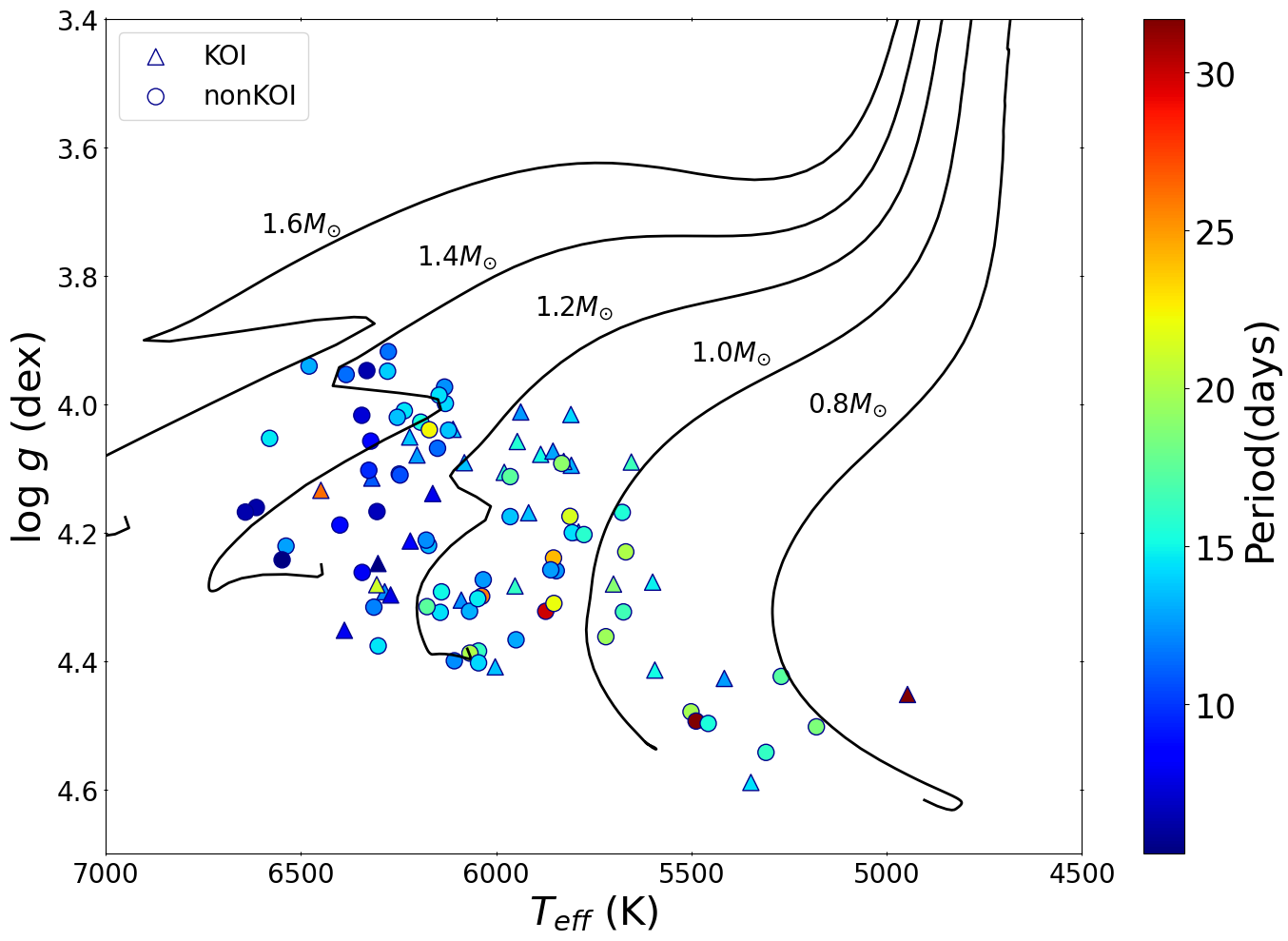}
\caption{Log surface gravity in function of effective
    temperature for the 92 stars of our sample. Colors indicate the
    period measured with the LS method ($\mPlsq$). Triangles are for
    KOI stars and circles for stars without planets. Indicative
    evolution tracks are at solar metallicity (solid black lines) for
    masses between 0.8$M_{\odot}$ and 1.6$M_\odot$.}
\label{fig:HR}
\end{figure}

We use the sample of stars selected by \citet{Kamiaka2018}. The sample
includes the \textit{Kepler} dwarfs LEGACY sample which consists of 66
stars with the best asteroseismic data so far in solar-type category
\citep[see e.g.][]{Lund2017, Aguirre2017}, and additional 28 KOI
(Kepler Object of Interest) with detectable oscillation
amplitudes. \rev{The Kepler }\rev{targets in the LEGACY sample have 
  magnitudes around 9, which \rev{are} in general brighter than those for
  the KOI sample (Kepler magnitude around 11). Thus, the LEGACY
  sample has spectra with higher height to background ratio and hence
  their periods can be estimated more precisely. The additional 28 KOI
  targets are useful to examine whether the stellar inclinations are
  correlated to the orbital axis of their transiting planets.}
\cite{Kamiaka2018} performed an asteroseismic analysis of this sample,
and estimated their stellar rotation periods and inclinations. Figure
\ref{fig:HR} plots the location of our targets on their surface
gravity and effective temperature plane.  Stellar evolutionary tracks
are computed with MESA code (Modules for Experiments in Stellar
Astrophysics; \citet{Paxton:2011aa}) for stars with initial mass of
0.8, 1.0, 1.2, 1.4, and 1.6 $M_{\odot}$ from their pre-main-sequence
phase to white dwarf phase, assuming the solar metallicity.

For the photometric analysis, we use the \textit{Kepler} long-cadence
PDC light curve \citep[PDC-msMAP,][]{Stumpe2014}, Data Release 25,
from Mikulski Archive for Space Telescope (MAST). The data consist of
quarters of $\sim 90$day duration with a cadence of $29.4$ minutes
over four years. We choose the light curves for Q2-Q14
quarters.

KOI-268 (KIC 3425851) has one of the best-quality light curves in our
sample and a precisely determined photometric period (see Figures
\ref{fig:LS-ACF-WA} and \ref{fig:koi268-LSQ} below).  We first compute
the \rev{duty cycle} for KOI-268 in each quarter, and find that it
varies between 81 and 96 percents, with an average $\sim$90 percents.
Then, we remove those quarters for other stars if the \rev{duty
  cycle} is less than $75\%$ of the reference value for KOI-268, in
order to secure the statistical significance of each
quarter. Furthermore, we require that the targets must have at least 7
quarters of available data, which removes 16 \rev{Cyg} A (KIC
12069424) and 16 \rev{Cyg} B (KIC 12069449) from our sample. Thus,
the number of our final targets for subsequent analyses becomes 92.

\subsection{Preprocessing stellar light curves}\label{subsec:lcprocessing}

\begin{figure}[t]
\centering
\includegraphics[width=16cm]{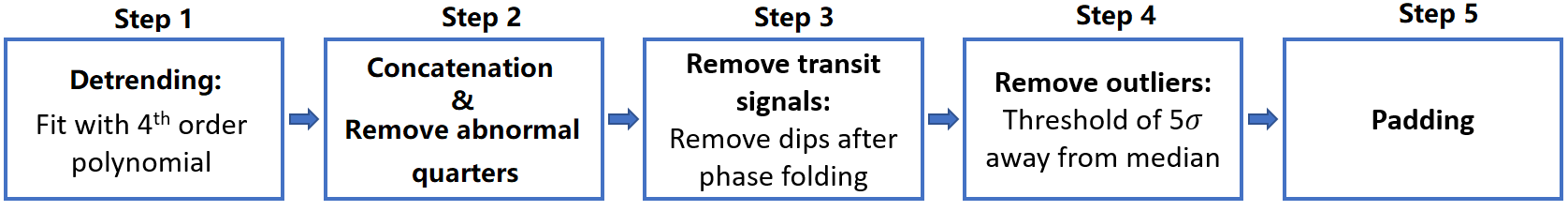}
\caption{Flow chart of our preprocessing of stellar light curves.}
\label{fig:flow-process}
\end{figure}

We preprocess the light curves of the 92 stars before performing the
photometric analysis, which is summarized in Figure
\ref{fig:flow-process}.

In order to remove the systematic trend in the light curves, we first
fit the original data $L_{\rm init}(t)$ in each quarter to a 4-th
order polynomial function $p_4(t)$. Then, we obtain the detrended
light curves in units of the fitted polynomials separately in each
quarter: $L(t) = L_{\rm init}(t)/p_4(t)-1$ (Step 1 of Figure
\ref{fig:flow-process}). \rev{In order to remove a long-term trend in a
  light curves, it is fairly common to fit using polynomials. We make
  sure that detrending with a third-order \rev{or} fourth-order
  polynomials does not change the estimate of the rotation periods
  unless they are longer than $40$ days.  Since the duration of each
  quarter is 90 days} \rev{and the Kepler instrument performs pointing corrections every 30 days which are known to generate a recurrent noise around that period}\rev{, periods longer than $40$ days are not reliable
  in any case. Therefore, we decide to use the fourth polynomials in
  detrending.}

Next, we concatenate the detrended light curves in available quarters
between Q2 and Q14 for each target star into a single array of time
series while preserving gaps between the quarters (Step 2 of Figure
\ref{fig:flow-process}). We set the starting time of this series to be
$t=0$.

\begin{figure}[t]
\centering
\includegraphics[width=14cm]{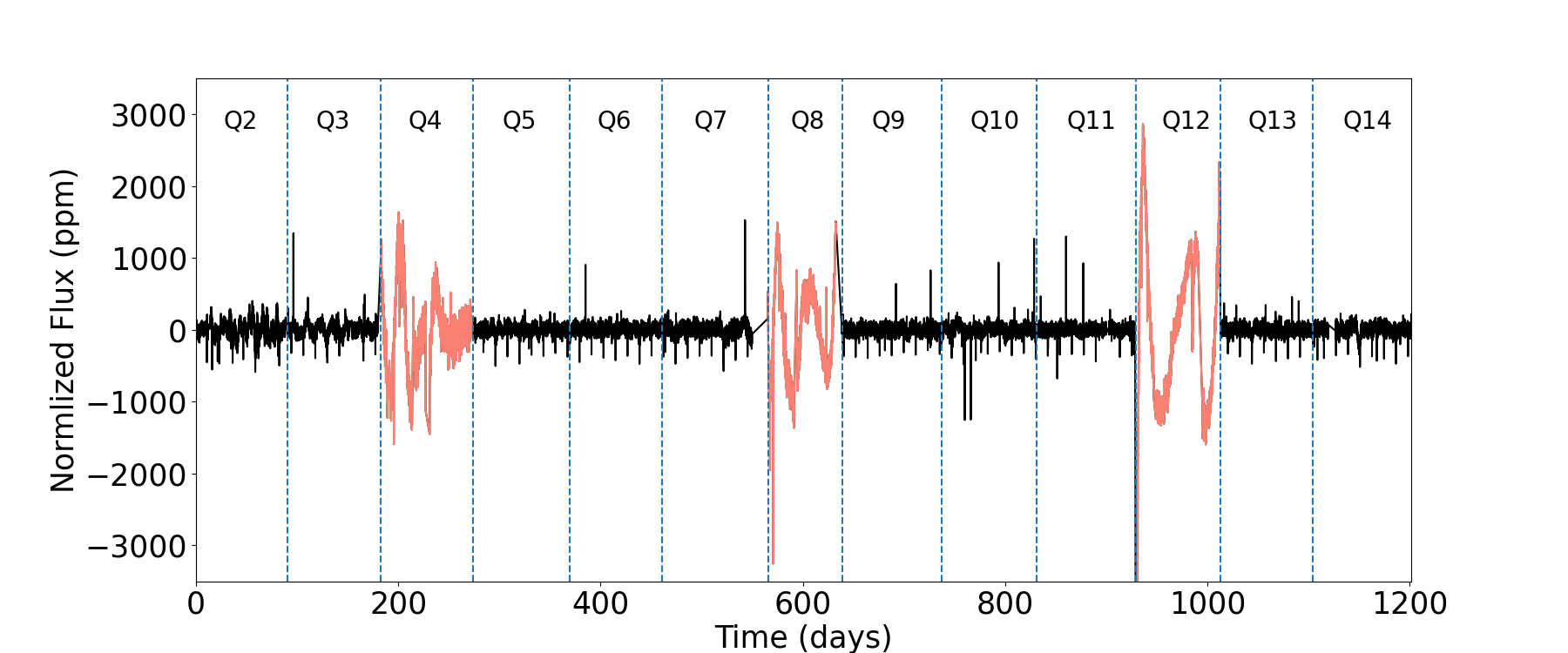}
\caption{The detrended light curve of \textit{Kepler}-1655 (KIC
  4141376, KOI-280) for Q2 to Q14 quarters.  We remove the three
  quarters, Q4, Q8, and Q12, in the subsequent photometric analysis,
  which exhibit anomalous variations illustrated in red.}
\label{fig:KIC4141376}
\end{figure}

For several stars, the detrended light curves still exhibit large
anomalous variations in some quarters. Figure \ref{fig:KIC4141376}
illustrates such an example for \textit{Kepler}-1655 (KIC 4141376,
KOI-280). The light curves in quarters Q4, Q8, and Q12 (plotted in
red) show strange behavior that seems to be uncorrelated with nearby quarters.  We suspect that they are due to unknown
contamination or instrumental effects.  Thus, we decided to remove
those quarters if the median absolute flux variation in a particular
quarter is three times larger than those of the neighboring
quarters. We removed such anomalous quarters from six targets:
\textit{Kepler}-1655 (KIC 4141376, KOI-280), \textit{Kepler}-100 (KIC 6521045),
\textit{Kepler}-409 (KIC 9955598, KOI-1925), KOI-72 (KIC 11904151), KIC
8694723, and KIC 10730618.

As shown in Figure \ref{fig:KIC4141376}, the detrended light curves
  also contain the dips due to the planetary transits (for KOI stars)
  and other possible outliers. We remove them according to the
  following procedure (Steps 3 and 4 of Figure
  \ref{fig:flow-process}).

In order to remove the periodic dips due to the planetary transit, we
\rev{first fold the light curves of the KOI sample according to the
  orbital period of \rev{the} planet candidates. Then we remove the data
  corresponding to the transit and occultation phases.  The
  occulations might not be visible for all targets. For those
  targets we remove the expected occultations by assuming a circular
  orbit. The duration of transits and occultations is of the order of
  hours as opposed to the orbital period of the order of days. Thus,
  the amount of data removed is relatively negligible, and does not
  change the estimate of the stellar rotation period in practice.}
In addition, \rev{we
  remove the other outliers by $\sigma$-clipping with a threshold
  of $5\sigma$ and the center value being the median of concatenated
  light curve. The data removed in this step is less than 1\%, and
  thus do not affect the period estimation either.}

Incidentally, we are not able to detect the transit signal for
KOI-5.02 and \textit{Kepler}-37b after folding, and we do not remove any
transit dips nor the expected eclipse dips.  We suspect that KOI-5.02
is a false-positive, and \cite{Barclay2013} reported that the transit
signal of \textit{Kepler}-37b is less than 20 ppm.  Thus possible transiting
signals of these two targets would have negligible influence on the
determination of stellar rotation period.

We note that the 90 day duration of each Kepler quarter effectively
puts an upper limit of detectable periods around 45 days. Indeed, the
rotation periods of main-sequence stars similar to our sample (Figure
\ref{fig:HR}) are usually less than 50 days
\citep[e.g.][]{Kamiaka2018}. So, we apply our own box-car filter of a
50-day width even though the PDC-SAP light curves has applied
smoothing to reduce long-term period systematics.  Since the \textit{Kepler}
data is nearly evenly sampled, we map the light curve to a uniformly
sampled grid at an interval of $\delta t = 29.4$ minutes.

Finally, we pad all removed portions in the light curves (removed
quarters, planetary transits, secondary eclipses, outliers, and
missing data) by Gaussian noise of zero mean and the variance of
$\sigma_{\rm noise}$ that is computed from the detrended light curve
for each target (Step 5 of Figure \ref{fig:flow-process}).  We also
tried the zero padding for comparison, and made sure that the two
padding schemes hardly change the estimated values of the photometric
rotation period, implying that the padding does not affect the result
in reality.

\subsection{Identification of Possible Binary/Multiple-star Systems
\label{subsec:binarity}}

A significant fraction of stars are in multiple systems. The
multiplicity may lead to a systematic bias on the rotation period
estimated from the photometric variation of the target star.  Thus, we
try to identify possible members in binary and multiple-star systems
in our sample.

For that purpose, we primarily adopt Renormalized Unit Weight Error
(RUWE) value from Gaia DR3, which measures the goodness of fit with
respect to a single-star model derived from the Gaia astrometry
\citep{Fabricius2021}.  A star with RUWE exceeding 1.4 is commonly
suspected to have an unresolved companion\citep{Evans2018,
  Wolniewicz2021}. We list 14 target stars with RUWE $\geq 1.4$ from
the Gaia DR3 catalog as possible candidates of binary/multiple-star
systems.

In addition, we identify three eclipsing binaries from \textit{Kepler}
Eclipsing Binary Catalogue \citep{keplerEBurl},
two binary systems from NASA Exoplanet Archive \citep{NEA6},
four multiple-star systems and two binaries from Table 4 of
\cite{Garcia2014}, one binary from \cite{VanEylen2014}, and one
binary from \cite{LilloBox2014}. Among the 13 systems, five
stars have RUWE $\geq 1.64$, while the other eight stars have RUWE
$\geq 1.05$.

In total, we identify 22 stars with possible companions, and summarize them in Table \ref{tab:multiplestar}. We compute their
photometric rotation periods in the same manner as other 70 stars
with no reported stellar companion (Appendix \ref{sec:method}), but
they may be interpreted with caution. We separately discuss seven KOI
stars out of them in \S \ref{subsec:discuss-multi} since they may be
potentially important targets for planetary studies.  In what follows,
however, we mainly focus on the other 70 stars; their estimated
rotation periods are summarized in Tables \ref{tab:singlestarAa},
\ref{tab:singlestarAb}, \ref{tab:singlestarBa} and
\ref{tab:singlestarBb}.

\section{Classification of stars} \label{sec:classification}

We measure rotation periods using three different photometric methods,
including the Lomb-Scargle (LS) periodogram, Auto-correlation function
(ACF) and wavelet analysis (WA). The details of the photometric
methods as well as a brief introduction to the asteroseismic analysis
by \citet{Kamiaka2018} are provided in Appendix \ref{sec:method}. In
this section, we classify the 70 stars based on photometric and
asteroseismic results, in order to examine the reliability of the
measured rotation period and/or to constrain the differential
rotation.  We adopt the LS measurement for the photometric
classification, and use ACF and WA as complementary information to
test the robustness of the derived rotation period.

\subsection{Photometric classification based on the variance of
  rotation periods in different
  quarters \label{subsec:LS-classification}}

The rotation period estimated using the LS method for the $q$-th
quarter $\Plsq$ may be different from the period $P_{\rm LS}$ computed for the
entire observed duration $T_0$.  Distribution of starspots at
different quarters should change with time due to the
creation/annihilation processes over a range of latitudes.  Thus,
combined with their possible latitudinal differential rotation, each
quarter may exhibit the fractional variation of the period on the
order of ten percent.

If the typical lifetime of starspots is less than $90$ days, their
pattern on the surface at different quarters can be regarded as an
uncorrelated realization drawn from the same statistical distribution.
Even if the lifetime of starspots has a broad distribution, the
variance of $\Plsq$ among different quarters would reflect the
differential rotation for stars. This possibility has been
theoretically studied by \citet{SSNB2022} using analytic model and
mock observations.  It is also possible that the large variance is
simply due to the low signal-to-noise ratio ({\it e.g.,} due to a very
low number of spots or to the fact that the star is very faint) of the
light curve for some quarters. These two possibilities could be
distinguished by comparing the LS result with
those based on ACF, WA, and asteroseismology on an individual basis (see
\S\ref{subsec:astero-classification} below). 

In order to quantify the fractional variance of the rotation periods
among different quarters, we introduce the following parameter $\Dls$
for each star:
\begin{equation}
 \label{eq:Dls}
  \Dls = \frac{1}{N_Q \mPlsq} \sum_{q}\left|\Plsq
  -\mPlsq\right|,
\end{equation}
where $N_Q$ is the number of quarters with measured $\Plsq$, and
$\langle\cdots\rangle$ denotes the median value operator.

\citet{Nielsen2013} proposed a similar indicator called MAD (Median
Absolute Deviation) $\langle\left|\Plsq -\mPlsq\right|\rangle$.  Our
indicator, equation (\ref{eq:Dls}), differs from MAD in the following
two ways. We use the {\it fractional} variance, which would be more
appropriate in classifying stars with a broad range of rotation
periods. In addition, we average the deviation over all the quarters
because the median of deviation sometimes underestimates the real intrinsic
variation of the distribution. This will be
discussed below individually for each target.

\begin{figure}[tbh!]
	\centering
	\subfigure{
		\begin{minipage}[t]{0.40\linewidth}
			\centering
			\includegraphics[width=1.0\linewidth]{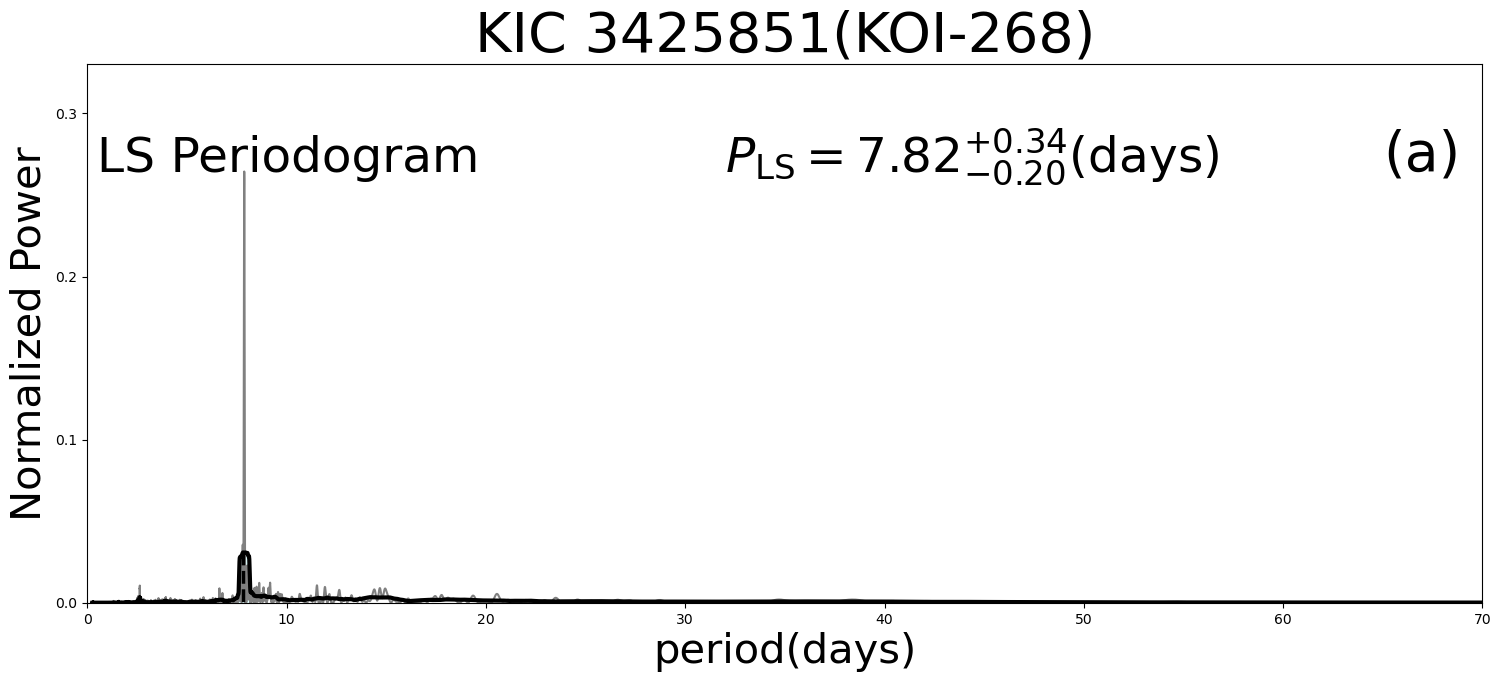}  
			\label{fig:spectrab}
		\end{minipage}
	}\vspace{-0.5em}
	\subfigure{
		\begin{minipage}[t]{0.40\linewidth}
			\centering
			\includegraphics[width=1.0\linewidth]{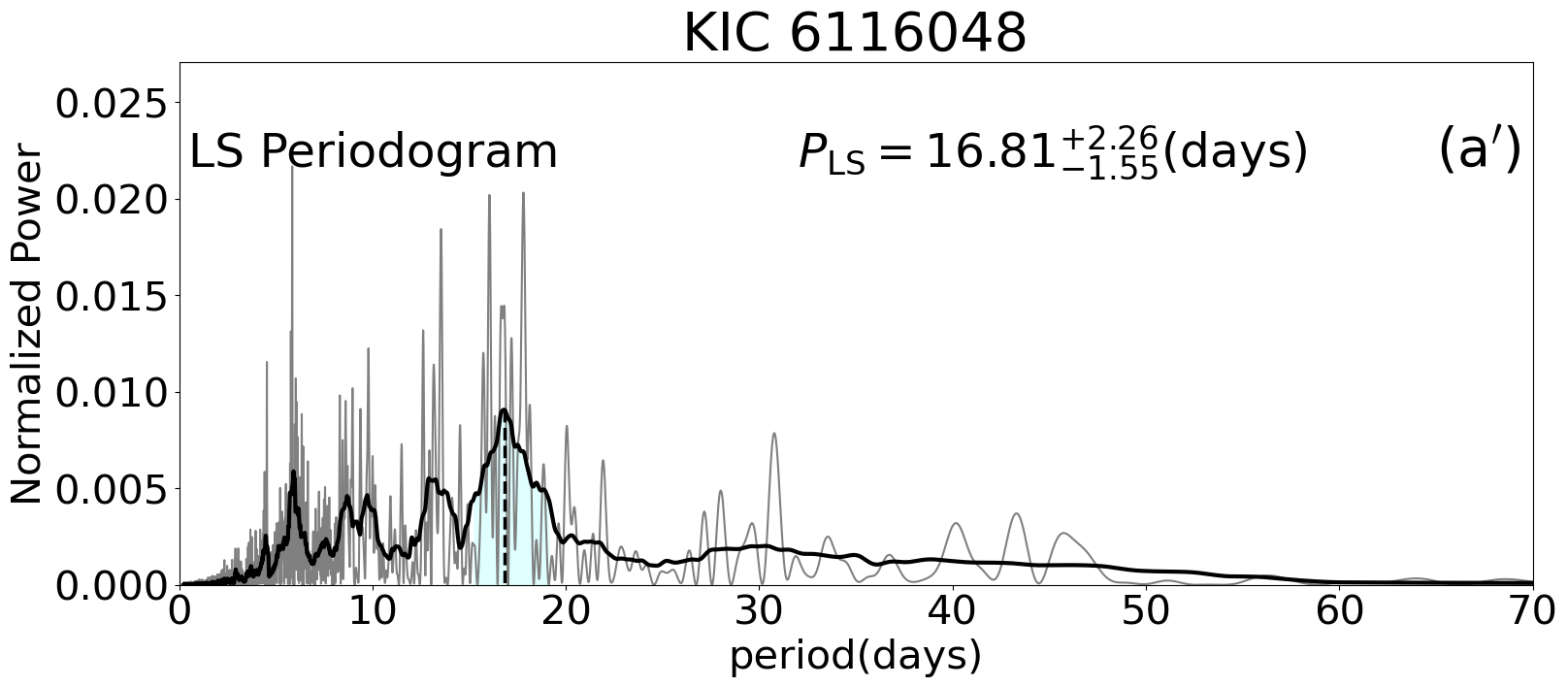}  
			\label{fig:spectrac}
		\end{minipage}
	}\vspace{-0.5em}
	\subfigure{
		\begin{minipage}[t]{0.40\linewidth}
			\centering
			\includegraphics[width=1.0\linewidth]{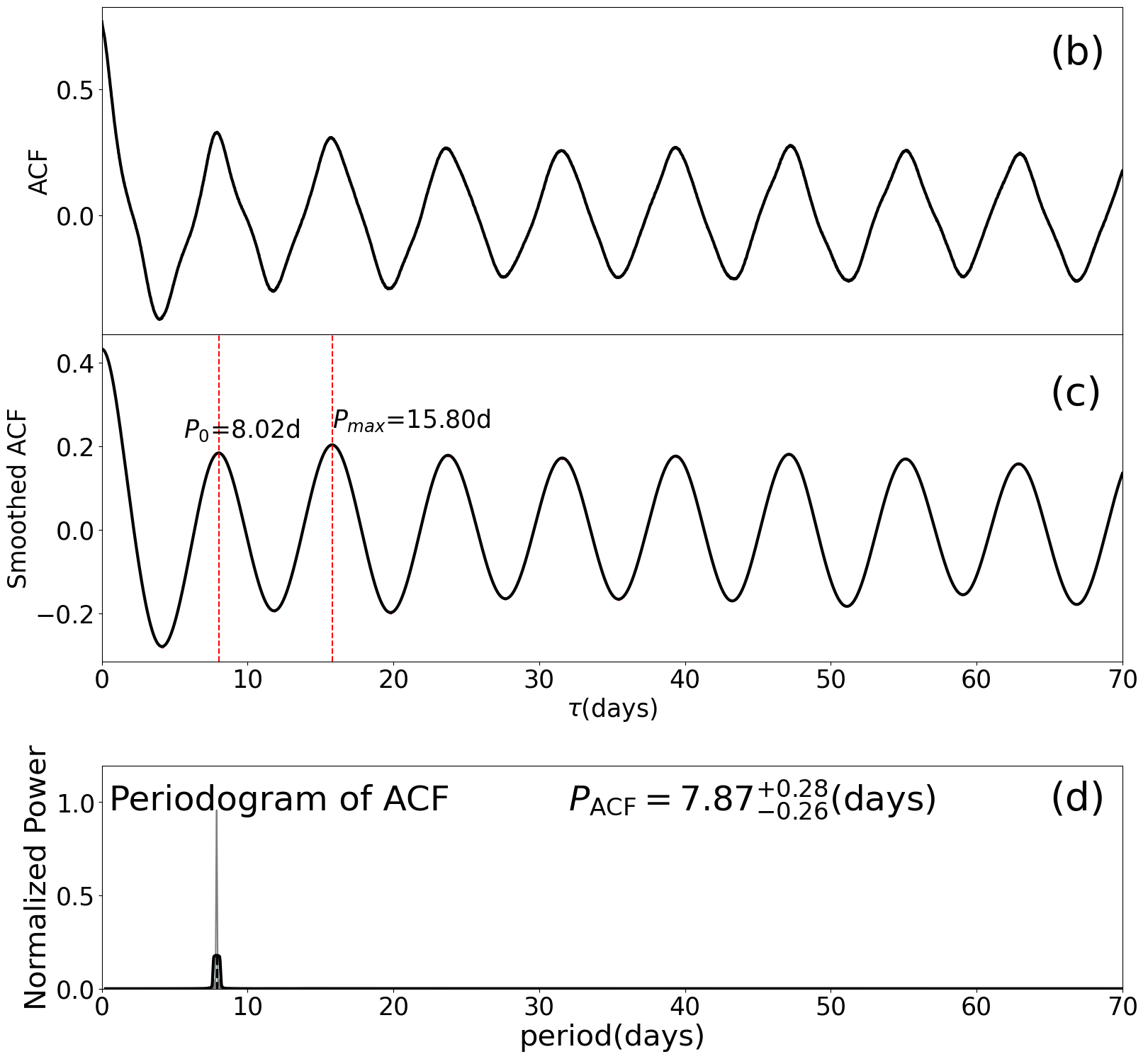}  
			\label{fig:spectrae}
		\end{minipage}
	}
	\subfigure{
		\begin{minipage}[t]{0.40\linewidth}
			\centering
			\includegraphics[width=1.0\linewidth]{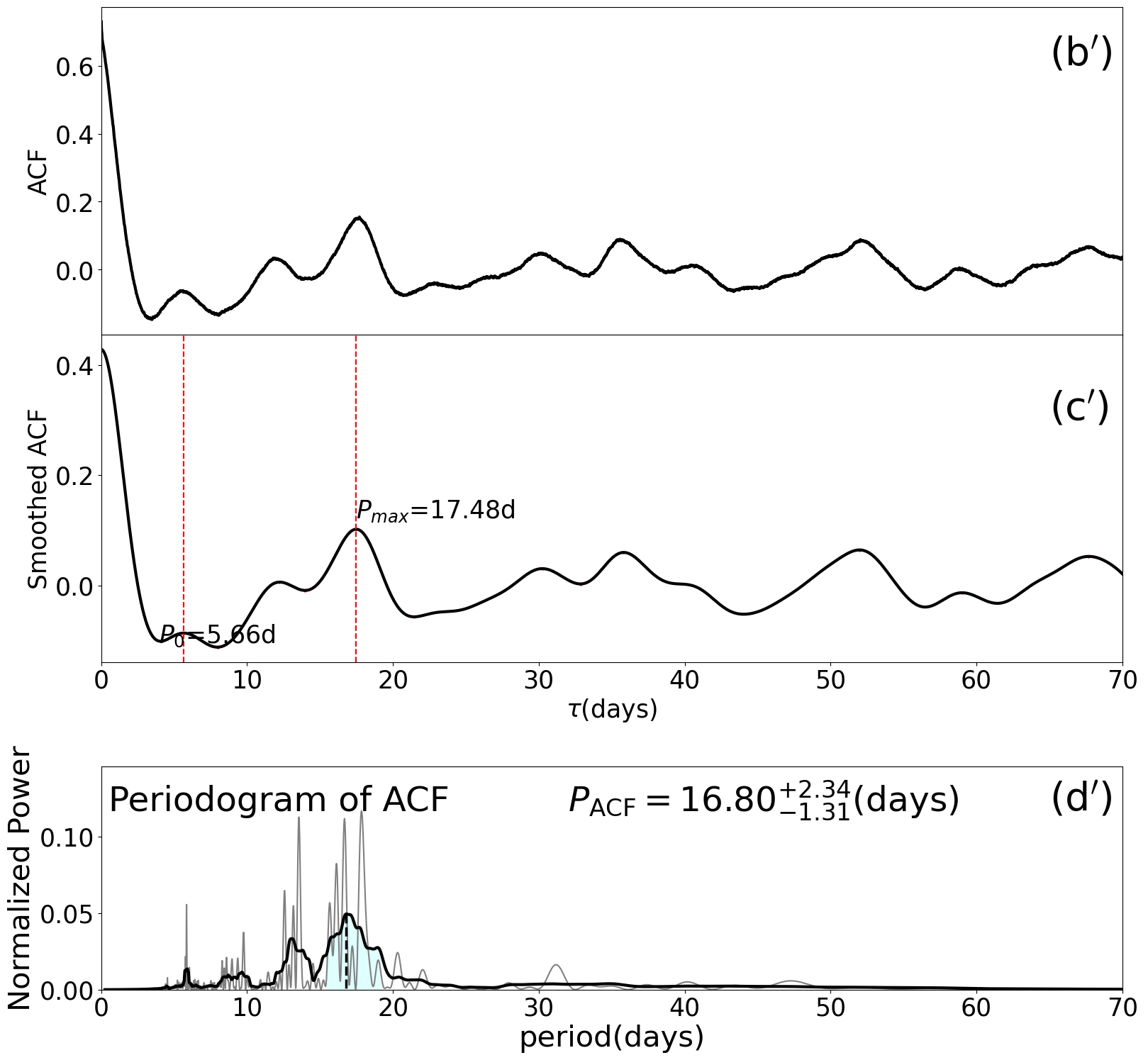}  
			\label{fig:spectraf}
		\end{minipage}
	}\vspace{-0.5em}
	\subfigure{
		\begin{minipage}[t]{0.44\linewidth}
			\centering
		\includegraphics[width=1.0\linewidth]{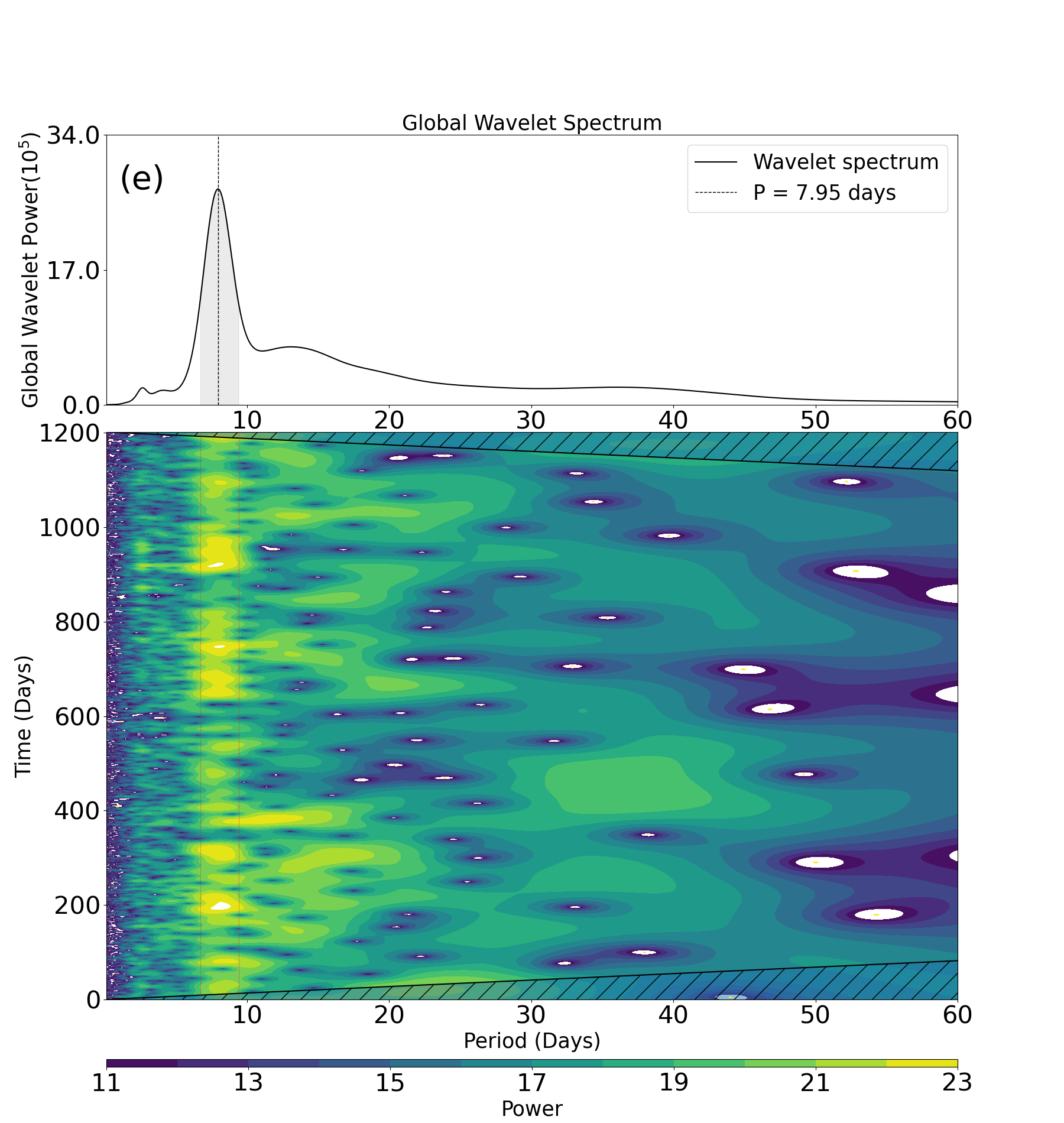}   
			\label{fig:spectrah}
		\end{minipage}
	}
	\subfigure{
		\begin{minipage}[t]{0.44\linewidth}
			\centering
		\includegraphics[width=1.0\linewidth]{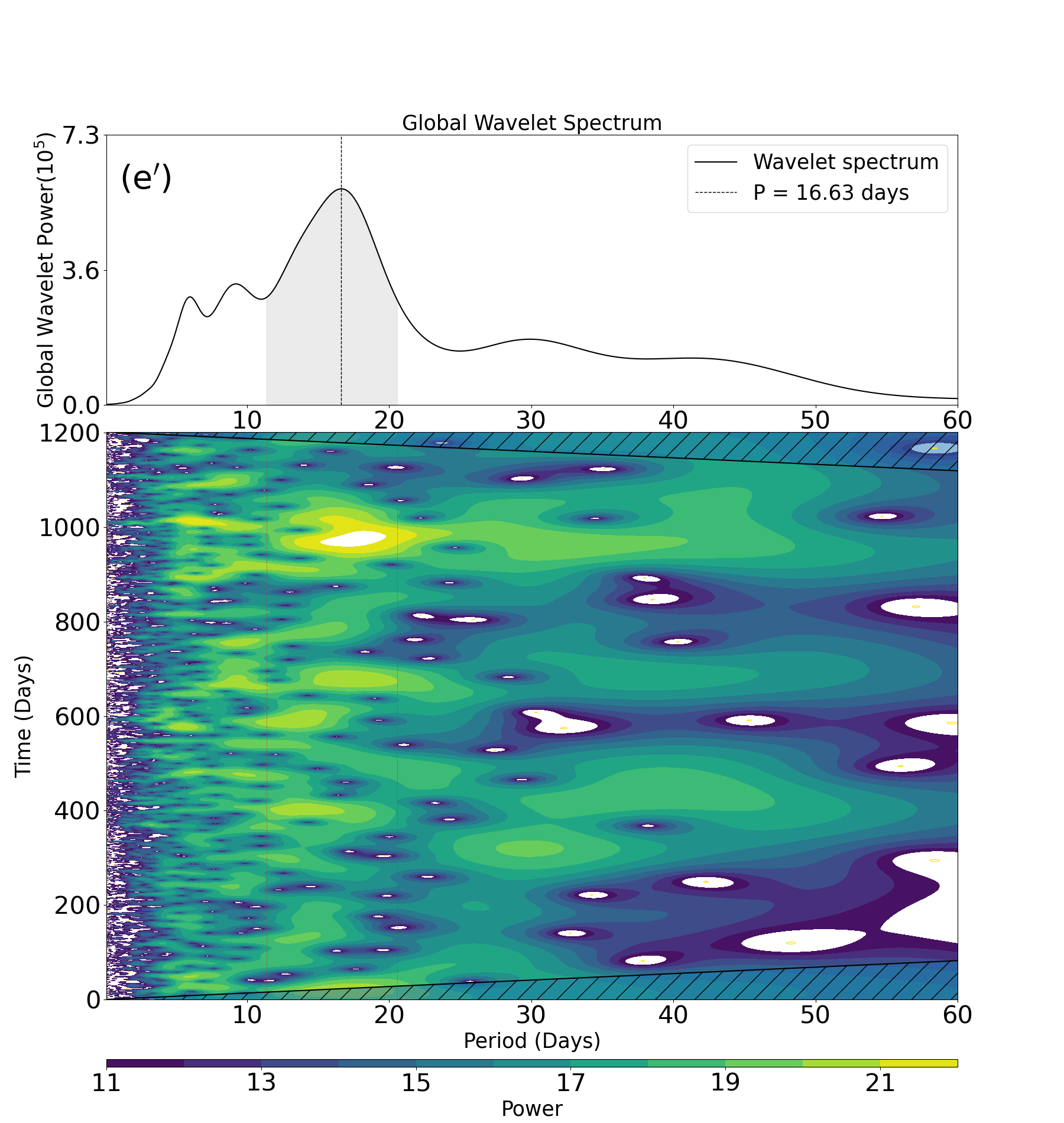}   
			\label{fig:spectrai}
		\end{minipage}
	}
\caption{Results for KOI-268 (KIC 3425851; Group A with $\Dls =
  0.014$; panels a to e) and KIC 6116048 (Group B with $\Dls = 0.372$;
  panels a' to e').  Panels a and a' show the LS periodogram. The
  middle panels are the ACF (b,b'), smoothed ACF (c,c'), and the
  LS periodogram of the smoothed ACF (d,d'), where $P_0$ denotes the
  period associated to the first maximum and $P_{max}$ labels the
  period associated to the highest maximum. Panels e and e' show the
  WPS and GWPS from wavelet analysis. }
    \label{fig:LS-ACF-WA}
\end{figure}

After trial-and-errors, we classify the stars into two groups: Group A
with $\Dls \leq 0.2$ and Group B with $\Dls > 0.2$. \rev{To select the
  threshold, we first plot the rotation period against $\Dls$, and
  find that a group of targets with $P_{\rm LS} >$ 30 days and significantly large error-bars start to occur from $\Dls \sim 0.2$. Thus, we decide to set a
  threshold of $\Dls$= 0.2 for a reliable group just for
  convenience}. There are 23 stars in Group A and 47 stars in Group B,
out of the 70 stars with no reported stellar companion; see
\S\ref{subsec:binarity}.  Figure \ref{fig:LS-ACF-WA} presents results
of our photometric analysis for the reference star KOI-268 (KIC
3425851; see section \ref{sec:sample}) that is classified in Group A
with $\Dls = 0.014$, as well as for KIC 6116048 (Group B with $\Dls =
0.372$).

Group A contains targets with a relatively robust periodic component
in their light curves, and their $P_{\rm LS}$ values are reliably estimated
from the prominent single peak as shown in the left panels of Figure
\ref{fig:LS-ACF-WA}.

In contrast, Group B, with $\Dls > 0.2$, consists of two different
types in general. The first type shows either a variation of measured
periods among different quarters or multiple co-existing periodic
signals (see {\it e.g.,} the right panels of Figure
\ref{fig:LS-ACF-WA}). Possible scenarios for the latter include
differential rotation on stellar surface, harmonics (see \S
\ref{subsubsec:LS}), unresolved companions, residuals of a systematic
trend, and other unknown contaminations.  The second type
shows very weak periodic signals, whose estimated periods are less
reliable and subject to significant uncertainties.

Note that the large variation of period measurements for the first
type of targets in Group B may not necessarily indicate that their
rotation periods are unreliable, but could be due to the physical
variation on stellar surface such as the differential rotation.

\subsection{Classification based on asteroseismic analysis
\label{subsec:astero-classification}}

We also classify our targets based on asteroseismically estimated
periods by \cite{Kamiaka2018}, in a complementary manner to the
photometric classification.  For that purpose, we adopt the ratio
$\Delta P_{\rm astero}/P_{\rm astero}$, where $\Delta P_{\rm astero}$
is the sum of upper and lower error of period estimation, {\it i.e.,}
$\Delta P_{\rm astero}= (\Delta P_{\rm astero})_++(\Delta P_{\rm
  astero})_-$; see equations (\ref{eq:dPastero+}) and
(\ref{eq:dPastero-}) in \S \ref{subsec:asteroseismic}.

We set the threshold value of $0.4$ after trial-and-errors, and
introduce two groups; Group a with $\Delta P_{\rm astero}/P_{\rm
  astero} \leq 0.4$ has 33 stars, while Group b with $\Delta
P_{\rm astero}/P_{\rm astero}>0.4$ has 37 stars.

\begin{table}[th]
\centering
\begin{tabular}{|l|l|l|l|}\hline
  \diagbox[width=\dimexpr
    \textwidth/4+2\tabcolsep\relax, height=1.5cm]{photometry}{asteroseismology}
& Group $a$ & Group $b$ &Total \\ \hline
		Group $A$ & 14 (7)  & 9 (4)   & 23 (11) \\
		Group $B$ & 19 (3)  & 28 (12) & 47 (15) \\ \hline
		Total     & 33 (10) & 37 (16) & 70 (26) \\ \hline
\end{tabular}
\caption{Groups A and B include stars with $\Dls \le 0.2$ and
  $>0.2$ respectively from the LS result, while groups a and b contain stars
  with $\Delta P_{\rm astero}/P_{\rm astero} \le 0.4$ and $>0.4$ respectively from
  the asteroseismic result.  The numbers in the parenthesis indicate the number of stars
  with planetary candidates (KOI catalogue).
  \label{tab:fourgroups} }
\end{table} 

\begin{figure}[ht]
\centering
\includegraphics[width=12cm]{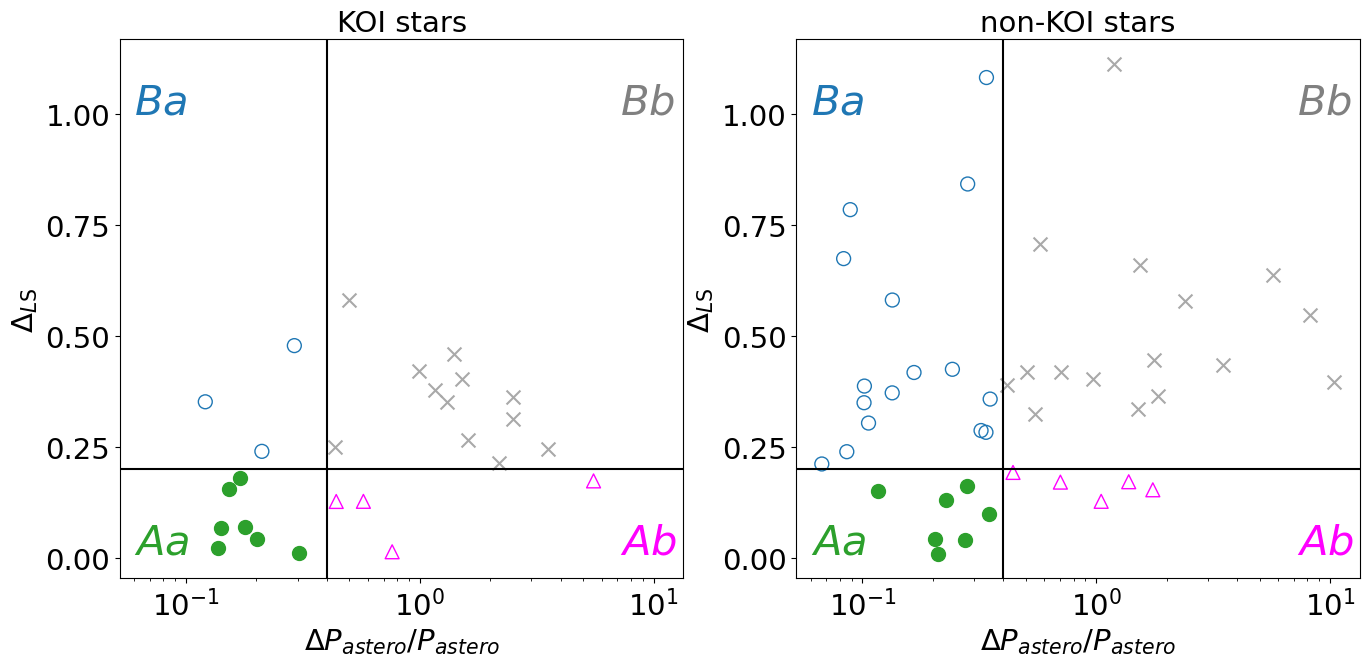}   
\caption{Classification of 70 targets on the $\Delta
  P_{astero}/P_{astero}$ -- $\Dls$ plane. The solid lines indicate
  the thresholds, $\Dls = 0.2$ and $\Delta P_{\rm astero}/P_{\rm
    astero}=0.4$, that we adopt in the classification.  Green filled
  circles, blue open squares, magenta open triangles, and gray
  crosses correspond to stars in Group Aa, Ba, Ab, and Bb,
  respectively.}
	\label{fig:classall}
\end{figure}


\begin{figure}[ht]
\centering
\includegraphics[height=5cm]{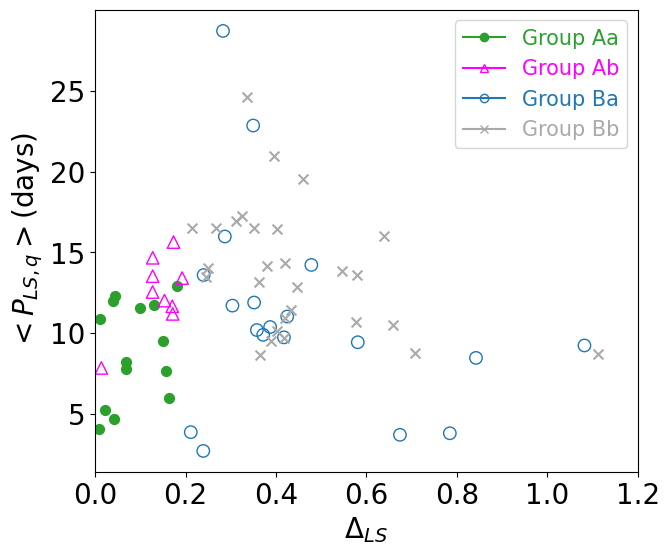}
\includegraphics[height=5cm]{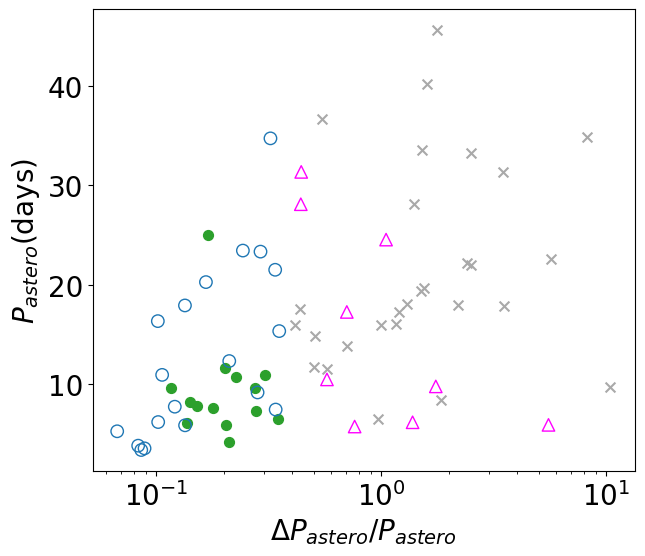}
\caption{$\mPlsq$ against $\Dls$ (left) and $P_{\rm astero}$ against
  $\Delta P_{\rm astero}/P_{\rm astero}$ (right) . Green, blue,
  magenta, and gray symbols correspond to Groups Aa, Ba, Ab, and Bb,
  respectively.
	\label{fig:PLS-DLS}}
\end{figure}

Combining the photometric and asteroseismic classification yields four
groups labeled as Aa, Ab, Ba, and Bb, which are summarized in Table
\ref{tab:fourgroups}.  Figure \ref{fig:classall} plots the locations
of 70 stars on the $\Delta P_{\rm astero}/P_{\rm astero}$ -- $\Dls$
plane, and the left and right panels of Figure \ref{fig:PLS-DLS} show
the correlations between $\mPlsq$ and $\Dls$, and between $P_{\rm
  astero}$ and $\Delta P_{\rm astero}/P_{\rm astero}$, respectively.

\section{Remarks on individual stars \label{sec:remarks}}

We plot the rotation periods for 70 stars with no obvious stellar
companion in Figure \ref{fig:Aa-KOI} (7 KOI stars in Group Aa), Figure
\ref{fig:Aa-nonKOI} (7 non-KOI stars in Group Aa), Figure
\ref{fig:Ab-KOI} (4 KOI stars in Group Ab), Figure \ref{fig:Ab-nonKOI}
(5 non-KOI stars in Group Ab), Figure \ref{fig:Ba-KOI} (3 KOI stars in
Group Ba), Figure \ref{fig:Ba-nonKOI} (16 non-KOI stars in Group Ba),
Figure \ref{fig:Bb-KOI} (12 KOI stars in Group Bb), and Figure
\ref{fig:Bb-nonKOI} (16 non-KOI stars in Group Bb).

In those figures, gray crosses
indicate the values of $\Plsq$ for different quarters with the
orange vertical lines being their median value $\mPlsq$, while the
blue, green, red, and black dots with error-bars represent the
values of $P_{\rm LS}$, $P_{\rm ACF}$, $P_{\rm WA}$, and $P_{\rm
astero}$ with their uncertainties defined in section
\ref{sec:method}. Those values are summarized in appendix; Tables
\ref{tab:singlestarAa} (Group Aa), \ref{tab:singlestarAb} (Group
Ab), \ref{tab:singlestarBa} (Group Ba), and \ref{tab:singlestarBb}
(Group Bb). In those Tables, we quote the upper and lower error-bars
with respect to the median value $\mPlsq$ by
introducing
\begin{equation}
\label{eq:medianPls-error}
\Delta \mPlsq_+
\equiv \left<\Plsq-\mPlsq \right>_+,
\quad
\Delta \mPlsq_- \equiv \left<\mPlsq - \Plsq \right>_-, 
\end{equation}
where $\langle \cdots\rangle_+$ and $\langle \cdots\rangle_-$ denote
the median operators for quarters with $\Plsq > \mPlsq$ and $\Plsq <
\mPlsq$, respectively. For a majority of stars, equation
(\ref{eq:medianPls-error}) gives a very similar value of the Half
Width at Half Maximum (HWHM) of the highest peak computed from the LS
periodogram over the entire observation span. In some cases, however,
it becomes much larger reflecting the significant skewness of the
distribution of $\Plsq$. We adopt equation (\ref{eq:medianPls-error})
as a complementary measure to emphasize the presence of such skewness,
which are listed in Tables \ref{tab:singlestarAa},
\ref{tab:singlestarAb}, \ref{tab:singlestarBa} and
\ref{tab:singlestarBb}.

Before considering the statistical comparison and implications, we
comment on individual results for a few targets in different groups.
In the following discussion, we assume solid rotation for
simplicity, and estimate the stellar rotation period $P_{\rm rot}$
from the spectroscopically determined rotation velocity $v\sin i_*$:
\begin{equation}
\label{eq:Prot-vsini}
P_{\rm spec}= \frac{2\pi R}{v\sin i_*} \sin i_*
\approx 50.6 \sin i_* \left(\frac{R}{R_\odot}\right)
\left(\frac{1\, {\rm km~ s}^{-1}}{v\sin i_*}\right) ~{\rm days},
\end{equation}
and from the asteroseismically determined rotation frequency
$\delta\nu_*$:
\begin{equation}
\label{eq:Prot-dnu}
P_{\rm astero} \approx
11.6 \left(\frac{1\, \mu {\rm Hz}}{\delta\nu_*}\right) ~{\rm days}.
\end{equation}

\subsection{KOI stars in Group Aa: \textit{Kepler}-100 \label{subsec:kepler100}}
  
\begin{figure}[ht]
\centering
\includegraphics[width=10cm]{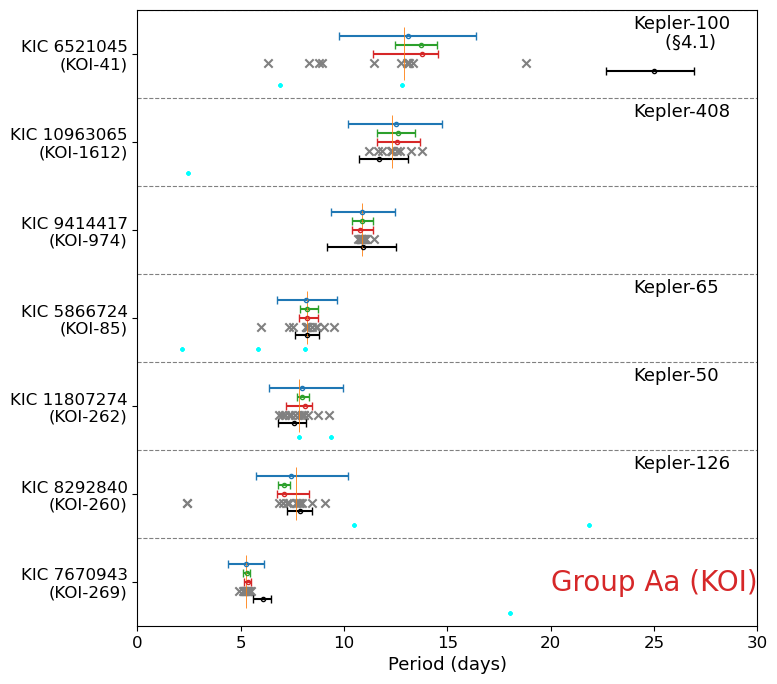}   
\caption{Rotation periods for seven KOI stars classified in Group
  Aa. We plot the values of $\Plsq$ for different quarters in gray
  crosses, with the orange vertical lines indicating their median
  value $\mPlsq$ for each target.  For comparison, we show the other
  three photometric rotation periods, $P_{\rm LS}$, $P_{\rm ACF}$,
  $P_{\rm WA}$, in red, green, and blue dots with error-bars
  (corresponding to their HWHMs), respectively.  Black dots indicate
  the asteroseismic results $P_{\rm astero}$ with quoted errors
  representing the 16 and 84 percentiles. For reference we plot the
  orbital periods of transiting planets in each system as cyan dots.}
	\label{fig:Aa-KOI}
\end{figure}

As Figure \ref{fig:Aa-KOI} clearly indicates, KOI stars in Group Aa
show good agreement among $\mPlsq$, $P_{\rm LS}$, $P_{\rm ACF}$,
$P_{\rm WA}$, and $P_{\rm astero}$.  Indeed, the major part of the
scatter around $P_{\rm LS,q}$ for those stars is dominated by a small
number of quarters with low signal-to-noise ratios.

The only exception is \textit{Kepler}-100 (KIC 6521045, KOI-41) with
$\Dls=0.18$ that is close to our threshold value of 0.2.  Its
photometrically estimated periods ($\mPlsq$, $P_{\rm LS}$, $P_{\rm
  ACF}$, and $P_{\rm WA}$) consistently point to $\approx 13$ days,
while $P_{\rm astero}\approx 25$ days. Intriguingly,
\citet{McQuillan2013} and \citet{Ceillier2016} concluded $P_{\rm ACF}
\approx 25$ days. Thus, we conduct a further examination of  \textit{Kepler}-100.

The LS periodograms of \textit{Kepler}-100 for different quarters (Figure
  \ref{fig:kepler100-LSQ}) shows apparent variations in measured period.  While it is not easy to define a robust period, six quarters
  (Q2, Q4, Q7, Q8, Q9, Q11) out the 12 quarters (except Q6) have the
  highest peaks around $13$ days. Our analysis excludes Q6
  because its median absolute flux variation is more than three times
  of those of the neighboring quarters (\S \ref{subsec:lcprocessing}).
  Nevertheless, the quarter has a very strong peak at $\Plsq \approx 25$
  days. We found that our own $P_{\rm ACF}$ becomes $\approx 25$ days
  if we include the Q6 data, which recovers the result
  of \citet{McQuillan2013} and \citet{Ceillier2016}.

We refer to spectroscopic measurement of rotational velocity.
According to Table 1 of \citet{Marcy2014}, \textit{Kepler}-100 have
$v\sin i_*=3.7 {\rm km s}^{-1}$ and $R=1.49\pm0.04R_\odot$. Thus
equation (\ref{eq:Prot-vsini}) implies that $P_{\rm spec} \approx
20.4\sin i_*$ days. A value of $\sin i_* \sim 0.6$ is required to
agree with $\mPlsq \sim 13$days, but it is inconsistent with the
asteroseismic result of $\sin i_* \approx 1$ and $P_{\rm
  astero}\approx 26.4^{+1.2}_{-1.7}$ days
($\delta\nu_*=0.44^{+0.03}_{-0.02} \, \mu$Hz; see Figure
\ref{fig:seismickepler100}).

\begin{figure}[ht]
\centering
\includegraphics[width=17cm]{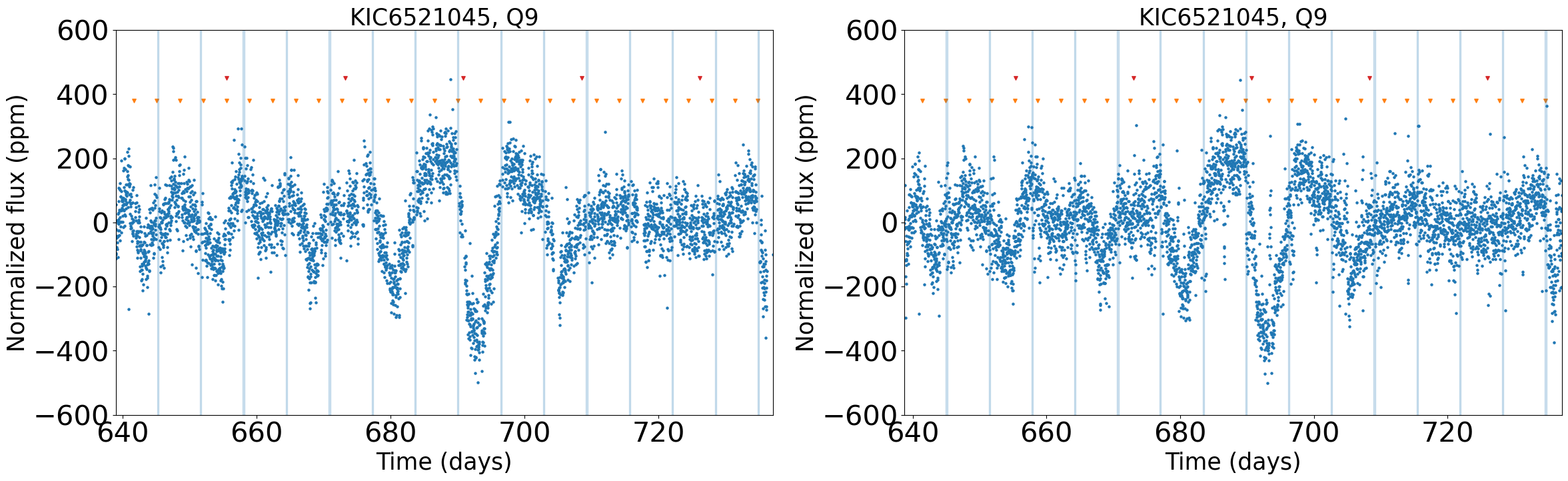} 
\caption{The lightcurve of \textit{Kepler}-100 for Q9. Left and right
  panels show the lightcurves with no padding and Gaussian
  noise-padding. \rev{This system hosts three planet candidates of
    orbital period $P_{\rm orb}$= 6.9, 12.8, and 35.3 days. We use
    vertical bars to indicate the padded epochs of transits and
    occultations corresponding to $P_{\rm orb}$=12.9 days, similar to
    the detected rotation period $\sim12.92$ days. We also mark the
    padded locations of transits and occultations for the remaining
    two planet candidates using triangles of different colors.}}
\label{fig:kepler100-LC-Q9}
\end{figure}

We note that the orbital periods of its three transiting planets
\citep{Marcy2014} are $6.9$ days (\textit{Kepler}-100 b), $12.8$ days
(\textit{Kepler}-100 c), and $35.3$ days (\textit{Kepler}-100 d), one
of which are close to our measured rotation period ($\sim$13 days).
We visual inspect the light curves to check whether our photometric
estimate is somehow contaminated by the removal of the transit signal
of \textit{Kepler}-100 b. As Figure \ref{fig:kepler100-LC-Q9}
suggests, however, the lightcurves exhibit clear periodic oscillations
that are not aligned with the location of the transit dip. Therefore,
it is unlikely to be an artifact of the transit signal
removal. \rev{We also confirm that different padding schemes (with
  and without noise) give consistent measurements of the rotation
  period}.

A possible explanation is that the 13-day period is the second
harmonics of 25 days. The variation of the period in different
quarters may suggest that \textit{Kepler}-100 has multiple large
active spots over the surface.  Indeed, our visual inspection of light
curve (Figure \ref{fig:kepler100-LC-Q9}) shows a pattern of a deeper
trough accompanied by a shallower trough. This pattern may be a
signature of multiple spots, which leads to the presence of the
harmonics (peak at $P_{\rm rot}$/2) in periodogram. Without visual
inspection, the automatic period detection method may the period
associated with harmonics as the actual rotation period. In any case,
we conclude that \textit{Kepler}-100 remains as one of the intriguing
systems that deserve for further investigation.

\subsection{non-KOI stars in Group Aa: KIC 5773345}

\begin{figure}[ht]
\centering
\includegraphics[width=10cm]{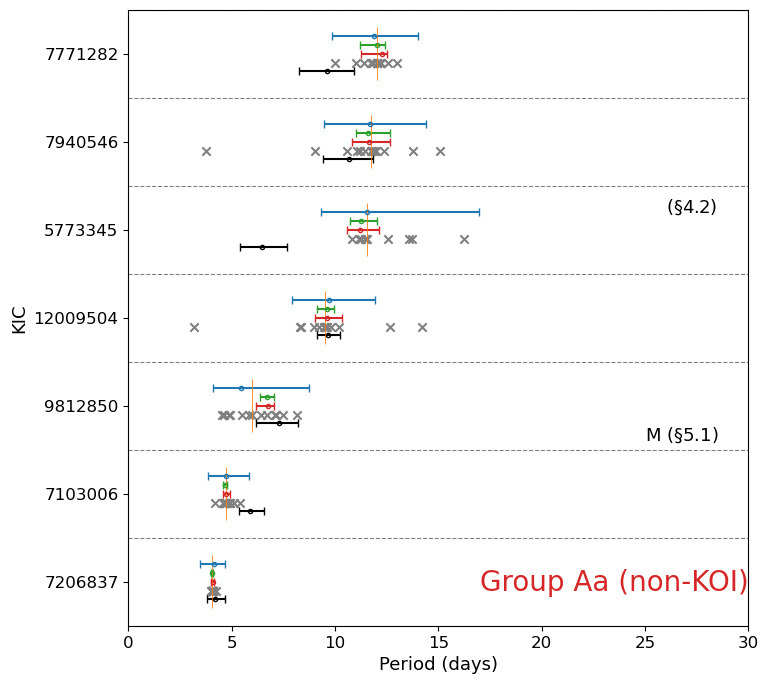}   
\caption{Same as Figure \ref{fig:Aa-KOI} but for seven non-KOI stars
  classified in Group Aa.}
	\label{fig:Aa-nonKOI}
\end{figure}

Photometric rotation periods $\mPlsq$, $P_{\rm LS}$, $P_{\rm ACF}$, and
$P_{\rm WA}$ agree within error-bars for all the non-KOI targets in
Figure \ref{fig:Aa-nonKOI}. In contrast, $P_{\rm astero} \approx 6.5$
days of KIC 5773345 is a factor of two smaller than the
photometrically derived values of $12$ days.  Thus, we examine KIC
5773345 in detail below.

Figure \ref{fig:5773345-LSQ} indicate that KIC 5773345 has
  persistent periodic signals of P$\sim12$ days over the entire
  observed span (except for the removed quarters), while there is no
  significant signal around $6.5$ days.  Asteroseismic constraints on
  $i_*$ and $\delta\nu_*$ for KIC 5773345 are shown in Figure
  \ref{fig:seismic5773345}.  

Another complementary constraint on the rotation period $P_{\rm
    rot}$ is provided from the spectroscopic data. For KIC 5773345,
  \cite{Bruntt2012} and \cite{Molenda2013} obtained $v\sin
  i_*=6.6\pm1.46$ km/s and $v\sin i_*=3.4\pm1.1$ km/s,
  respectively. The factor of two difference likely comes from the
  difficulty of modeling the turbulence as pointed out by
  \citet{Kamiaka2018}; see their Figure 8.  Adopting
  $R=2.00^{+0.05}_{-0.07}R_\odot$ \citep{Chaplin2013}, equation
  (\ref{eq:Prot-vsini}) gives $P_{\rm spec}/\sin i_* \approx
  15.2^{+4.3}_{-2.8}$ days and $29.5^{+14.1}_{-7.2}$ days,
  respectively.

Our photometric estimate of $P_{\rm rot} \sim 12$ days becomes
  consistent with $P_{\rm spec}/\sin i_*$ of \cite{Bruntt2012} and
  \cite{Molenda2013} if $\sin i_* \sim 0.8$ and $\sin i_* \sim 0.4$,
  respectively.  Incidentally, it is known that the asteroseismology
  constrains the parameter $\delta\nu_* \sin i_*$ primarily, instead
  of $\delta\nu_*$ or $\sin i_*$ \citep{Kamiaka2018}. Substituting
  $\delta\nu_* \sin i_*=0.93^{+0.1}_{-0.09} \,\mu$Hz (Figure
  \ref{fig:seismic5773345}) into equation (\ref{eq:Prot-dnu}) implies
  $P_{\rm astero}/\sin i_* = 12.5^{+1.3}_{-1.2}$ days, which agrees with
  the result of \cite{Bruntt2012} within their 1$\sigma$ uncertainty
  while inconsistent with that of \cite{Molenda2013}.

Therefore, the discrepancy between our photometric and asteroseismic
periods for KIC 5773345 may come from a intrinsic problem of asteroseismic analysis in separating
$\delta\nu_* \sin i_*$ into $\delta\nu_* (=1/P_{\rm astero})$ and
$\sin i_*$ in asteroseismic analysis. In such case, the photometric estimate of $P\sim12$ is likely the true rotation period. Nevertheless, it may be premature to conclude at this point, and KIC
5773345 should be regarded as another interesting target for follow-up observation.

\subsection{KOI and non-KOI stars in Group Ab:
  \textit{Kepler}-409,  KIC 7970740, and KOI-268 \label{subsec:kepler409}}

\begin{figure}[ht]
\centering
\includegraphics[width=10cm]{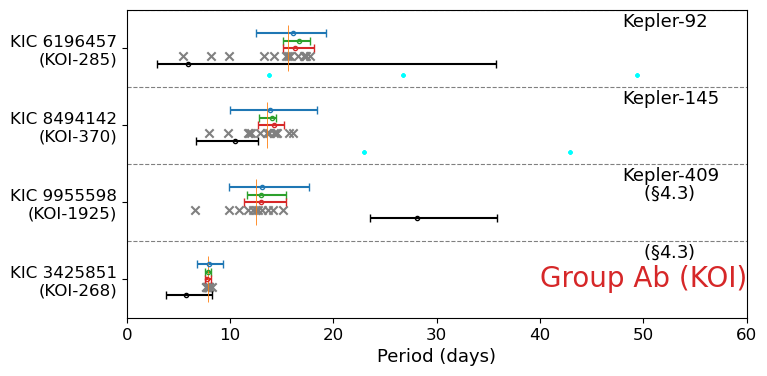}   
\caption{Same as Figure \ref{fig:Aa-KOI} but for six KOI stars
    classified in Group Ab.}  
	\label{fig:Ab-KOI}
\end{figure}

In general, stars classified as Group Ab have consistent estimates for
the photometric rotation period as shown in Figure \ref{fig:Ab-KOI}. The asteroseismic measurements for group b targets have large uncertainties so that targets do not show statistically significant discrepancy between photometric and asteroseismic periods, except for \textit{Kepler}-409 (KIC 9955598, KOI-1925) and KIC 7970740.

\begin{figure}[ht]
\centering
\includegraphics[width=10cm]{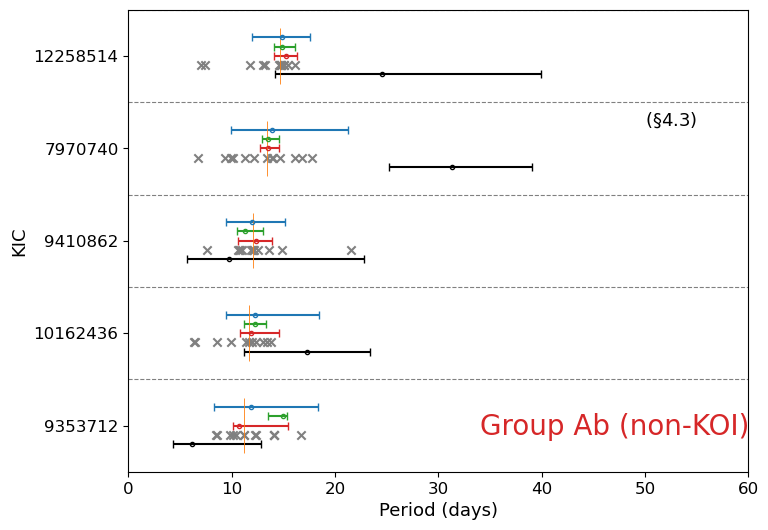}   
\caption{Same as Figure \ref{fig:Aa-KOI} but for five non-KOI stars
    classified in Group Ab.}  
	\label{fig:Ab-nonKOI}
\end{figure}

For \textit{Kepler}-409, photometric results consistently point to $P_{\rm rot}
\sim 13$ days, while $P_{\rm astero} \sim 29$ days (Figure
\ref{fig:kepler409-LSQ}). We also refer to spectroscopic estimations for comparison. Equation (\ref{eq:Prot-vsini})
and (\ref{eq:Prot-dnu}) suggest that $P_{\rm spec}\approx 22.3\sin
i_*$ days if $v\sin i_* =2$ km/s \citep{Marcy2014} while $P_{\rm
  astero}\approx 37.4^{+4.0}_{-3.3} \sin i_*$ days for
$\delta\nu_*\sin i_* =0.31\pm 0.03\, \mu$Hz (Figure
\ref{fig:seismickepler409}).

For KIC 7970740, $\Plsq$ varies from 9 to 18 days (Figure
  \ref{fig:7970740-LSQ}), while $P_{\rm astero} \sim
  41.1^{+5.0}_{-4.0} \sin i_*$ days for $\delta\nu_*=0.28\pm 0.03 \,
  \mu$Hz (Figure \ref{fig:seismic7970740}).

Another interesting target is KIC 3425851 (KOI-268) for which the
photometric period is precisely measured (Figure
\ref{fig:koi268-LSQ}). If we perform the asteroseismic analysis with
the prior for $\delta\nu_*$ from photometric estimate, one may be able to constrain the stellar inclination more tightly; see Figure \ref{fig:seismickoi268}.

\subsection{KOI stars in Group Ba: \textit{Kepler}-25, 93, and 128
  \label{subsec:KOI-Ba}}

\begin{figure}[th]
\centering
\includegraphics[width=10cm]{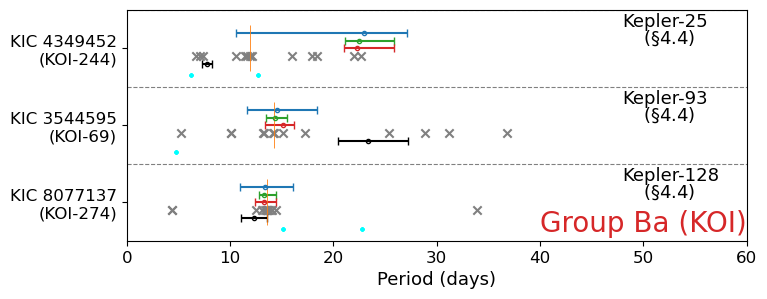}   
\caption{Same as Figure \ref{fig:Aa-KOI} but for three KOI stars
    classified in Group Ba.}  
	\label{fig:Ba-KOI}
\end{figure}

There are three planet-hosting stars in Group Ba (Figure
\ref{fig:Ba-KOI}). \textit{Kepler}-25 (KIC 4349452, KOI-244) is one of the two transiting
planetary systems for which the real spin-orbit misalignment angle has been estimated
jointly from the Rossiter-McLaughlin effect and asteroseismology
\citep{Benomar2014b,Campante2016}.

Figure \ref{fig:kepler25-LSQ}
indicates that the photometric rotation period for \textit{Kepler}-25 varies
significantly from quarter to quarter and not  reliable. On the
other hand, the asteroseismic rotation period seems to be very
precisely estimated; $P_{\rm astero} \approx 7.9^{+0.5}_{-0.4}$ days.
Adopting $v\sin i_* \approx 9.5 $km/s and $R=(1.31\pm 0.02)R_\odot$
\citep{Marcy2014}, $P_{\rm spec} \approx 6.9\sin i_*$ which is in
rough agreement with $P_{\rm astero}$.

\textit{Kepler}-93 (KIC 3544595, KOI-69) also exhibits large variation
in $\Plsq$ (Figure \ref{fig:kepler93-LSQ}).  Referring to
spectroscopic measurement from \citep{Marcy2014}, we obtain $P_{\rm
  spec} \approx 92\sin i_*$ days with $v\sin i_* \approx 0.5 $km/s and
$R=(0.92\pm 0.02)R_\odot$. However, such a small value of $v\sin i_*$
is not so reliable as it is subject to a large uncertainty in
turbulence modeling. On the contrary, asteroseismic analysis provides
a relatively precise estimation of $P_{\rm astero} \approx
23.3^{+0.5}_{-0.4}$ days (Figure \ref{fig:seismickepler93}).

Finally, we note that \textit{Kepler}-128 (KOI-274, KIC 8077137) is classified
as Group B because of three apparent outliers, which are $\Plsq \approx 4$ days for Q11 and Q14 and $\Plsq
\approx 32$ days for Q2. Except for the three quarters, $\Plsq \approx
(13-14)$ days persistently, and indeed agrees well with $P_{\rm
  astero}=12.3 \pm 1.3$ days. Thus, \textit{Kepler}-128 had better be
classified as Group A in reality.

\subsection{non-KOI stars in Group Ba}

Sixteen non-KOI stars in Group Ba exhibit a diverse distribution of
periodicities in the LS periodogram at different quarters. Most of
them have multiple peaks whose relative heights vary among different
quarters so that it is not easy to determine the photometric rotation
period robustly. We show two examples (KIC 6225718 and 9965715) in
Figures \ref{fig:6225718-LSQ} to \ref{fig:9965715-LSQ}. As
examined in \S \ref{subsec:KOI-Ba}, $P_{\rm astero}$ is likely to provide better period estimate for stars in Group Ba.

\begin{figure}[ht]
\centering
\includegraphics[width=10cm]{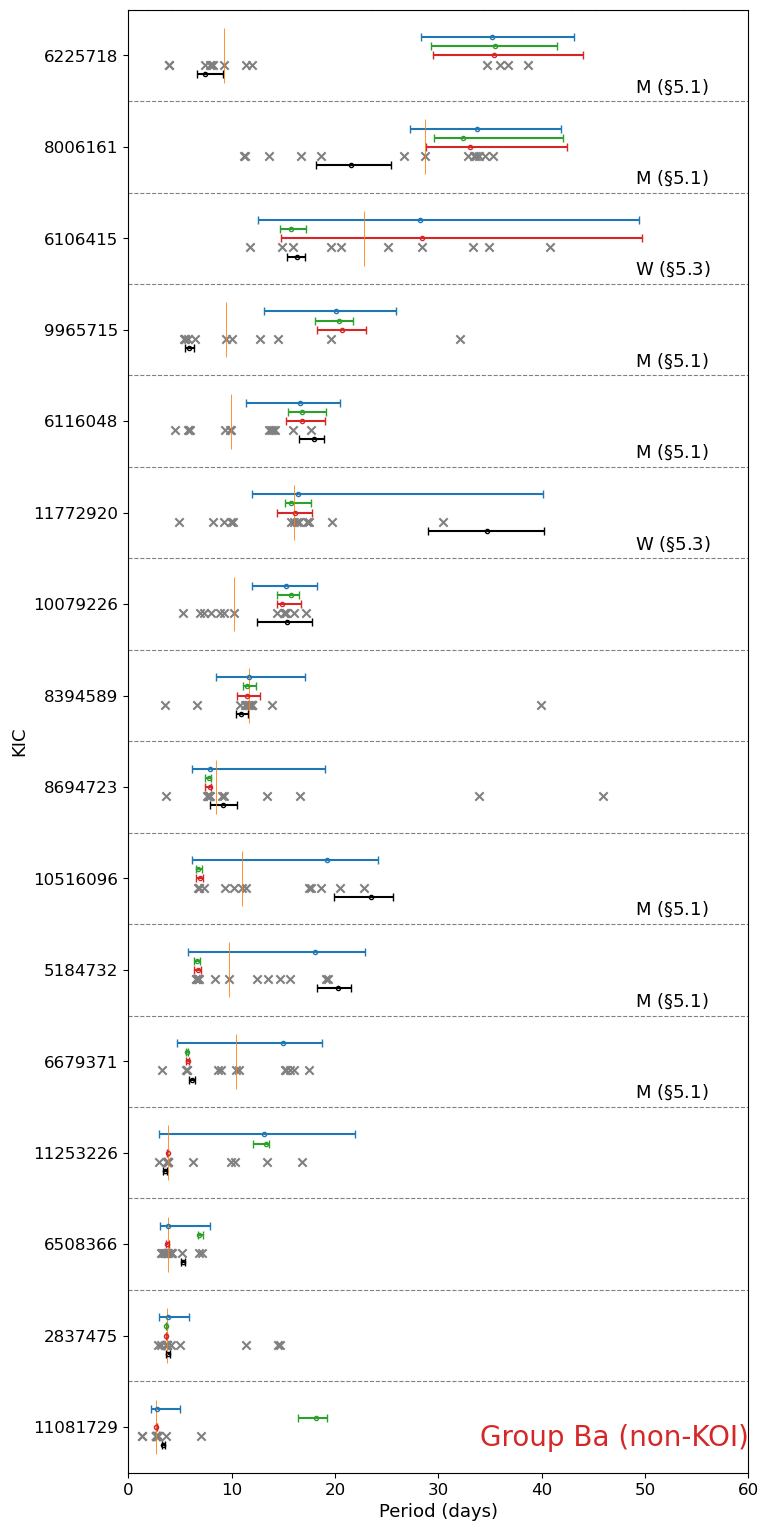}   
\caption{Same as Figure \ref{fig:Aa-KOI} but for 16 non-KOI stars
    classified in Group Ba.}  
	\label{fig:Ba-nonKOI}
\end{figure}

\begin{figure}[ht]
\centering
\includegraphics[width=10cm]{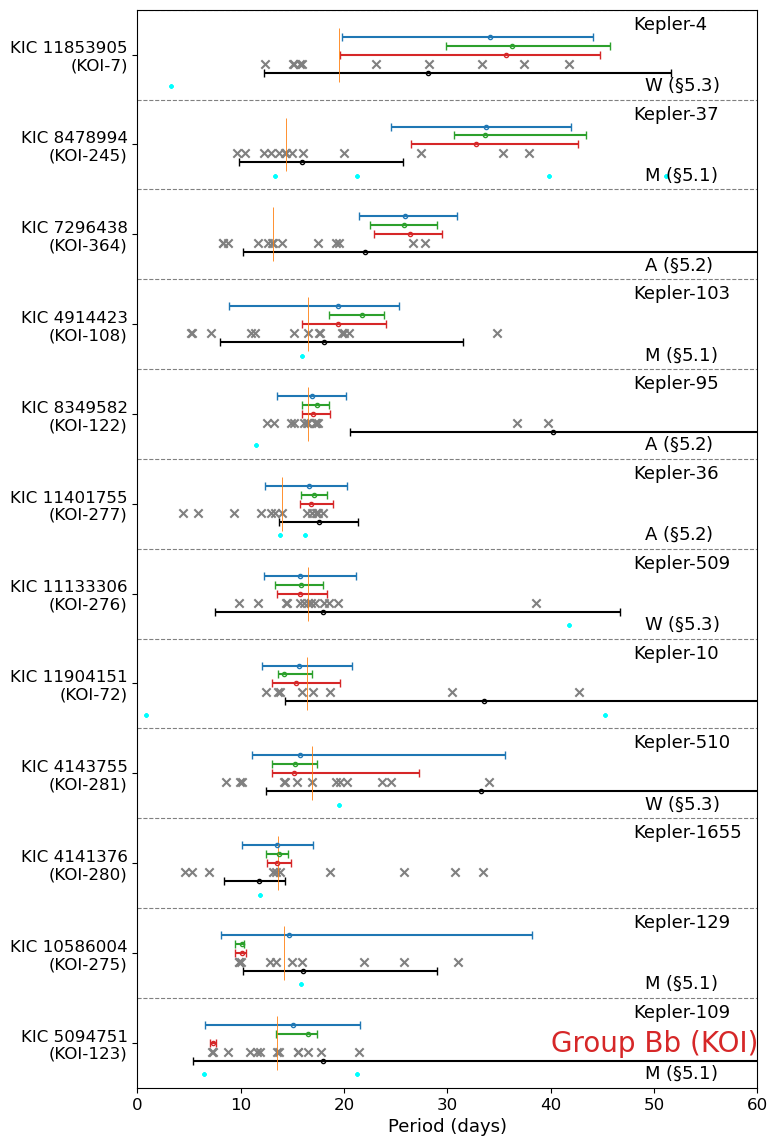}   
\caption{Same as Figure \ref{fig:Aa-KOI} but for 10 KOI stars
    classified in Group Bb.}  
	\label{fig:Bb-KOI}
\end{figure}

\begin{figure}[ht]
\centering
\includegraphics[width=10cm]{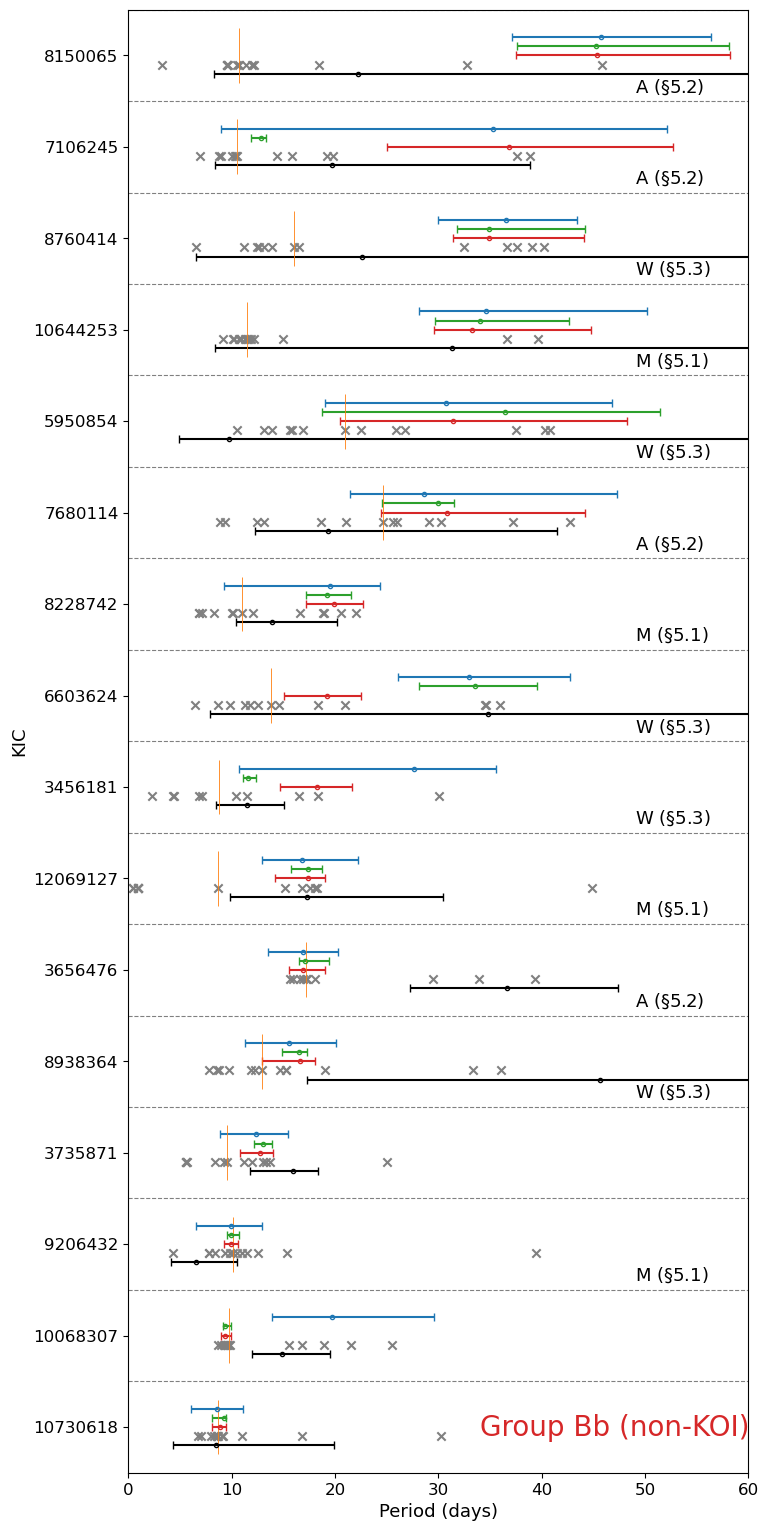}   
\caption{Same as Figure \ref{fig:Aa-KOI} but for 16 non-KOI stars
    classified in Group Bb.} 
	\label{fig:Bb-nonKOI}
\end{figure}

Finally, we show the results for KOI and non-KOI stars in Group Bb
(Figures \ref{fig:Bb-KOI} and \ref{fig:Bb-nonKOI}), without discussing
them individually.

\clearpage

\section{Comparison among photometric analyses\label{subsec:comparephoto}}

\begin{figure}[t]
\centering
\includegraphics[width=14cm]{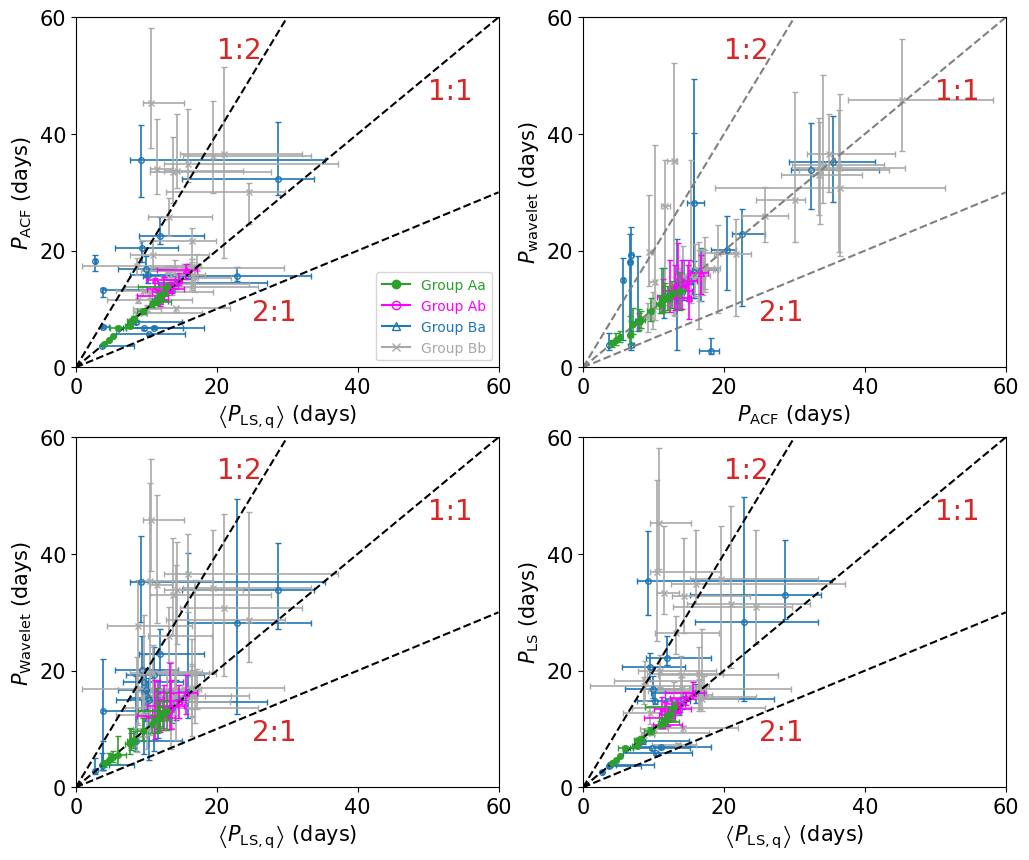}
\caption{Comparison among four different photometrically measured
  rotation periods: $P_{\rm ACF}$ vs. $\mPlsq$ (upper-left), $P_{\rm
    WA}$ vs. $P_{\rm ACF}$ (upper-right), $P_{\rm WA}$ vs. $\mPlsq$
  (lower-left), and $P_{\rm LS}$ vs. $\mPlsq$ (lower-right). We use
  green, magenta, blue and gray to represent stars in Groups Aa, Ab,
  Ba, and Bb, respectively.}
\label{fig:comparepphoto}
\end{figure}

Figure \ref{fig:comparepphoto} presents the comparison of the rotation
periods $P_{\rm LS}$, $P_{\rm ACF}$, $P_{\rm WA}$, and $\mPlsq$. The
quoted error-bars for each method are computed according to the
description in section \ref{sec:method}. Thus, they do not necessarily
represent the same confidence level, and should be interpreted with
caution.

We classified our targets based on $\Dls$ defined by equation
(\ref{eq:Dls}). Thus, the good agreement between $P_{\rm LS}$ and
$\mPlsq$ for Group A is as expected; see the lower-right
panel of Figure \ref{fig:comparepphoto}. On the other hand, agreement
among $P_{\rm LS}$, $P_{\rm ACF}$, and $P_{\rm WA}$ implies that those
photometrically measured periods are reliable and robust, at least for
23 stars in Group A. This should be kept in mind when we discuss the
comparison against $P_{\rm astero}$ in subsection
\ref{subsec:photo-astero} below.

The majority of 47 targets in Group B still exhibit consistent
rotation periods even though their error-bars are relatively large.
Figure \ref{fig:comparepphoto} shows that 8 stars have
inconsistent estimates among $P_{\rm LS}$, $P_{\rm ACF}$, and $P_{\rm
  WA}$; after careful examination, we find five targets with multiple
periodic components, two targets with an anomalous quarter, and one
target with no clear periodic signal. We discuss those eight targets
individually below.

\subsection{Multiple periodic components:
  KIC 1108172, 11253226,  6508366, 10068307, 5094751 \label{subsec:Bmultiple}}

When multiple periodic components exist in the lightcurve, photometric analyses are not enough to discern the true rotation period. In such case, three different analyses assign different relative power to candidate peaks in spectra which occasionally leads to inconsistencies in period detected. For instance,
the GWPS is known to increase the power of the fundamental period of a
signal \citep{Ceillier2017}, and the ACF method uses a Gaussian
smoothing to suppress the short-term modulation including the possible
influence of harmonics. Thus both of them may be biased toward a
longer period.  Here, we provide our interpretation for 5 stars that
exhibit multiple periodic components and have different estimates
among $P_{\rm LS}$, $P_{\rm ACF}$, and $P_{\rm WA}$.

\begin{itemize}
\item KIC 11081729 and 11253226: their $P_{\rm LS}$ is much shorter than
  $P_{\rm ACF}$. The visual inspection of their wavelet power spectra
  indicates that the short period component persists more than
  half of the entire observed time span. Thus,
   $P_{\rm LS}$ is more likely to represent their true rotation periods.
\item KIC 6508366 and 10068307: they both have two components roughly
  different by a factor of two.  For KIC 6508366, the shorter period
  component is observed in the wavelet spectra for all quarters,
  while the longer period is visible only for a few quarters.  For KIC
  10068307, in contrast, the two components exist
  simultaneously within all of the quarters. Thus, we interpret the
  longer period corresponding to $P_{\rm LS}$ as the true rotation
  period, while the shorter period being its second harmonics.
\item KIC 5094751 (\textit{Kepler}-109, KOI-123): It has multiple periodic
  components in Q4 and Q6 with $P_{\rm LS}$ being longer than both
  $P_{\rm ACF}$ and $P_{\rm WA}$; see also the right panel of Figure
  \ref{fig:LS-ACF-WA}. It could be due to a complex spot configuration
  and/or stellar surface differential rotation, and we can not be certain about
  which represents the true rotation period.
\end{itemize}

\subsection{Abrupt increase of photometric variation
  in a single quarter: KIC 3456181,
  7106245 \label{subsec:suddenvariation}}

\begin{figure}[htb!]
  \centering \subfigure{
		\begin{minipage}[t]{0.45\textwidth}
			\centering
			\includegraphics[width=1\linewidth]{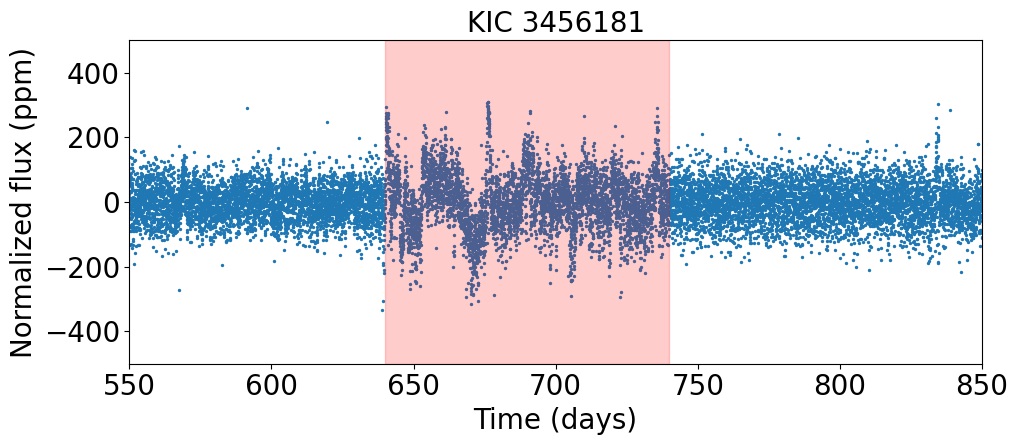}   
			\label{fig:Bsudden1}
		\end{minipage}
	}\vspace{-0.5em}
	\subfigure{
		\begin{minipage}[t]{0.45\linewidth}
			\centering
			\includegraphics[width=1\linewidth]{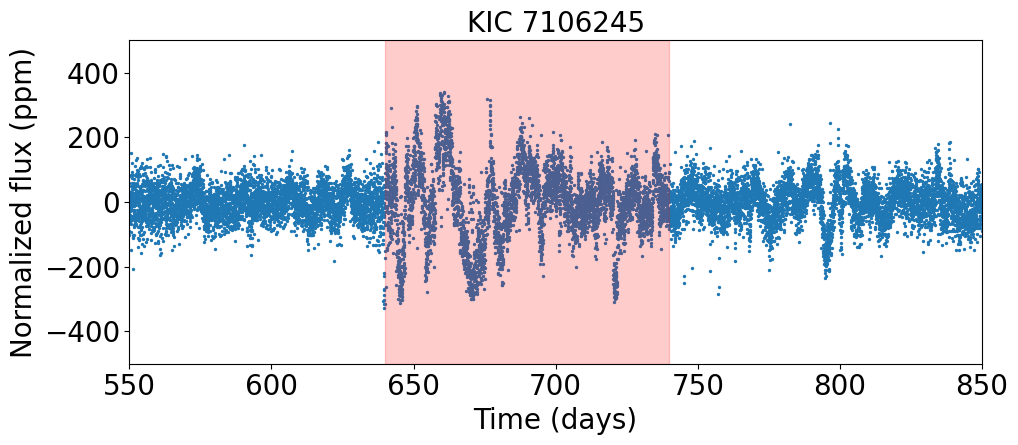}  
			\label{fig:Bsudden2}
		\end{minipage}
	}\vspace{-0.5em}
\caption{Part of quarters of light curves of KIC 3456181 and KIC
  7106245. The shaded region mark the anomalous quarters where the
  flux variation suddenly increases.}
	\label{fig:Bsudden}
\end{figure}

As described in Section \ref{subsec:lcprocessing}, using a systematic
method, we removed anomalous quarters whose median absolute flux
variations are three times larger than those of the neighboring
quarters for six targets in total; \textit{Kepler}-1655 (KIC 4141376,
KOI-280), \textit{Kepler}-100 (KIC 6521045), \textit{Kepler}-409 (KIC
9955598, KOI-1925), KOI-72 (KIC 11904151), KIC 8694723, and KIC
10730618. The goal is to reduce the risk of having a biased
measurement. However, among the stars for which the procedure does not
detect anomalous quarters (that is, stars for which no quarter removal
is performed), there still remains some irregular and abrupt increases
in flux, though less prominent, which may affect the estimation of the
rotation period.

These abnormal parts might be partly due to residual systematics, or a
contamination from other neighbor sources. Such abnormal variation
usually give a dominating peak in the power spectrum, over-weighing
other weaker peaks that are more likely associated with the actual
rotational signals. Examples of this type of light curves are given in
Figure \ref{fig:Bsudden}, whose shaded region shows a
strong but transient flux variation. We visual inspect the wavelet
spectra of the two given targets and notice that measured low
frequency peaks ($\sim$ 27 days in $P_{\rm WA}$ for KIC 3456181 and
$\sim$ 35 days in $P_{\rm WA}$ and $P_{\rm LS}$ for KIC 7106245) are
associated with the abnormal regions instead of periodic modulation
that we look for.  In such circumstance, $\left<P_{\rm LS, q}\right>$
is less biased by those abnormal quarters.

For all targets in our sample with abrupt increase in flux variation,
we check the module and channel number of the abnormal quarters to see
whether it comes from a specific CCD plate. However, we notice that
such abrupt variations are found in various modules and channels. In
addition, we check the position of image on the plate and ensure that
the location of image is not around the edge. It is natural to think
of the increase in variation as the modulation from a large
spot. Nevertheless, known that the lifetime of a spot scales with its
size, abrupt variations which last for only 30-40 days seems less
likely to be the lifetime of a large spot. Alternatively, such
abnormal variation could be the residual of systematics like 30-day
re-orientation of telescope. Especially, for relatively 'quiet' stars
with few active regions, there is barely any rotational modulation in
the light curve. The residual of systematics would become significant
and act like a sudden increase in flux variation.

\subsection{Weak or ambiguous periodic variation: KIC 6603624}

\begin{figure}[tbh!]
\centering
\includegraphics[width=10cm]{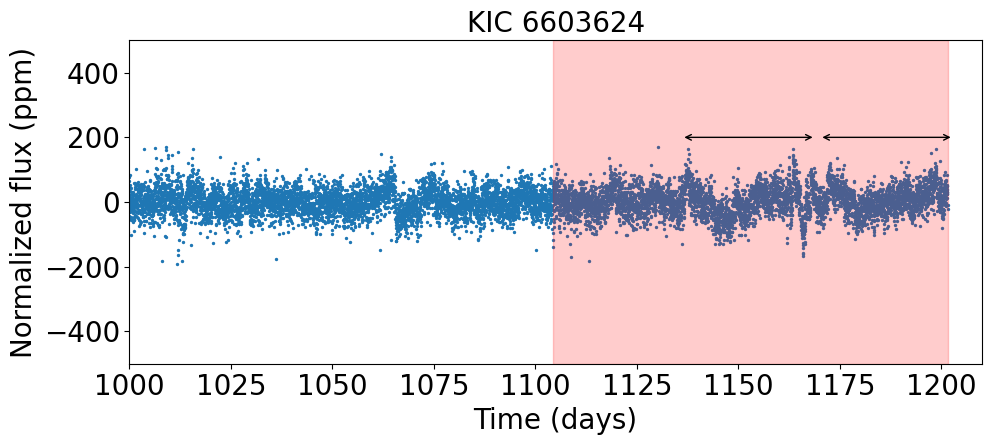}
\caption{Part of light curves for KIC 6603624, and the shaded region
  corresponds to Q14. Two black arrows indicate the locations of
  possible 30-day periodic signal.}
\label{fig:lcnonoise}
\end{figure}

Figure \ref{fig:lcnonoise} illustrates part of light curves for
  KIC 6603624 as an example of stars exhibiting a weak flux variation.
  The WA power spectrum for this type of targets is generally very
  noisy, and we find that it is simply dominated by the peak at $P
  \sim 32$ days in Q14 alone for KIC 6603624.  As Figure
  \ref{fig:lcnonoise} indicated, however, it lasts just for two cycles
  and it is not clear if it should be interpreted as a rotational
  feature. Thus, for this type of targets, the {\it measured} period
  is very unreliable, and should be taken with caution. We identify
  eight such targets with weak or ambiguous periodic components, and
  label them by ``W'' in the 'Remark' column of Table
  \ref{tab:singlestarAa}, \ref{tab:singlestarAb},
  \ref{tab:singlestarBa}, and \ref{tab:singlestarBb}.


\section{Comparison between photometric and asteroseismic estimates
\label{subsec:photo-astero}}

\begin{figure}[ht]
  \centering
  \includegraphics[width=14cm]{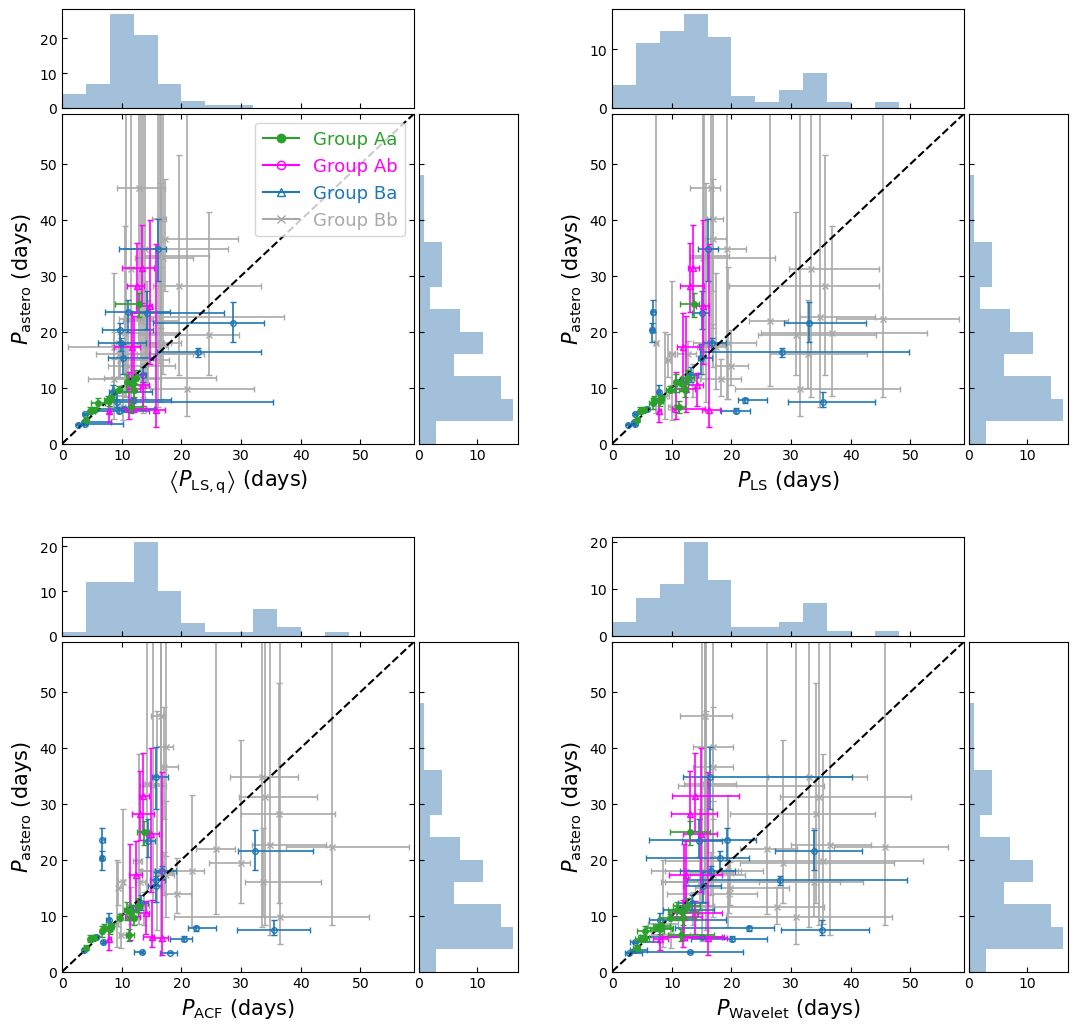}
  \caption{Comparison of $P_{\rm astero}$ against photometrically
      derived rotation periods; $\mPlsq$ (upper-left), $P_{\rm LS}$
      (upper-right), $P_{\rm ACF}$ (lower-left), and $P_{\rm WA}$
      (lower-right).  Green, blue, magenta, and gray symbols with
      error-bars correspond to Groups Aa, Ba, Ab, and Bb,
      respectively.}
	\label{fig:asteroall}
\end{figure}

\begin{figure}[htb!]
  \centering
\includegraphics[width=8cm]{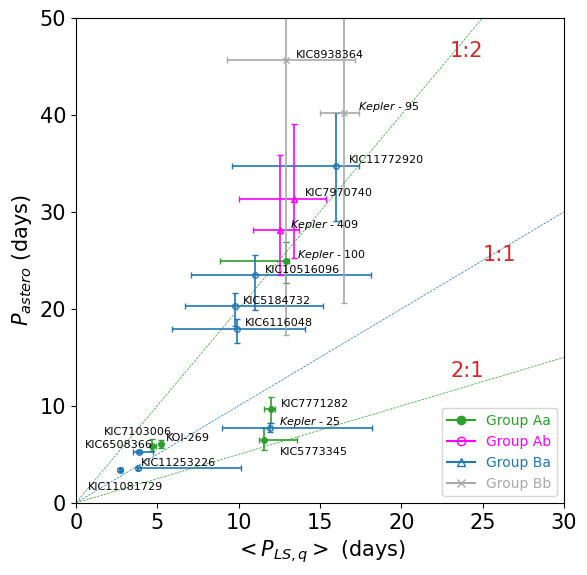}   
\caption{$\mPlsq$ versus $P_{astero}$ for 17 targets.
  There are 19 targets that do not agree
  within their uncertainties, but we do not plot
  \textit{Kepler}-100 and \textit{Kepler}-409 here because
  part of their quarters are removed during the lightcurve processing.
  Green, blue, magenta, and gray symbols with error-bars
  correspond to Groups Aa, Ba, Ab, and Bb, respectively.}
	\label{fig:incon}
\end{figure}

Figure \ref{fig:asteroall} compares the rotation period of our targets
derived from asteroseismic analysis by \citet{Kamiaka2018} against our
four photometric estimates: $\mPlsq$ (upper-left), $P_{\rm LS}$
(upper-right), $P_{\rm ACF}$ (lower-left), and $P_{\rm WA}$
(lower-right).  Figure \ref{fig:incon} plots 17/19 stars whose
$\mPlsq$ and $P_{\rm astero}$ do not agree within their uncertainties
defined by equations (\ref{eq:medianPls-error}), (\ref{eq:dPastero+})
and (\ref{eq:dPastero-}).  Since we have already discussed most of the
17 targets in \S \ref{sec:remarks}, we do not repeat such discussion
here on an individual basis, but present a few general conclusions below.

First, stars classified as Group Aa are in good agreement except
\textit{Kepler}-100 (\S \ref{subsec:kepler100}). While it is quite expected
from the definition of our classification scheme, it is still
encouraging. Second, there exist a population of stars whose $P_{\rm
  astero}$ is approximately twice of the corresponding photometric
estimate. For those stars in Group Ba, $P_{\rm astero}$ is likely to
be the true period.  For those in Group Ab, on the other hand, it may
not necessarily imply that $\mPlsq$ is closer to the true value due to
a possible bias toward shorter period for $\mPlsq$.

Of course, the difference between $\mPlsq$ and $P_{\rm astero}$ may
point to an interesting physical effect, instead of an unknown
artifact.  Photometric estimates are based on the stellar surface
variation, and thus sensitive to the latitudinal and size distribution
of starspots, as well as other surface activities, while
asteroseismic analysis is supposed to provide an averaged rotation
period over a range of latitudes and depths inside the star. On the other hand,
Figure \ref{fig:incon} seems to suggest a tendency of $P_{\rm astero}
\sim 2 \mPlsq$. This may be interpreted that $\mPlsq$ represents the
second harmonics of the true rotation period of $P_{\rm astero}$, as
already pointed out by \cite{Garcia2013}. If confirmed, it implies the
importance of the complementary visual inspection of automatic
photometric analyses.

\section{Discussion}

\subsection{Possible implications for differential rotation
\label{subsec:diffrot}}

Figures \ref{fig:Aa-KOI} to \ref{fig:Bb-nonKOI} show that a majority
of our targets have a range of different values of $\Plsq$ in
different quarters. The origin of those variations is not easy to
identify, and may be a combination of stellar surface activity,
instrumental effect, and unknown noise/contamination.

Of course, the variation of $\Plsq$ could be of physical origin, at
least for some targets. The most likely possibility includes an effect
of the latitudinal differential rotation coupled with the surface
distribution and dynamics of starspots.  For instance, the presence of
$n$ active regions along the same latitude would result in significant
peaks with the $n$-th harmonics of the rotation period in the LS
periodogram. Such alias would outweigh the fundamental rotation period
if such active regions are equally spaced along the same latitude band
\citep[e.g.][]{Chowdhury2018}. Furthermore, the co-existence of dark
spots (the central darker part, {\it umbra}, and surrounding lighter
part, {\it penumbra}) and brighter spots ({\it faculae}) may cancel
out part of the photometric variation, which complicates the
extraction of the true rotational signal. Our Sun is an example of
stars with complex surface configuration (frequent appearance of
multiple active regions) which leads to unstable and occasionally low
variability in both chromospheric and photometric measurements
\citep[see e.g.][]{Donahue1996, Reinhold2020}. As \cite{Donahue1996}
pointed out, it is very likely solar-like star with similar surface
configuration to our Sun may not pass the threshold for reliable
detection of rotation period.

Consider a simple model of the latitudinal differential rotation
\citep{Reiners2003,Hirano2012}:
\begin{eqnarray}
\label{eq:omega-latitude}
\omega(\ell)\approx\omega_\mathrm{eq}(1-\alpha\sin^2 \ell),
\end{eqnarray}
where $\omega(\ell)$ is the angular velocity of the stellar surface at
latitude $\ell$, $\omega_\mathrm{eq}$ is its value at the equator
($\ell=0$), and $\alpha$ is a parameter characterizing the strength of
the differential rotation.  For the Sun, $\alpha \approx 0.2$.

In order to proceed further, we need to assume the size and latitude
distribution functions of starspots and their dynamics (migration over
the stellar surface and creation/annihilation timescales), which is unknown. \citet{SSNB2022} examined the variation of $\Plsq$
assuming that the spot distribution in each quarter is independently
drawn from the empirical distribution to match the observed flux
modulation of the Sun.

For simplicity, we may equate our $\Dls$ to the variation of
$\alpha\sin^2 \ell$ in equation (\ref{eq:omega-latitude}).
Furthermore, if the spot latitude $\ell_{\rm spot}$ varies between
$0^\circ$ and $45^\circ$, $\Dls \sim \alpha/2$.  Thus, in principle,
we may estimate $\alpha$ individually for our targets.  In reality,
however, most of the variation of $\Plsq$ in Figures \ref{fig:Aa-KOI}
to \ref{fig:Bb-nonKOI} are very diverse, implying that the
differential rotation combined with the spot distribution may not be
dominant except for stars exhibiting modest variations.  While we do
not conduct further quatitative examination, \textit{Kepler}-408, KOI-974, \textit{Kepler}-50,
KOI-269, KIC 7103006, KIC 7206837, and KOI-268 may be potentially
interesting targets for tightly constraining their differential
rotation.

\subsection{KOI stars in stellar binary/multiple systems
  \label{subsec:discuss-multi}}

\begin{figure}[ht]
\centering \includegraphics[width=10cm]{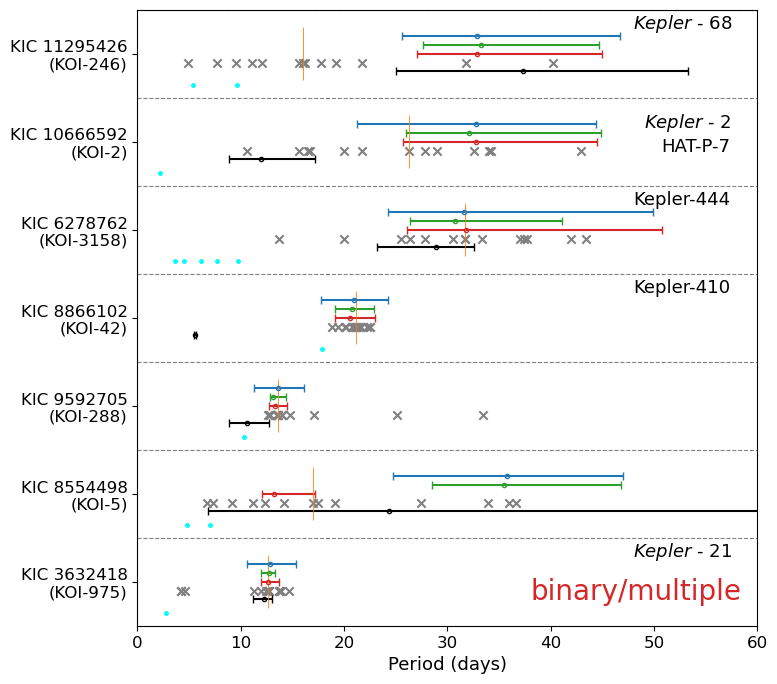}
\caption{Same as Figure \ref{fig:Aa-KOI} but 
for 7 KOI stars in stellar binary/multiple systems.}  
	\label{fig:binary-KOI}
\end{figure}

We have 22 stars with possible stellar companions (\S
\ref{subsec:binarity}). Since their lightcurves might be contaminated
by the companions, we exclude them from our main
analysis. Nevertheless, we examine seven KOI stars in this subsection
because they may have interesting implications for their planetary
architecture.  Figure \ref{fig:binary-KOI} plots their rotation
periods in the same manner as Figure \ref{fig:Aa-KOI}.  Among them, we
select the following three stars that are potentially interesting,
the LS periodogram and asteroseismic constraints of which are shown in \S
\ref{app:kepler2}, \S \ref{app:kepler410}, and \S \ref{app:koi288}.

\textit{Kepler}-2 (HAT-P-7, KOI-2, KIC 10666592) is one of the two transiting
planetary systems along with \textit{Kepler}-25 for which the real spin-orbit
misalignment angle has been estimated jointly from the
Rossiter-McLaughlin effect and asteroseismology
\citep{Benomar2014b,Campante2016}.  In reality, however, it is not
easy to identify its rotation period photometrically from Figure
\ref{fig:kepler2-LSQ}. On the other hand, $P_{\rm astero} \approx 12$ days from
Figure \ref{fig:seismickepler2} seems to be a reliable estimation of rotation
period. This is a good example of the complementary power of
asteroseismology.

\textit{Kepler}-410 (KOI-41, KIC 8866102) has precise estimates
for both photometric and asteroseismic rotation periods, but their
values are very different ($\mPlsq \sim 4 P_{\rm astero}$).
Intriguingly, Figure \ref{fig:kepler410-LSQ} indicates that the LS
periodograms for all quarters have persistent secondary peaks at
$\approx 10$ days, which is very close to the second harmonics of the
starspot modulation analytically predicted by \citet{SSNB2022}. Thus,
it is possible that $\mPlsq \approx 20$ days, instead of $P_{\rm
  astero} \approx$ 6 days, is the true rotation period of \textit{Kepler}-410.
On the other hand, \textit{Kepler}-410 is a binary system\citep{Van2014}, and
it is also likely that either asteroseismic or photometric analyses
may be contaminated by the companion star to some extent.

Finally, $\mPlsq$ and $P_{\rm astero}$ of KOI-288 (KIC 9592705) agree
within their error-bars. Figure \ref{fig:koi288-LSQ} shows that its
photometric periods are very robust. Thus, the joint photometric and
asteroseismic analysis of KOI-288 may put a tighter constraint on its
inclination. Incidentally, this is an example of systems whose $P_{\rm
  astero}$ is close to the orbital period of the planetary companion
\citep{Suto2019}, although it may be just by chance.

\subsection{Photometric and asteroseismic rotation periods
  as a function of stellar effective temperature
  \label{subsec:P-Teff}}

Stars lose the angular momentum through magnetic winds, and the loss
rate is supposed to be sensitive to the depth of the convective zone,
and therefore to their mass and effective temperature.  Indeed, it is
known observationally that stars with $T_{\rm eff} < 6200$K have
a thicker convective envelope and much slower rotation rate than those
with $T_{\rm eff} > 6200$K \citep{Kraft1967}.  \citet{Winn2010b} found
that hot Jupiters around stars with $T_{\rm eff} < 6200$K prefer the
orbital axis well-aligned to the stellar spin axis, and proposed that
the spin-orbit alignment results from the stronger tidal interaction
between close-in giant planets and their host stars with a thicker
convective envelope.

\begin{figure}[ht]
\centering \includegraphics[width=8cm]{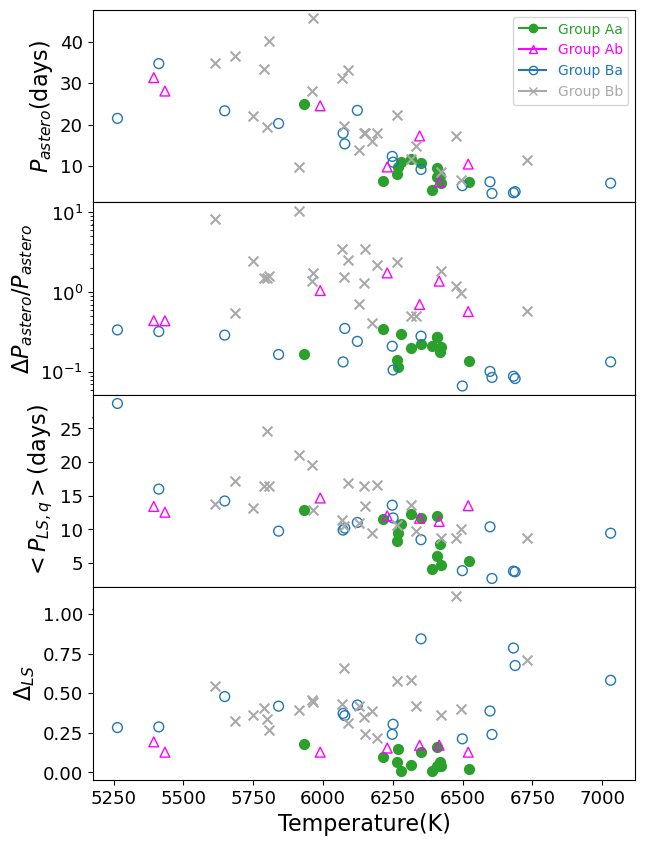}
\caption{Stellar rotation periods against the effective temperature.
  From top to bottom, $P_{\rm astero}$, $\Delta P_{\rm astero}/P_{\rm
    astero}$, $\mPlsq$, and $\Dls$ are plotted as a function of the
  stellar effective temperature. Green, blue, magenta, and gray
  symbols correspond to Groups Aa, Ba, Ab, and Bb, respectively.
	\label{fig:P-Teff}}
\end{figure}

Figure \ref{fig:P-Teff} plots the photometric and asteroseismic
rotation periods against the stellar effective temperature.  We adopt
the SDSS $T_{\rm eff}$ from Table 7 of \cite{Pinsonneault2012} for 60
stars. Our result is basically consistent with Figure 3 of
\citet{Hall2021} and qualitatively reproduce the Kraft break.  On the
other hand, our targets with $P_{\rm astero}>30$ days often have
smaller values of $\mPlsq$ and may not properly represent the surface
rotation period.  Furthermore, the uncertainties of the asteroseismic
measurements seem to be anti-correlated with $T_{\rm eff}$ (see Figures
\ref{fig:asteroall} and \ref{fig:incon}).

\subsection{Stellar inclinations as a function of stellar effective
    temperature \label{subsec:incl-Teff}}

Finally, we study the stellar inclination as it could be an
  indicator of misalignment for planetary systems observed by transit
  and also has an imprint of the genesis of the stars. The stars are
  considered as a single population, from which we seek to identify
  the underlying distribution of stellar inclination and
  its dependence with temperature. The adopted criteria is the $\sin
  i$, which is obtained following the
  method described below.

For the asteroseismic sample of stars, we directly convert the
posterior distribution of $i_*$ by \citet{Kamiaka2018}, and compute
the distribution of $\sin i_{*, \rm{astero}}$. For the spectroscopic determination of $\sin i_{*, \rm{spec}}$, it is required to use equation
(\ref{eq:Prot-vsini}), which depends on $R$, $P_{\rm spec}$
(substituted by the photometric rotation period $P_{\rm LS}$ for the
current purpose) and $v\sin i_*$. Furthermore, it is necessary to consider the distribution
functions of $R$ and  $P_{\rm LS}$ and $v\sin i_*$. Here, we assume uncorrelated Gaussian distributions9
for $R$ and $v\sin i_*$ with their mean being the measured value and
standard deviation being its $1\sigma$ error. Finally, in order to evaluate the distribution of the period, we directly use the LS periodogram of $P_{\rm LS}$ as the underlying probability distribution. However, we only use a range of the LS periodogram that spans over a slice that is twice the HWHM around the highest peak. This distribution is also normalised by integration over the considered range mentioned above. All of the probability distributions of the variables being defined, we then compute the distribution of $\sin i_{*,  \rm{spec}}$ via Monte Carlo sampling of $R$, $P_{\rm LS}$ and $v\sin i_*$.

\begin{figure}[ht]
\centering
\includegraphics[width=10cm]{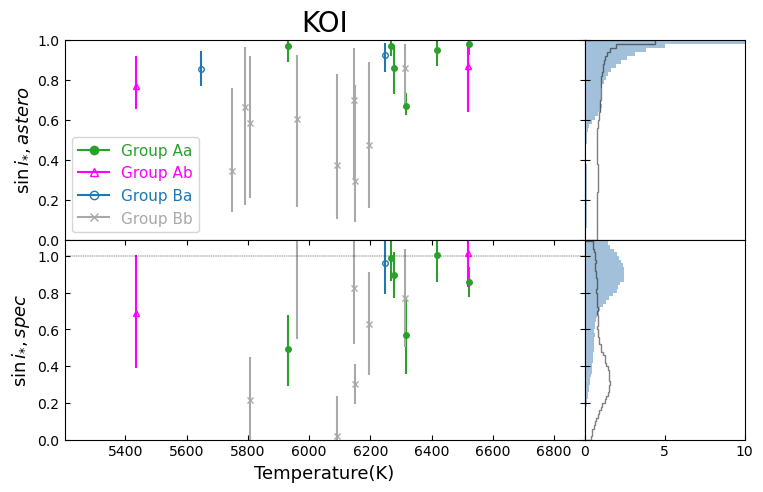}
\caption{Stellar inclination $\sin i_*$ against the effective
    temperature for KOI stars.  Left panels plot the values of $\sin
    i_*$ derived from asteroseismology \citep{Kamiaka2018} and the
    photo-spectroscopic method adopting spectroscopic $v \sin i_*$
    values from California-Kepler Survey \cite{Petigura2017} and
    \cite{Huber2013b}.  Green, blue, magenta, and gray symbols with
    error-bars correspond to Groups Aa, Ba, Ab, and Bb, respectively.
    Right panels show the distribution functions of $\sin i_*$; black
    lines consider all stars, while shaded regions correspond to stars
    in Groups Aa, Ba and Ab excluding the least reliable Group Bb. }
\label{fig:incl-Teff-koi}
\centering
\includegraphics[width=10cm]{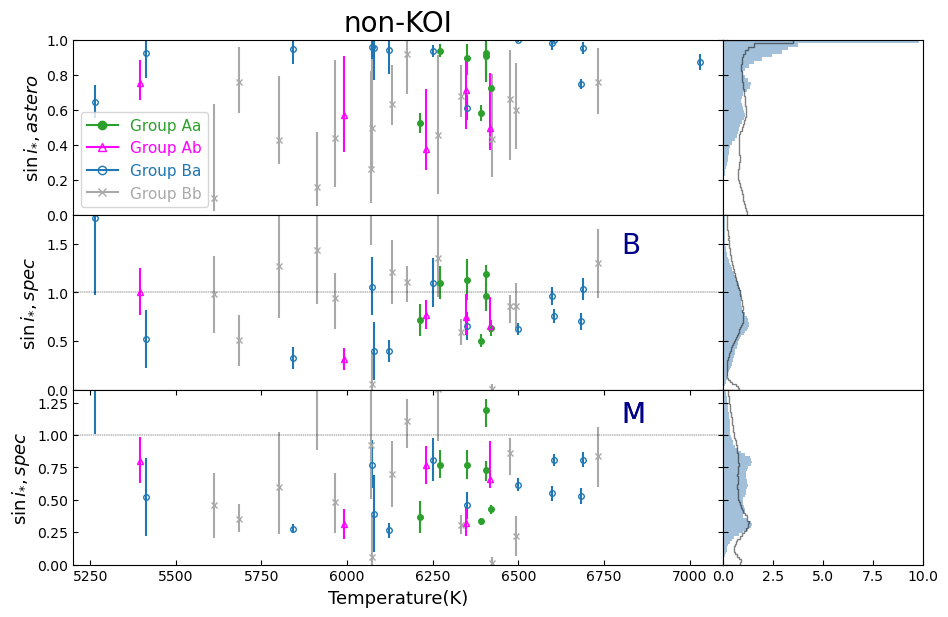}
\caption{Same as Figure \ref{fig:incl-Teff-koi} but for non-KOI
  stars. Spectroscopic $v\sin i$ values for deriving $\sin i_{*,
    spec}$ are adopted from \cite{Lund2017}, \cite{Molenda2013},
  \cite{Bruntt2012} and \cite{Brewer2016}. For some targets, there are
  more than one available $v\sin i$ values. In the second panel, we
  adopt \cite{Bruntt2012} if there are both $v\sin i$ from
  \cite{Bruntt2012} and from other literature. In the third panel, we
  use $v\sin i$ from \cite{Molenda2013} if multiple values are
  available.}
\label{fig:incl-Teff-nonkoi}
\end{figure}

Left panels of Figure \ref{fig:incl-Teff-koi} plot $\sin i_{*,
    \rm{astero}}$ and $\sin i_{*, \rm{spec}}$ for KOI stars against
  the stellar effective temperature following the same procedure as
  described in \cite{Kamiaka2018} (in their Section 5.3), while right
  panels show the corresponding distribution functions.

Although the sample of KOI stars is small, we note that if one
  excludes stars from Group Bb (the least reliable stars), KOIs tend to
  have $\sin i_*$ clustered around unity, and that spectroscopy and
  seismology are in good agreement. This indicates that a majority of
  stars with planets show a preference for a spin-orbit alignment, which is consistent with the previous conclusion by
  \cite{Kamiaka2018}.

Figure \ref{fig:incl-Teff-nonkoi} focuses on non-KOI
  stars. Excluding stars from Group Bb, $\sin i_{*, \rm{astero}}$
  exhibits a broader distribution compared to KOI stars. We note that
  the distribution is not uniform in $\sin i_*$ as it would be
  expected if there was no preference in their rotational spin
  direction on the sky. As shown by \cite{Kamiaka2018},
  asteroseismology overestimates inclinations below $i \simeq
  30^\circ$, corresponding to $\sin i_* \lessapprox 0.5$. This is
  likely the cause for a lack of population with reliable measurement
  when $\sin i_* < 0.4$.  In addition, as pointed out by
  \cite{Kamiaka2018}, available spectroscopic measurements are
  unreliable for some stars, in particular when the $v\sin i_*$ is low
  (due to the dominance of the stellar turbulence). Therefore,
  spectroscopic results from \cite{Molenda2013} and from
  \cite{Bruntt2012} sometimes lead to a mathematically inconsistent
  result of $\sin i_*>1$. If we exclude these impossible cases, there
  is no clear preferential stellar inclination. This is again
  consistent with an isotropic distribution of stars and contrasts
  with results from the KOI case shown in Figure
  \ref{fig:incl-Teff-nonkoi}.

\subsection{\rev{Complementarity of photometric and asteroseismic analyses}}

\rev{This paper has performed comprehensive photometric analyses of
  the 92 Kepler stars selected by \cite{Kamiaka2018} for the
  asterosesmic analysis.} \rev{The photometric intensity }
  \rev{variability measures the period of rotational modulation caused by
  the active regions at a specific latitude on stellar surface. The
  asteroseismic analysis, on the other hand, studies the pulsation
  pattern and derives an averaged rotation period over a range of
  latitudes and depths inside the star. Since stars are not ideal
  solid-body rotators, their radial and latitudinal differential
  rotations would affect the estimates of their {\it rotation periods}
  in different methods. Therefore, the comparison between the
  photometric and asteroseismic results provides unique and
  complementary insights into the nature of those targets.  One of the
  most important examples is the stellar inclination angle measured
  from asteroseismic analysis, which is a key ingredient for measuring
  a three-dimensional spin-orbit angle of exoplanetary systems
  \cite[see
    e.g.][]{Benomar2014b,Winn2015,Kamiaka2018,Kamiaka2019,Suto2019}.
  While the asteroseismic analysis provides a unique relation between
  the stellar inclination and the rotation period, its degeneracy is
  not easy to \rev{be} broken by asteroseismology alone
  \citep[see][]{Kamiaka2018}.  When a more precise estimate of the
  rotational period is available from the intensity variability, it is
  very useful to break the degeneracy.  Hence, a detailed examination
  of various photometric methods as presented in this work provides
  an important framework for those complementary approaches.}


\section{Summary and conclusions \label{sec:conclusion}}

We have performed a comprehensive photometric and asteroseismic
analysis of 92 solar-like main-sequence stars observed by
  \textit{Kepler}. We focus on 70 stars that do not have stellar companions, and classify them in four groups (Aa, Ab, Ba, and
Bb) according to the fractional variance and precision of the rotation
periods $\mPlsq$ and $P_{\rm astero}$; see \S
\ref{subsec:LS-classification} and
\ref{subsec:astero-classification}. We have presented detailed
comparison among photometric and asteroseismic constraints for those
stars on an individual basis.

Group Aa has 14 stars with robust estimates for both $\mPlsq$ and
$P_{\rm astero}$. Twelve out of the 14 stars have $\mPlsq \approx
P_{\rm astero}$ as expected. The remaining two include \textit{Kepler}-100
($\mPlsq\approx 13$ days and $P_{\rm astero} \approx 25$ days), and KIC
5773345 ($\mPlsq\approx 12$ days and $P_{\rm astero} \approx 6.5$
days); see \S \ref{app:kepler100} and \ref{app:5773345}. They are
potentially interesting targets for further study.  Group Ab has 9
stars. For these stars, $\mPlsq$ is consistent with $P_{\rm astero}$, except for
\textit{Kepler}-409 and KIC 7970740; see \S \ref{app:kepler409}
and\ref{app:kic7970740}.

There are 19 stars in Group Ba and 28 stars in Group Bb. Since their
photometric rotation periods vary from quarters to quarters, it is not
easy to assign definite rotation periods photometrically. For stars in Group Ba, $P_{\rm astero}$ provides a better estimate for the
true rotation period without being affected by the stellar surface
activity.  It is interesting to note that the stars for which
photometric and asteroseismic periods are different exhibit a tendency
of $P_{\rm astero} \sim 2 \mPlsq$ (Figure \ref{fig:incon}).  This may
imply that $\mPlsq$ represents the second harmonics of the true
rotation period of $P_{\rm astero}$, at least for some of those stars.

We find that there are 4 KOI stars in our sample whose rotation period
agrees with the orbital period of their planetary companion; \textit{Kepler}-65
(Group Aa), \textit{Kepler}-50 (Group Aa), \textit{Kepler}-1655 (Group Bb), and
KOI-288. Due to a limited number of our sample, this may be simply a
statistical fluke and not definitive, but could imply a possible
interaction between the star and planet in those systems
\citep{Suto2019}.

Most of stars exhibit significant quarter-to-quarter variations of
$\Plsq$, which come from the combined effects of the differential
rotation, various stellar surface activities, and other noises.  Among
them, 7 targets including \textit{Kepler}-408, KOI-974, \textit{Kepler}-50, KOI-269, KIC
7103006, KIC 7206837, and KOI-268 may be potentially interesting in
constraining their differential rotation.

Our main findings are summarized as follows.
\begin{enumerate}
\item Most of our targets exhibit significant quarter-to-quarter
  variances in the photometric periods. This is reasonable, given the
  complicated dynamics and activities of the stellar surface and
  starspots.  Thus, the estimated photometric period should be
  regarded as a simplified characterization of the true stellar
  rotation period, especially under the presence of the latitudinal
  differential rotation.
\item There is a fraction of stars with a relatively small
  quarter-to-quarter variance in the photometric periods (Group A; 23
  out of 70 stars in our sample), for which the three photometric
  methods (the Lomb-Scargle periodogram, autocorrelation function, and
  wavelet analysis) enable the precise determination of the stellar
  rotation period in a consistent manner.  Furthermore, it is
  encouraging that the asteroseismic rotation period for most of those
  stars agrees with the photometric rotation period within their
  uncertainties. Thus, rotation periods for stars satisfying our
  photometric classification condition should be reliable and accurate
  even without the asteroseismic estimate (that is not available in
  most cases).
\item Rotation periods more than $\sim$ 30 days either photometrically
  or asteroseismically are not reliable in our \textit{Kepler}
  sample. They may suffer from some residual systematics or correspond
  to the harmonics, and should be interpreted with caution. \rev{If
    the 30-day systematic trend is caused by the monthly re-pointing
    of \textit{Kepler}, it will vanish in other missions with less
    frequent re-orientation} \rev{such as PLATO \citep{Rauer2014_PLATO}.}
\item There are \rev{19} stars whose photometric and asteroseismic
  periods are not consistent. They should be potentially interesting
  targets deserved for further individual investigations.
\end{enumerate}

We would like to note that the large variability observed in some stars
  and reflected by our classification might be further explored in
  another study. Indeed, in some stars and through a spot modeling, it
  may be possible to estimate the latitudes of the spots, their
  migration over time and the stellar rotation profile.

\acknowledgments

We thank S. Aigrain for providing simulated Kepler light curves
introduced in \cite{Aigrain2015}.  \rev{Y.S. thanks Taksu Cheon and
  Kei Iida for their hospitality at Kochi University of Technology and
  at Kochi University, respectively.}  We gratefully acknowledge the
support from Grants-in Aid for Scientific Research by the Japan
Society for Promotion of Science (JSPS) No.18H012 and No.19H01947, and
from JSPS Core-to-core Program ``International Network of Planetary
Sciences''.

This research made use of Astropy,\footnote{http://www.astropy.org} a
community-developed core Python package for Astronomy
\citep{astropy2013, astropy2018}, Lightkurve, a Python package for
Kepler and TESS data analysis \citep{2018ascl.soft12013L}, Waipy, a
Python package for wavelet analysis \citep{waipy}, Matplotlib
\citep{Hunter2007}, Astroquery \citep{astroquery2019}, tesscut
\citep{tesscut2019} and pandas \citep{pandas2010}.

\appendix

\section{Methods to estimate the stellar rotation period
\label{sec:method}}

There are a variety of methods to identify the periodicity in the
photometric variations of the light curve.  Among them, the
Lomb-Scargle (LS) periodogram, autocorrelation function (ACF), and
wavelet analysis (WA) have been applied extensively to the
high-quality photometric data from space missions including
\textit{Kepler}.

For instance, \cite{Nielsen2013} applied the LS periodogram
to $12,151$ \textit{Kepler} main-sequence stars and detected their rotation
periods. \cite{McQuillan2014} measured rotation periods of $34,030$
\textit{Kepler} main-sequence stars using autocorrelation
function. \cite{Garcia2014} estimated the rotation period of 310
solar-like stars by combining autocorrelation function and wavelet
analysis.  Finally, we consider the asteroseismic rotation periods
obtained by \citet{Kamiaka2018}.

Even though those methods have been widely used in the literature,
they may have their own advantage and disadvantage.  Thus, we compute
the rotation periods separately derived from the four methods. The
detailed comparison of the resulting values is useful in understanding
their robustness and reliability.  In what follows, we briefly
describe the basic principle of the four methods, together with
typical examples of their outputs.

\subsection{Photometric analysis}

\subsubsection{The Lomb-Scargle periodogram \label{subsubsec:LS}}

The LS periodogram is one of the most popular methods to detect
periodic signals embedded in the data \citep{Lomb1976, Scargle1982}.
Specifically, we adopt the generalized LS periodogram by
\cite{Zechmeister2009}, which fits the data to a single -mode
sinusoidal function of frequency $f$ and a constant term.  We denote
the resulting best-fit squared residuals at each frequency as
$\chi^2(f)$. In the present paper, we adopt the standard
normalization:
\begin{equation}
\label{eq:LS-power}
  P(f)= \frac{\chi^2_0 -\chi^2(f)}{\chi^2_0},
\end{equation}
where $\chi^2_0$ is the best-fit squared residuals with respect to a
non-varying constant reference model.

We compute equation (\ref{eq:LS-power}) for each target at 
frequencies:
\begin{equation}
  f_n= f_{\rm min} + n~\delta f (< f_{\rm max}), 
\end{equation}
where $n$ is an integer, and we set $f_{\rm min}=(71{\rm day})^{-1}
=0.16\mu {\rm Hz}$ and $f_{\rm max}=(0.24{\rm day})^{-1}=48\mu {\rm
  Hz}$ so as to cover the range of the rotation period of the
solar-type stars. The frequency interval $\delta f$ is set to be
$(n_0T_0)^{-1}$, where $T_0$ is the total time span of the light curve
for each star (typically $\sim 1200$days), the oversampling factor
$n_0$ is chosen to be $20$ \citep[see][e.g.,]{VanderPlas2018}.

Finally, we smooth the periodogram sampled at $f_n$ using a box-car
filtering over 0.1 $\mu$ Hz $=(116{\rm days})^{-1}$ to suppress the
spurious discreteness effect in the frequency space.

\begin{figure}[t]
\centering
	\centering
	\subfigure{
		\begin{minipage}[t]{0.40\linewidth}
			\centering
			\includegraphics[width=1.0\linewidth]{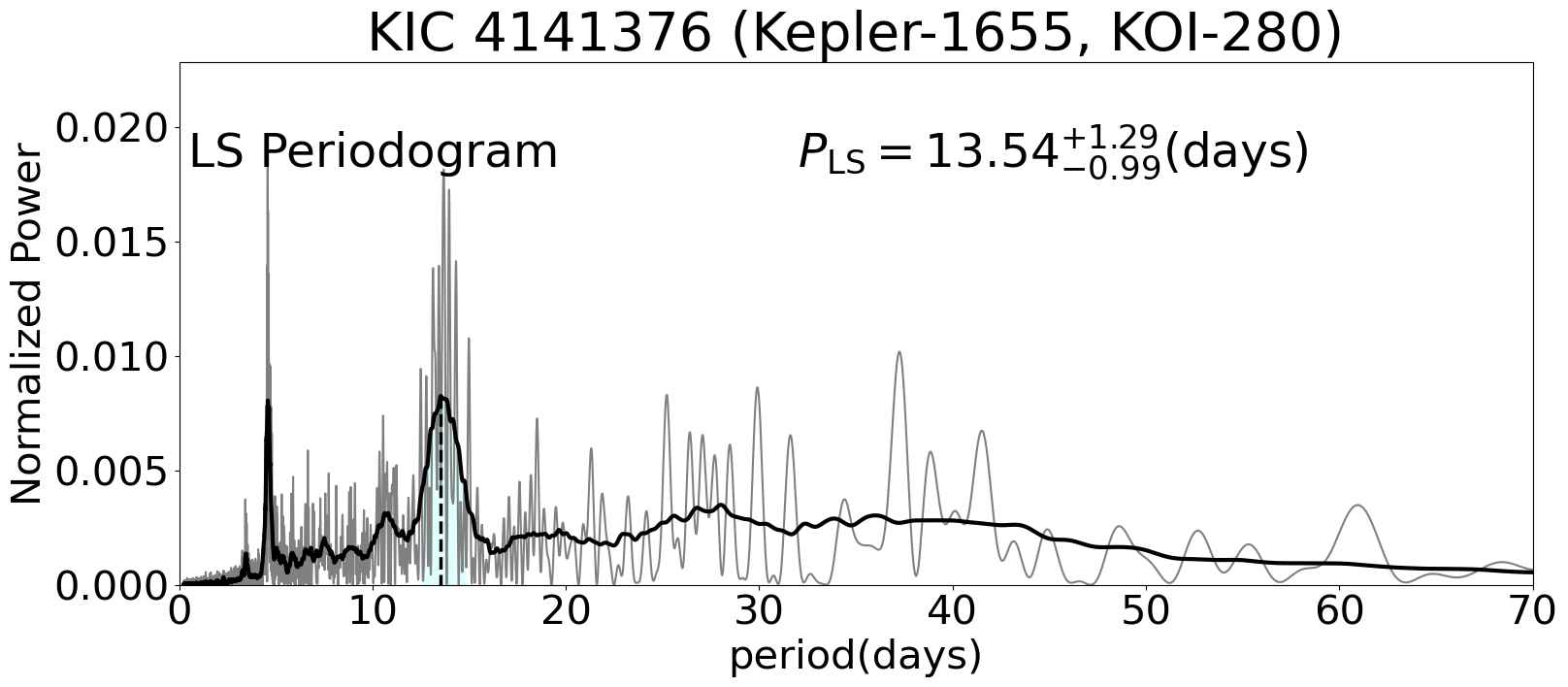}  
			\label{fig:spectrab4141376}
		\end{minipage}
	}\vspace{-0.5em}
	\subfigure{
		\begin{minipage}[t]{0.40\linewidth}
			\centering
			\includegraphics[width=1.0\linewidth]{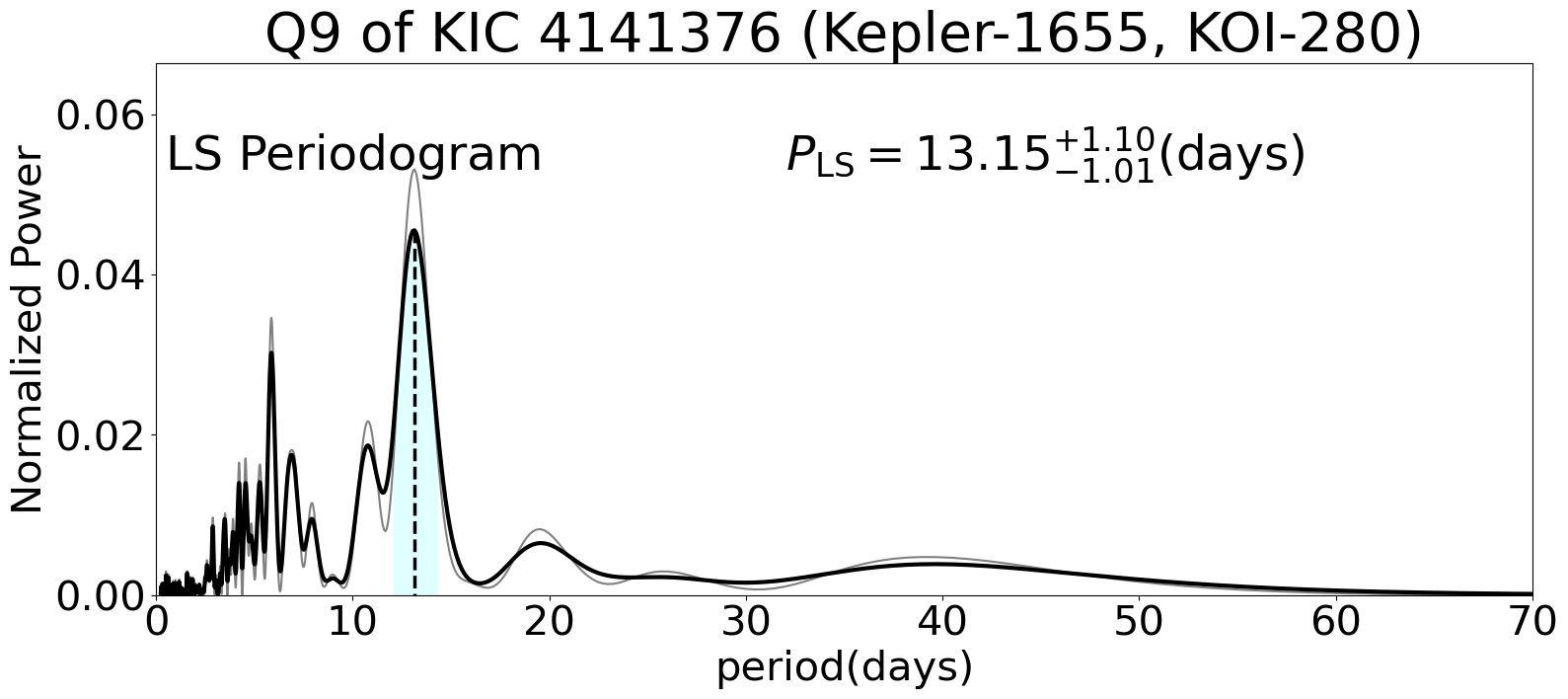}  
			\label{fig:spectrac4141376}
		\end{minipage}
	}\vspace{-0.5em}
\caption{An example of the LS periodogram for \textit{Kepler}-1655 (KIC
  4141376, KOI-280).  Left: LS periodogram computed from entire light curve. Right: periodogram computed from Q9. The thick black line indicates the
  boxcar-smoothed result (over 0.1 $\mu$Hz) of the original LS
  periodogram (thin gray curves).  The vertical blue dashed line
  indicates the location of $P_{\rm LS}$, and the associated blue
  shaded area corresponds to HWHM in both sides.}
\label{fig:LS-power4141376}
\end{figure}

Figure \ref{fig:LS-power4141376} shows an example of the LS
periodogram in time domain, {\it i.e.,} $P(T) \equiv P(f=1/T)$.  We
identify the highest peak in the smoothed periodogram, and estimate
the LS  rotation period $P_{\rm LS}$ and the associated errors
$(\Delta P_{\rm LS})_+$ and $(\Delta P_{\rm LS})_-$ from the peak
location and the corresponding HWHM (half width at half maximum) as
illustrated in the vertical line and boundaries in the blue shaded
region of Figure \ref{fig:LS-power4141376}.

We repeat the same procedure using the light curve for the $q$-th
quarter (each with $T_0\approx 90$ days) separately, and obtain
$\Plsq$ ($q =2,\cdots, 14$). As we discuss extensively below,
the distribution of $\Plsq$ for different $q$ and the
comparison against $P_{\rm LS}$ are useful in examining both the
reliability of the measurements and the degree of the latitudinal
differential rotation due to the non-static nature of starspots.

We note here a fairly common problem in selecting the the highest peak
in the LS periodogram as the stellar rotation period $P_{\rm LS}$.
Figure \ref{fig:LS-power4141376}, for example, shows another
significant peak around 4 days, and it could correspond to a true
rotation period, a higher harmonic, or a false positive due to stellar
surface activity, instrumental noise, and/or contamination from
possible companions
\citep[e.g.][]{Reinhold2013,Santos2017,Chowdhury2018}. It could also
come from the symmetric distribution of starspots
\citep{SSNB2022}. Hence, we examine the reliability of
the highest peak in the LS periodogram by repeating the analysis in
individual quarters and also through the close comparison against
other complementary methods, in particular a time-localized
spectrum-wavelet power spectrum.

\subsubsection{Autocorrelation function}

We compute auto-correlation function \citep[see
  e.g.][]{McQuillan2013,McQuillan2014} of the light curve $L(t)$:
\begin{equation}
  \label{eq:ACF}
  A(\tau) = \frac{\int_0^{T_0/2} dt L(t)L(t+\tau)}
  {\int_0^{T_0/2} dt L(t)L(t)}.
\end{equation}
Then, we smooth $A(\tau)$ using the Gaussian filter of $50\delta
t\approx 1 {\rm day}$ in order to suppress the short-term modulations
unrelated to, and/or the harmonics of,  the stellar rotation.

\begin{figure}[t]
\centering
\includegraphics[width=10cm]{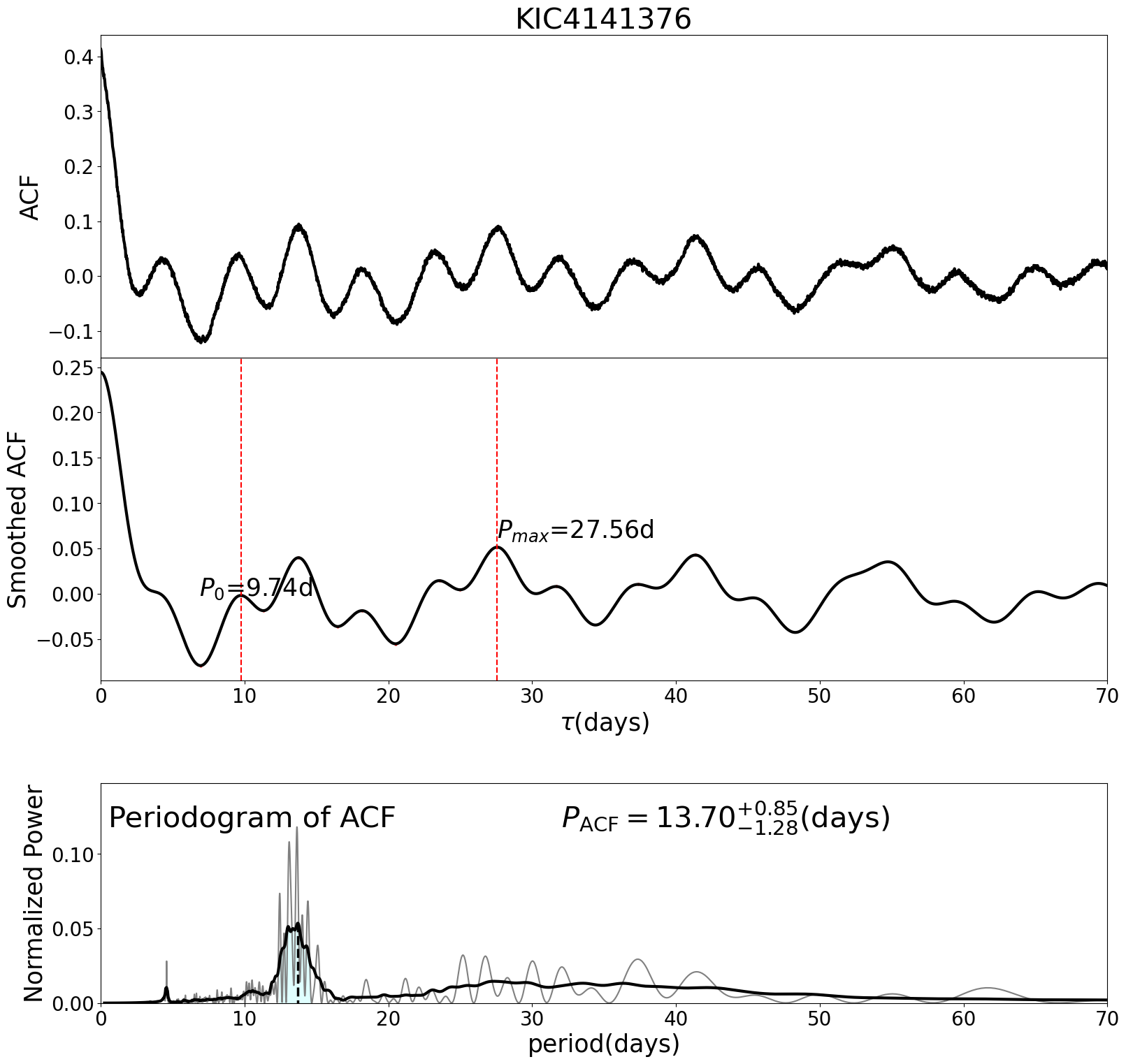}
\caption{Examples of ACF analysis for KIC 4141376 (KOI-280,
    \textit{Kepler}-1655).  Top, middle and bottom panels show ACF,
  smoothed ACF, and LS periodogram of the smoothed ACF. In the middle
  panel, $P_0$ is the period associated with the first maximum and
  $P_{max}$ is the one associated with the highest maximum.}
\label{fig:acf4141376}
\end{figure}

Figure \ref{fig:acf4141376} shows examples of the ACF analysis for
\textit{Kepler}-1655 (KIC 4141376, KOI-280).  The periodic patterns in
the top and middle panels are the clear signature of the stellar
rotation.  A substantial fraction of our current targets, however,
exhibits a more complicated pattern. So we apply the generalized LS
periodogram to the smoothed ACF, and extract the embedded period.
Following the procedure described in \S \ref{subsubsec:LS}, we define
the stellar rotation period $P_{\rm ACF}$ and the associated
uncertainties from the ACF method.

\subsubsection{Wavelet}

Wavelet analysis (WA) is one of the widely applied methods that
identify time-dependent signals in a time-frequency domain. Since the
photometric variations in the stellar light curves are time-dependent
over the observed duration ($T_0$) reflecting the dynamical nature of
spots on the stellar surface, WA is naturally suited for the detection
of the underlying stellar rotation period including the possible
effect of the differential rotation.

Following \citet{Torrence1998,Garcia2014}, we adopt the Morlet wavelet
that corresponds to a sinusoidal wave localized in the time domain
using a Gaussian window of the form:
\begin{equation}
  \psi(t;s) = \frac{1}{\pi^{1/4}}e^{iw_0t/s}
  \exp\left(-\frac{t^2}{2s^2}\right),
\end{equation}
and search for the best-fit value of the scale $s$.  The
  dimensionless parameter $w_0$ in introduced to control the
  resolution in the time-domain, and we adopt the common value of
  $w_0=6$. The scale $s$ is related to the period of the signal in the
  standard Fourier mode $P$ as
\begin{equation}
  \label{eq:P-s}
  P = \frac{4\pi}{w_0 + \sqrt{2+w_0^2}} s,
\end{equation}
\citet{Torrence1998}.  In practice, we search for $s$ in the range
between 0.1 days and 65 days in a linearly equal bin of 0.1 days.

\begin{figure}[t]
  \centering
  \includegraphics[width=10cm]{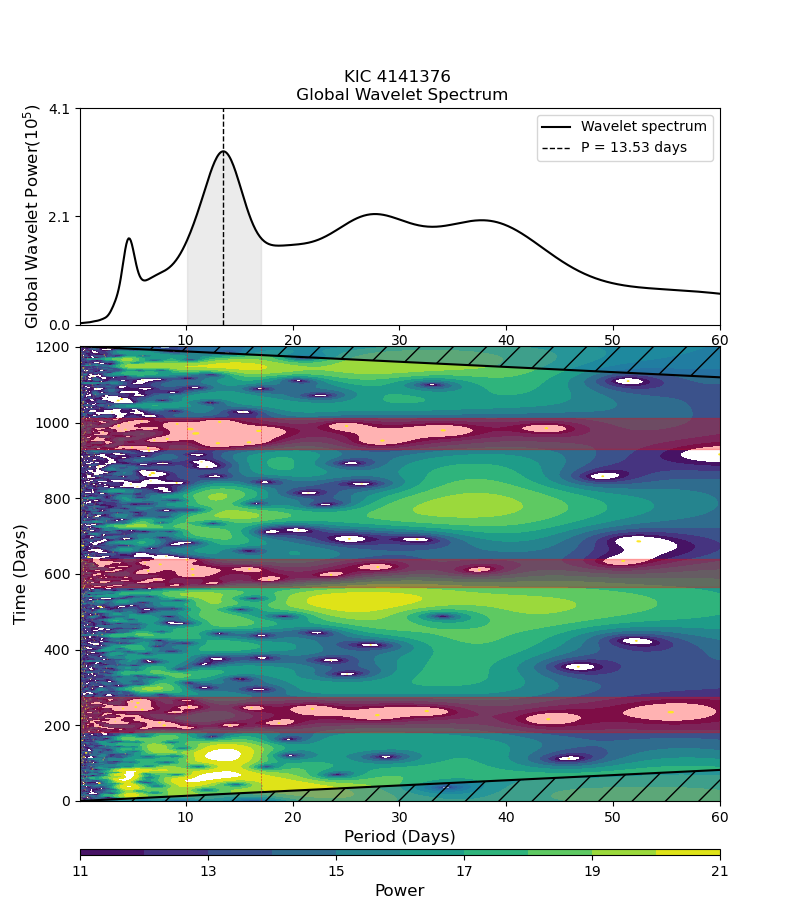}
\caption{Wavelet power spectrum (lower panel) and global wavelet power
  spectrum (upper panel) of \textit{Kepler}-1655 (KIC 4141376,
  KOI-280). The brighter color in wavelet power spectrum indicate a
  larger power. The cross-hatched region near the bottom edge of the
  waver power spectrum is the cone of influence within which power is
  less reliable. The dotted line in the right panel marks the maxima
  of GWPS, which corresponds to the measured period. The removed
    quarters are indicated in magenta. The shaded region in the upper
    panel covers the HWHM in both sides of the peak.}
	\label{fig:waveletkic4141376}
\end{figure}

Figure \ref{fig:waveletkic4141376} presents an example of WA for
\textit{Kepler}-1655 (KIC 4141376, KOI-280). The left panel shows the
wavelet power spectrum (WPS), while the right panel is the
corresponding global wavelet power spectrum (GWPS). GWPS is defined as
the average of WPS over time axis multiplied by the variance of the
time series. Note that the wavelet spectra in the present paper are
plotted against $P$, equation (\ref{eq:P-s}), instead of $s$.
Finally, we define the WA rotation period $P_{\rm WA}$ from the
highest peak in the GWPS, and compute the uncertainties based on its
HWHM.

While the above definition of the uncertainties in GWPS turns out to
be generally larger than those in LS and ACF, this may be simply due
to a matter of definition, given the fact that a majority of their
peak profiles is far from Gaussian nor symmetric. Thus, their
  quantitative interpretation and comparison against other results
  should be made with caution.

Instead, we would like to emphasize that WPS provides useful
information on the time-dependence of the measured stellar rotation
period over the entire observing time, which is likely related to the
strength of the differential rotation of each star. This is the most
important and complementary aspect of WA.  For instance, the left
panel of Figure \ref{fig:waveletkic4141376} indicates that the
measured rotation period is relatively stable and does not show any
detectable variation, implying that the differential rotation is very
weak and/or that those spots responsible for the photometric
variations are static over a timescale of 1200 days.

It is also important to note that Figure \ref{fig:waveletkic4141376}
clearly indicates that the peak around 4 days exhibited in Figures
\ref{fig:LS-power4141376} and \ref{fig:acf4141376} mainly come from
the contribution of Q2. While it is not clear why Q2 behaves very
differently from the other quarters, WA strongly indicates that the 4
day period does not correspond to the true stellar rotation
period. This illustrates the complementary role of the WA relative to
other methods.

\subsection{Asteroseismic analysis \label{subsec:asteroseismic}}

Asteroseismic analysis estimates the stellar rotation frequency
$\delta\nu_*$ and inclination $i_*$ by modeling the pulsation pattern
of stars around a few thousands $\mu$Hz \citep[see][e.g.,]{Ledoux1951,
  Tassoul1980, Gizon2003, Mosser2013}.

\begin{figure}[t]
  \centering
  \includegraphics[width=10cm]{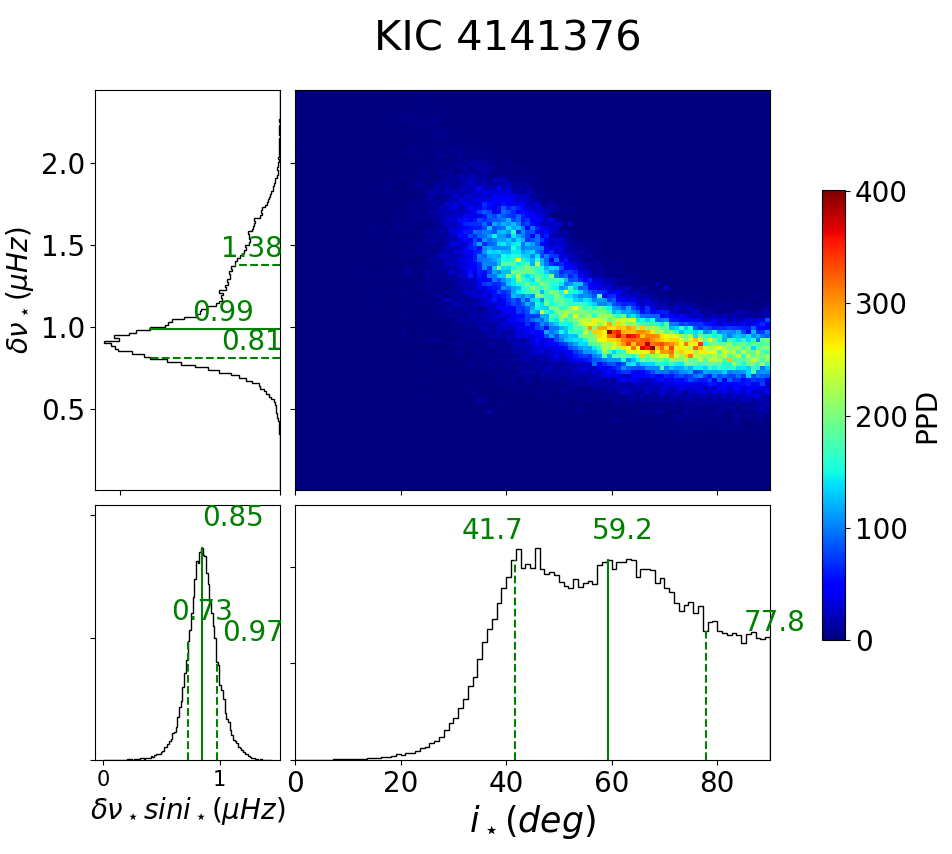}
  \caption{Posterior probability density (PPD) on the stellar
    inclination $i_*$ and rotation frequency $\delta\nu_*$ for
    \textit{Kepler}-1655 (KIC 4141376, KOI-280) marginalized over all
    other parameters. Top-left, bottom-left, and bottom-right panels
    plot one-dimensional marginalized densities of $\delta\nu_*$,
    $\delta\nu_*\sin{i_*}$, and $i_*$, respectively. Solid and dashed
    lines in those panels indicate the median and 16\%--84\% intervals
    for those parameters derived from asteroseismology alone.  }
\label{fig:seismic4141376}
\end{figure}

Figure \ref{fig:seismic4141376} shows an example of constraints on
$i_*$ and $\delta\nu_*$ for \textit{Kepler}-1655 (KIC 4141376,
KOI-280), derived from asteroseismic analysis by
\citet{Kamiaka2018}. Following them, we define the asteroseismic
rotation period and their uncertainty as
\begin{eqnarray}
\label{eq:Pastero}
  P_{\rm astero} &=& \frac{1}{\delta\nu_*(50\%)}, \\
\label{eq:dPastero+}
  (\Delta P_{\rm astero})_+ &=& \frac{1}{\delta\nu_*(16\%)} - P_{\rm astero},\\
\label{eq:dPastero-}
  (\Delta P_{\rm astero})_- &=&  P_{\rm astero} - \frac{1}{\delta\nu_*(84\%)},
\end{eqnarray}
where $\delta\nu_*(p)$ refers to the value at the percentile $p$ in its
one-dimensional marginalized density (top-left panel in Figure
\ref{fig:seismic4141376}). We emphasize that the above definitions are not
necessarily unique, and that the comparison against the photometric
rotation period needs to be interpreted with caution especially when
$\Delta P_{\rm astero}/P_{\rm astero}$ is large.  We adopt the
asteroseismic result by \citet{Kamiaka2018} for all the target stars.
Further details of their analysis may be found in
\cite{Kamiaka2018}, \cite{Kamiaka2019}, and \cite{Suto2019}.

\section{Comparison with previous photometric analysis}

\begin{figure}[ht]
\centering
\includegraphics[width=14cm]{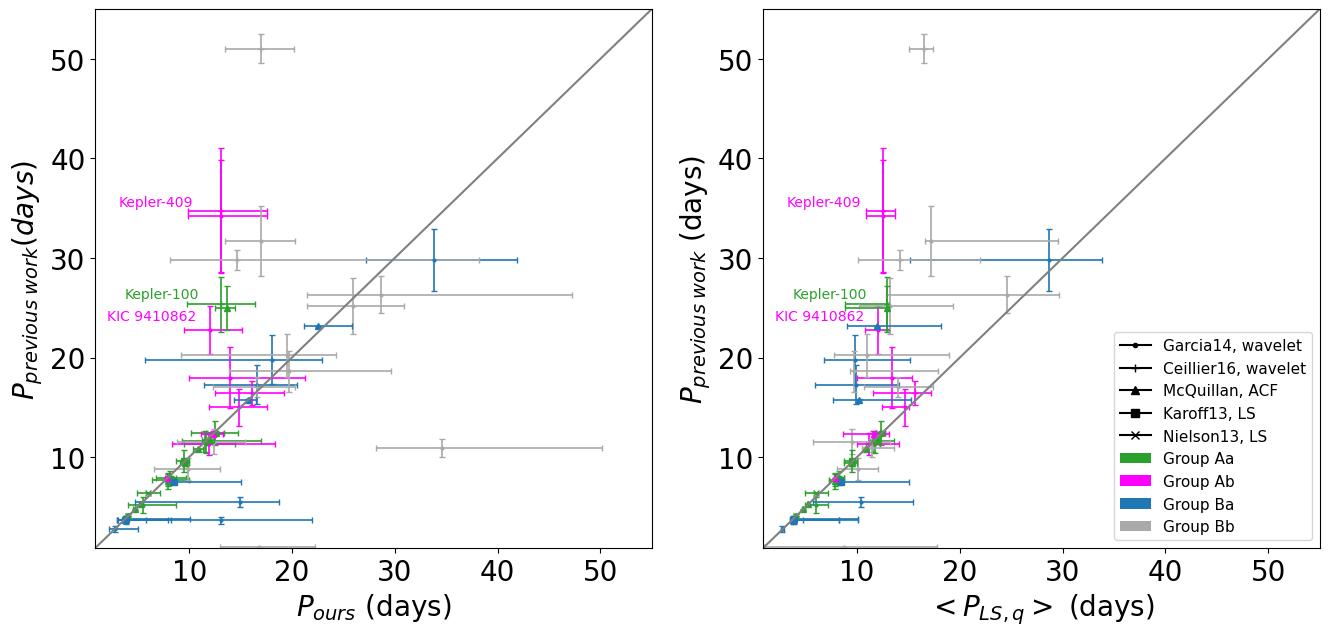}
\caption{Comparison of the rotation periods for 44 stars overlapped in
  the previous literature and our sample.  Left panel shows the
  comparison based on the same photometric method, while right panel
  plots their rotation periods against our $P_{\rm LS}$.
  Different colors indicate four groups that we introduced
  in the previous section, while different symbols indicate different
  papers; $P_{\rm LS}$ from \cite{Nielsen2013}
  and \cite{Karoff2013}, $P_{\rm ACF}$ from \cite{McQuillan2013} and
  \cite{McQuillan2014}, and $P_{\rm WA}$ from \cite{Ceillier2016} and
  \cite{Garcia2014}.}
\label{fig:PL}
\end{figure}

Among our 70 targets discussed in the previous section, 44 have their
photometric rotation periods published in previous literature. Left
panel of Figure \ref{fig:PL} plots the comparison of the rotation
periods derived from the same photometric method.  We use different
colors to distinguish the four groups that we introduced in the
previous section, while different symbols indicate different papers
shown in the caption; $P_{\rm LS}$ from \cite{Nielsen2013} and
\cite{Karoff2013}, $P_{\rm ACF}$ from \cite{McQuillan2013} and
\cite{McQuillan2014}, and $P_{\rm WA}$ from \cite{Ceillier2016} and
\cite{Garcia2014}. Right panel of Figure \ref{fig:PL} plots the
rotation periods in previous literature against our $P_{\rm LS}$ on
which our photometric classification is based.  In what follows, we
present detailed discussion of the comparison against those papers
individually.

\subsection[]{\citet{Nielsen2013}}

\cite{Nielsen2013} first conducted the LS analysis for Q2 to Q9
separately, using the \textit{Kepler} PDC-SAP light curve (SAP is
light curve without removal of systematic effect. msMAP is light curve
which has been processed, e.g. removal of systematic effect).  They
assume that $P_{\rm LS,q}$ should be within $[1, 30]$ days and define
the median of those $P_{\rm LS,q}$ as as the rotation period, $P_{\rm
  rot}$.  Among their final sample of 12151 stars, we have 7 targets
in common; 5 targets belong to our category of likely
binary/multiple-star systems (see Table \ref{tab:multiplestar}) which
we exclude in the comparison, and the other two stars (KIC 7206837 and
9812850) belong to our Group Aa and their rotation periods are in good
agreement with our $\langle L_{\rm LS}\rangle$.

\subsection[]{\cite{Karoff2013}}

Similarly, \cite{Karoff2013} performed a quarter-wise LS analysis for
a sample of 20 stars using Q2-Q14 of \textit{Kepler} PDC-SAP.  They
defined $P_{\rm rot}$ from the peak with S/N higher than 4 in all 14
quarters.  We have 8 overlapped targets, 4 in the category of likely
binary/multiple-star systems. The remaining 4 targets in Group Aa (KIC
12009504) and Group Ba (KIC 2837475, KIC 11253226, KIC 8694723) have
rotation periods consistent with our results.

Incidentally, we note that KIC 11253226 is one of the 8 targets for
which our LS, ACF and WA methods indicate different rotation periods
(Figure \ref{fig:Ba-nonKOI}). As we discuss in Section
\ref{subsec:Bmultiple}, when multiple peaks occur in spectra, the LS analysis tends to prefer the high frequency signal which lasts for over half of the entire observation,
while ACF and WA select a lower frequency signal. Since \cite{Karoff2013} required the signal to exist
clearly in all quarters of Q2-Q14, their value for KIC 11253226 is
consistent with our estimates of $\mPlsq \approx P_{\rm astero} \approx
4$ days, instead of $P_{\rm ACF}\approx P_{\rm WA}\approx 13$ days.

\subsection[]{\cite{McQuillan2013} and \cite{McQuillan2014}}

\cite{McQuillan2013} and \cite{McQuillan2014} applied ACF method on
Q3-Q14 of \textit{Kepler} PDC-MAP lightcurve. PDC-MAP is an earlier
version of data release for \textit{Kepler} PDC light curve. It
differs from PDC-msMAP in that the later one conducts a
stricter correction of the ~30 days Earth-point recovery
artifact\cite{Stumpe2014}. We have 20 stars in common, including 10
KOI stars and 10 non-KOI stars; ten stars belong to Group Aa,
two stars to Group Ba, two stars to Group Ab, and
the remaining six stars are in likely binary/multiple star systems.

Among 14 stars with no apparent stellar companion, \textit{Kepler}-100 is the
only case for which our $P_{\rm ACF}$ ($\approx 14$ days) is different
from their estimate ($\approx 25$days). This potentially interesting
case has been carefully examined in \S \ref{subsec:kepler100}.

\subsection[]{\cite{Garcia2014}}

\cite{Garcia2014} applied both wavelet analysis and
ACF\citep{McQuillan2013} method on concatenated Q0-Q14 \textit{Kepler} light
curves. When the two different methods estimate the rotation periods
in agreement within 20\%, they adopt the peak location of the GWPS as
$P_{\rm rot}$ and the HWHM of the corresponding peak as its
uncertainty.

We have 28 stars in common; three targets in Group Aa, eight in Group
Ba, five in Group Ab, nine in Group Bb, and three in the category of
possible binary/multiple star systems.  Among single star systems, two
stars in Group Ab (KIC 9410862 and \textit{Kepler}-409) and three
stars in Group Bb (KIC 12069127, 10644253 and 3656476) have rotation
periods inconsistent with their results.

The rotation periods for KIC 9410862, \textit{Kepler}-409, and 3656476
reported in \cite{Garcia2014} are about two to three times larger than
our results. We examined the GWPS spectra for these three targets, and
were unable to identify significant periodic components corresponding
to their values.

For KIC 9955598 (\textit{Kepler}-409, KOI-1925), we recover a period
  consistent with the value of $27.25$ days \citep{Garcia2014} if we
  include an anomalous quarter Q13 with sudden increase in photometric
  variation that is removed in our analysis (Figure
  \ref{fig:Ab-KOI}). Further discussion on \textit{Kepler}-409 is found in \S
  \ref{subsec:kepler409}.

For KIC 9410862 (Figure \ref{fig:Ab-nonKOI}) and 3656476 (Figure
\ref{fig:Bb-nonKOI}), the high-pass filter of a frequency of
  $1/20$ day$^{-1}$ applied to the latest PDC-SAP light curve (and
  PDC-msMAP) tends to suppress the signal of fundamental period for
  slow rotators \citep{Santos2019, Garcia2014}. This may explain why
  our result prefers shorter periods than theirs.  Our $P_{\rm
    astero}\approx 10$ days for KIC 9410862 is consistent with our
  photometric value, while $P_{\rm astero}\approx 40$ days for KIC
  3656476 is a factor of two larger than our photometric estimate and
  indeed consistent with the value of \cite{Garcia2014}.

As for the other two stars in Group Bb, KIC 12069127 (Figure
  \ref{fig:Bb-nonKOI}) is one of the targets with multiple significant
  periodic components in the GWPS spectra. The low frequency
  peak with P$\sim$17 days is selected by us while the high frequency
  peak with P$\sim$1 day is selected by \cite{Garcia2014}. Despite
  both peaks are detected in our GWPS spectra, smoothing applied
  to the raw power suppress the extremely high frequency which would
  lead to a different selection of period.  It would be important to
  note that a periodicity less than 1 day is close to the frequency
  range of stellar pulsation. In our sample, there are two targets
  showing significant peak at period less than 1 day (KIC 12069127 and
  9139163).

KIC 10644253 is also a target with multiple periodic components, a
continuous high frequency signal and a short low frequency signal. Our
quarter-wise measurements prefer $\left<P_{\rm LS,q}\right>$ over
$P_{\rm LS}$, $P_{\rm ACF}$ and $P_{\rm WA}$ (Figure
\ref{fig:Bb-nonKOI} and Table \ref{tab:singlestarBb}). Result from
\cite{Garcia2014} is consistent with our $\left<P_{\rm LS,q}\right>$
instead of $P_{\rm LS}$, $P_{\rm ACF}$ and $P_{\rm WA}$.

The rotation periods of \cite{Garcia2014} for the other 20 targets are
in agreement with our $P_{\rm WA}$. We note here, however, that there
are four targets whose values of $P_{\rm WA}$ disagree with our
$P_{\rm LS}$ and/or $P_{\rm ACF}$; their $P_{rot}$ for KIC 10068307
disagrees with our $P_{\rm LS}$ and $P_{\rm ACF}$ (Figure
\ref{fig:Bb-nonKOI}), and those for KIC 11081729, 11253226, and
6508366 agree with our $P_{\rm LS}$, but disagree with our $P_{\rm
  ACF}$ (Figure \ref{fig:Ba-nonKOI}).

\subsection[]{\cite{Ceillier2016}}

\cite{Ceillier2016} basically followed the method of
\cite{Garcia2014}, and measured the rotation period of 11 KOI stars,
all of which are included in our sample; four stars are in Group Aa,
two stars are in Group Ab, three in Group Bb, and the other two stars
are in likely binary/multiple star systems. There are three out of 11
stars for which their $P_{\rm rot}$ is different from our $P_{\rm
  WA}$; $P_{\rm rot}$ of \textit{Kepler}-100 is around twice of our
$P_{\rm WA}$, and $P_{\rm rot}$ of \textit{Kepler}-95, and
\textit{Kepler}-409 is around thee times larger than our $P_{\rm WA}$.

For \textit{Kepler}-100 and \textit{Kepler}-409, we confirmed that we could recover $P_{\rm
  WA}$ consistent with $P_{\rm rot}$ of \cite{Ceillier2016} if we
repeat the analysis with keeping somewhat anomalous quarters; see
discussion in \S \ref{subsec:kepler100} and \ref{subsec:kepler409},
respectively.  Their measured period for \textit{Kepler}-95 (KIC 8349582,
KOI-122) is longer than 50 days, which is beyond our detection limit
($\sim$50 days).

It is also useful to add the fact that the inconsistency for the
  above targets arise mainly due to the different choice of light
  curves; \cite{Ceillier2016} and \cite{Garcia2014} adopt KADACS,
  which tends to select longer period signals than PDC-msMAP light
  curve that is adopted in the present paper.

\clearpage

\section{Potentially interesting targets \label{sec:interesting-target}}

There are several targets that are potentially interesting and deserve
further studies. We show their LS periodogram for all quarters and
asteroseismic constraint in this section.

\subsection{\textit{Kepler}-100 (KIC 6521045, KOI-41) \label{app:kepler100}}

\begin{figure}[ht]
\centering
\includegraphics[width=14cm]{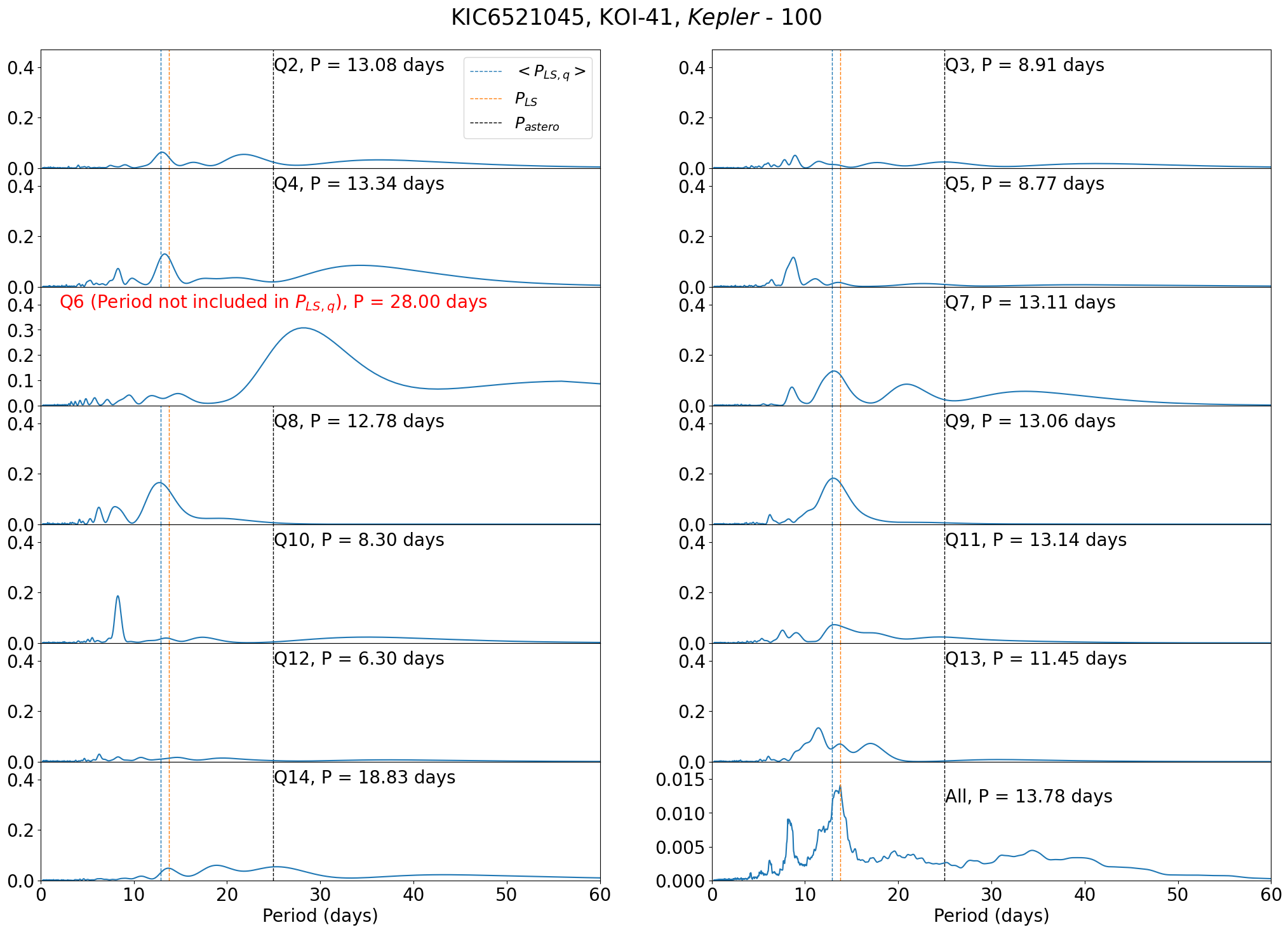}   
\caption{The LS periodograms of \textit{Kepler}-100 (Group Aa)
    for different quarters.}  
	\label{fig:kepler100-LSQ}
  \centering
  \includegraphics[width=8cm]{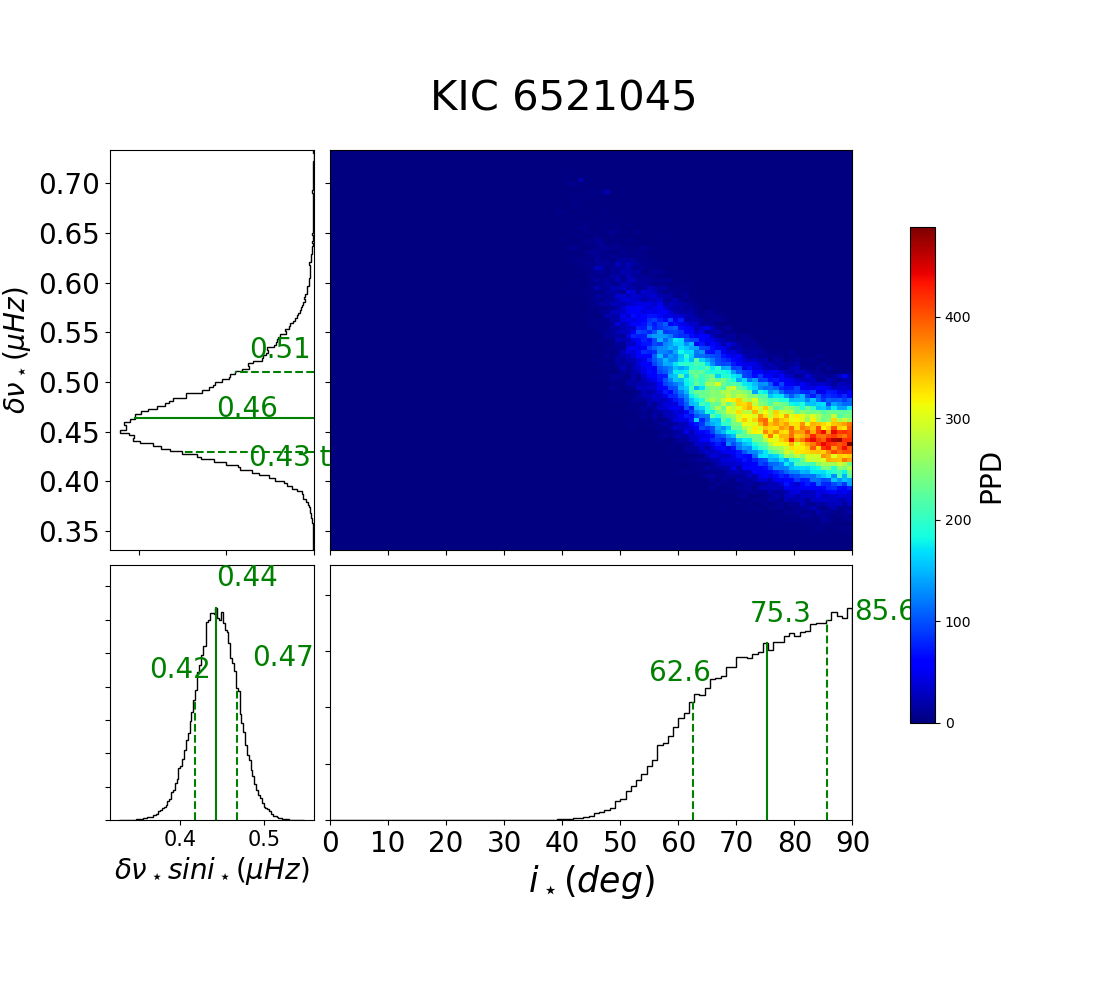}
  \caption{Asteroseismic constraints on $\delta\nu_*$ and $i_*$
      for \textit{Kepler}-100  (Group Aa).}
\label{fig:seismickepler100}
\end{figure}

\clearpage

\subsection{KIC 5773345 \label{app:5773345}}

\begin{figure}[ht]
\centering
\includegraphics[width=14cm]{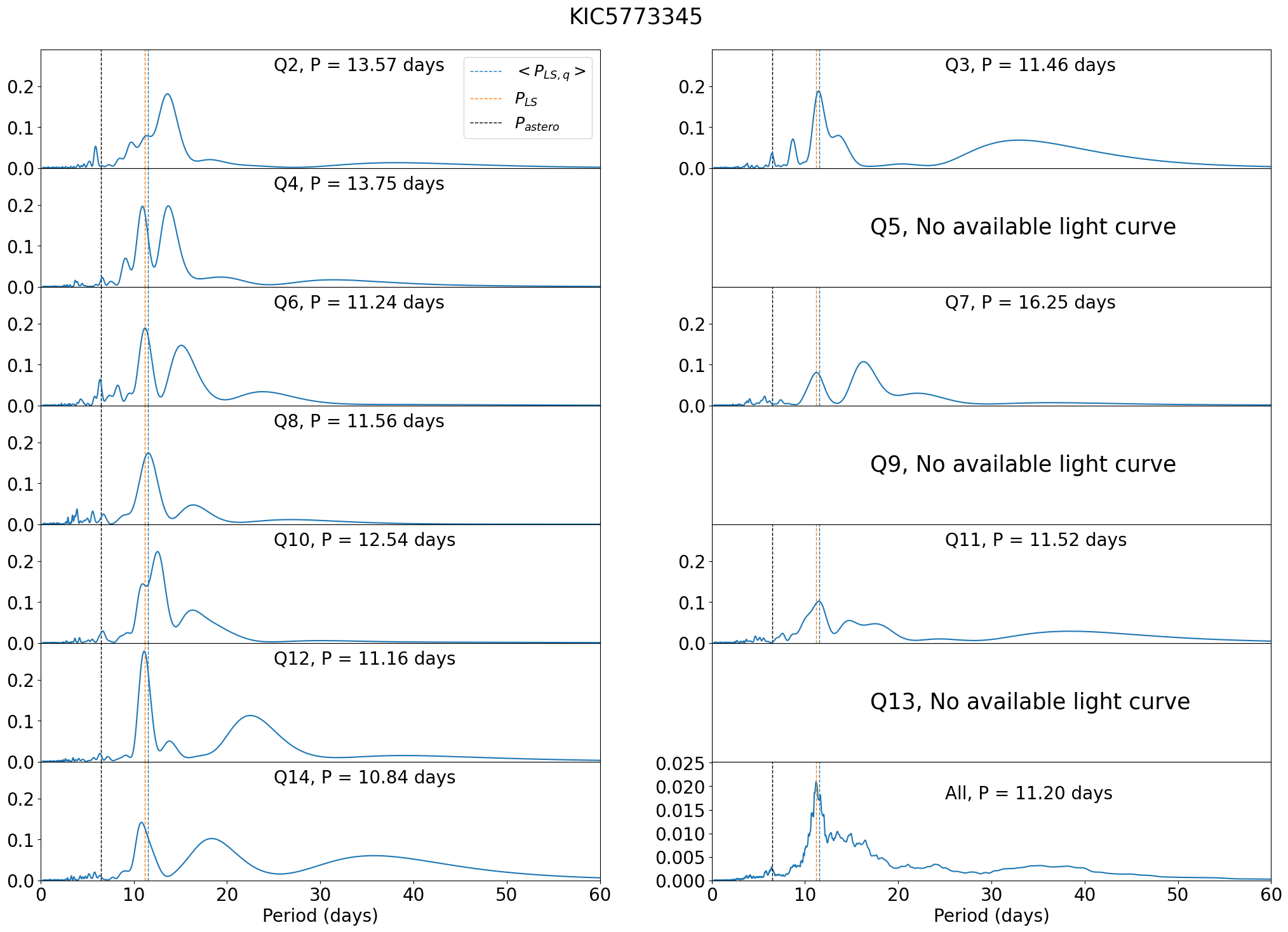}   
\caption{The LS periodograms of KIC 5773345  (Group Aa)
    for different quarters.}  
	\label{fig:5773345-LSQ}
\end{figure}

\begin{figure}[ht]
  \centering
  \includegraphics[width=8cm]{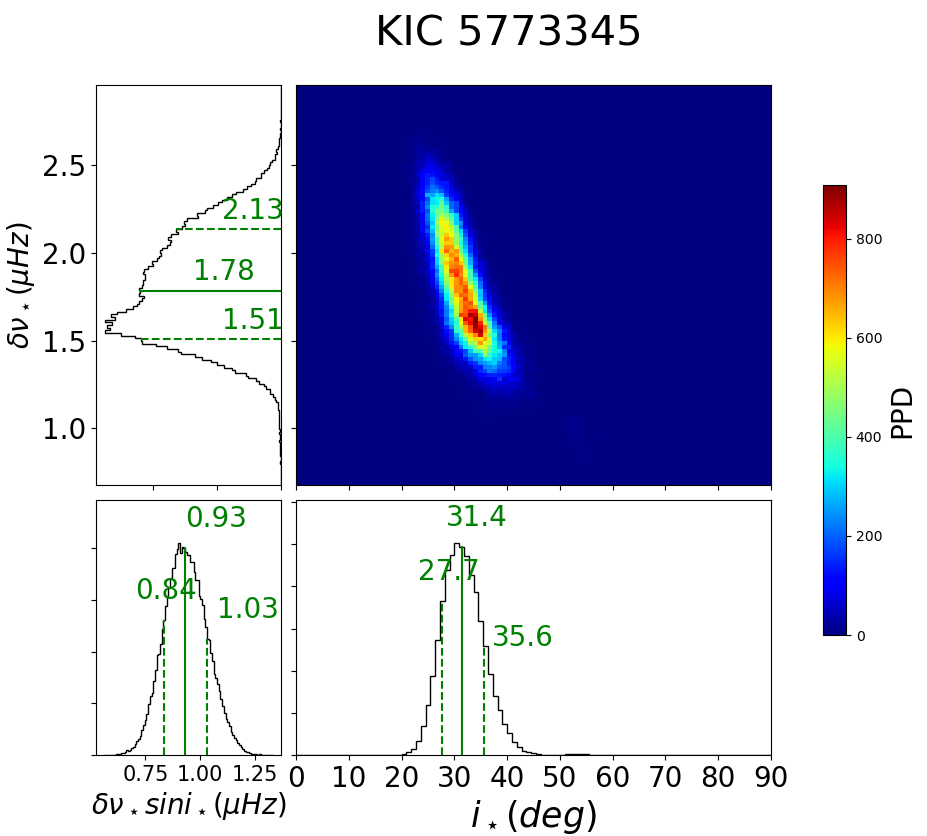}
  \caption{Asteroseismic constraints on $\delta\nu_*$ and $i_*$
      for KIC 5773345  (Group Aa).}
\label{fig:seismic5773345}
\end{figure}

\clearpage

\subsection{\textit{Kepler}-25  (KIC 4349452, KOI-244) \label{app:kepler25}}

\begin{figure}[ht]
\centering
\includegraphics[width=14cm]{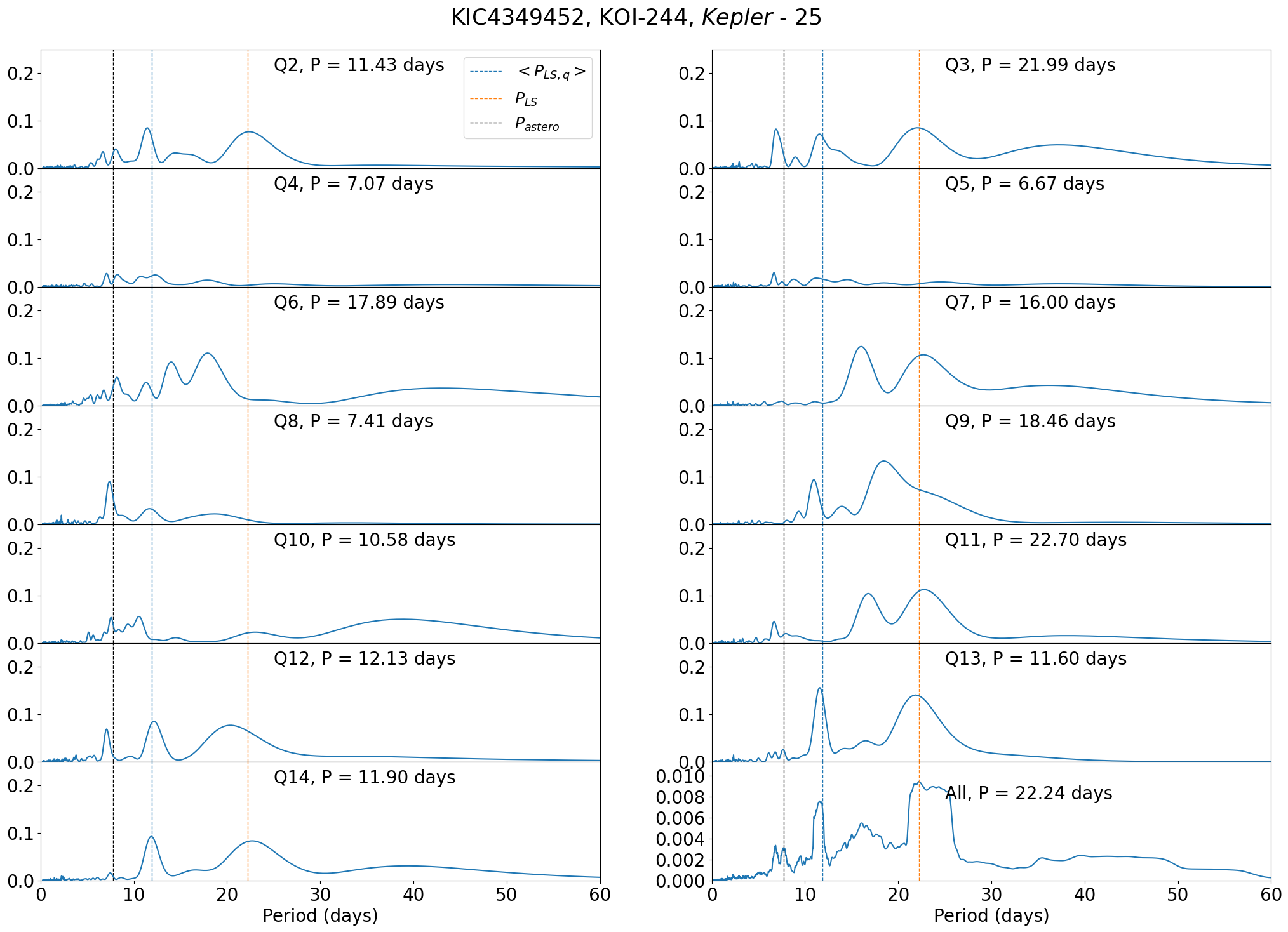}   
\caption{The LS periodograms of \textit{Kepler}-25  (Group Ba)
    for different quarters.}  
	\label{fig:kepler25-LSQ}
\end{figure}
\begin{figure}[ht]
  \centering
  \includegraphics[width=8cm]{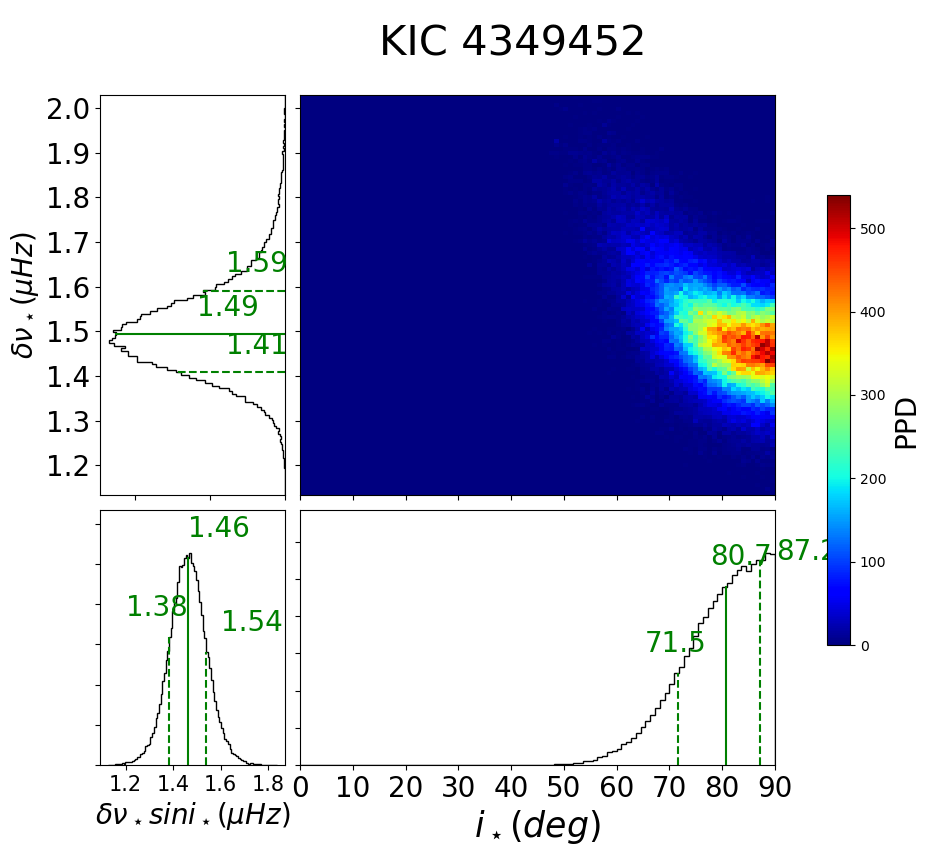}
  \caption{Asteroseismic constraints on $\delta\nu_*$ and $i_*$
        for \textit{Kepler}-25  (Group Ba).}
\label{fig:seismickepler25}
\end{figure}

\clearpage

\subsection{\textit{Kepler}-93 (KIC 3544595, KOI-69) \label{app:kepler93}}

\begin{figure}[ht]
\centering \includegraphics[width=14cm]{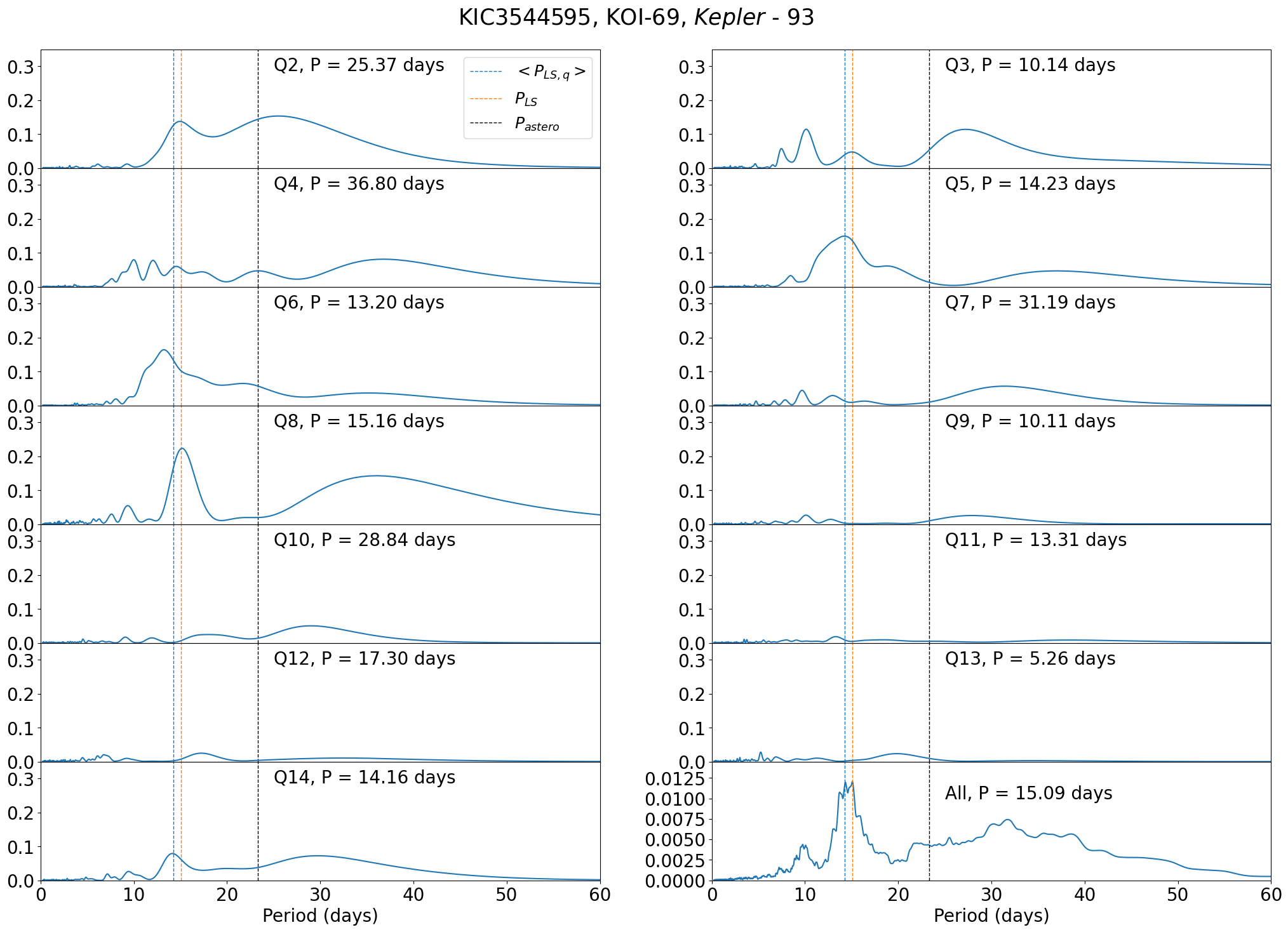}
\caption{The LS periodograms of \textit{Kepler}-93  (Group Ba)
    for different quarters.}  
	\label{fig:kepler93-LSQ}
\end{figure}
\begin{figure}[ht]
  \centering
  \includegraphics[width=8cm]{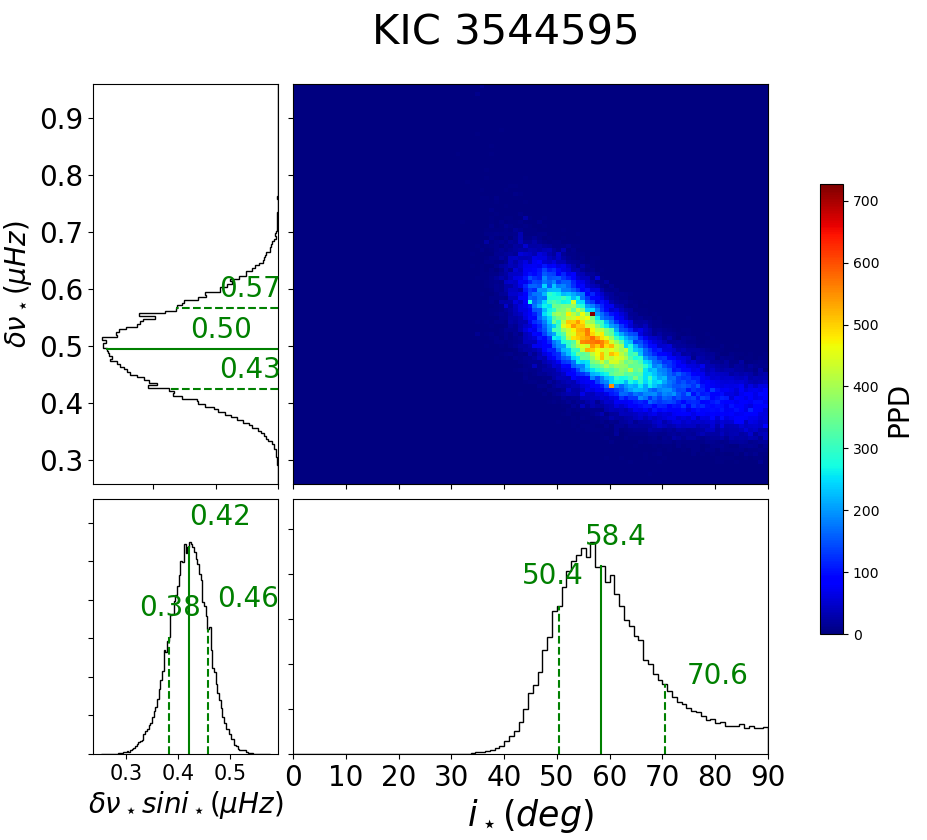}
  \caption{Asteroseismic constraints on $\delta\nu_*$ and $i_*$
        for \textit{Kepler}-93  (Group Ba).}
\label{fig:seismickepler93}
\end{figure}

\clearpage
\subsection{\textit{Kepler}-128  (KOI-274, KIC 8077137) \label{app:kepler128}}

\begin{figure}[ht]
\centering
\includegraphics[width=14cm]{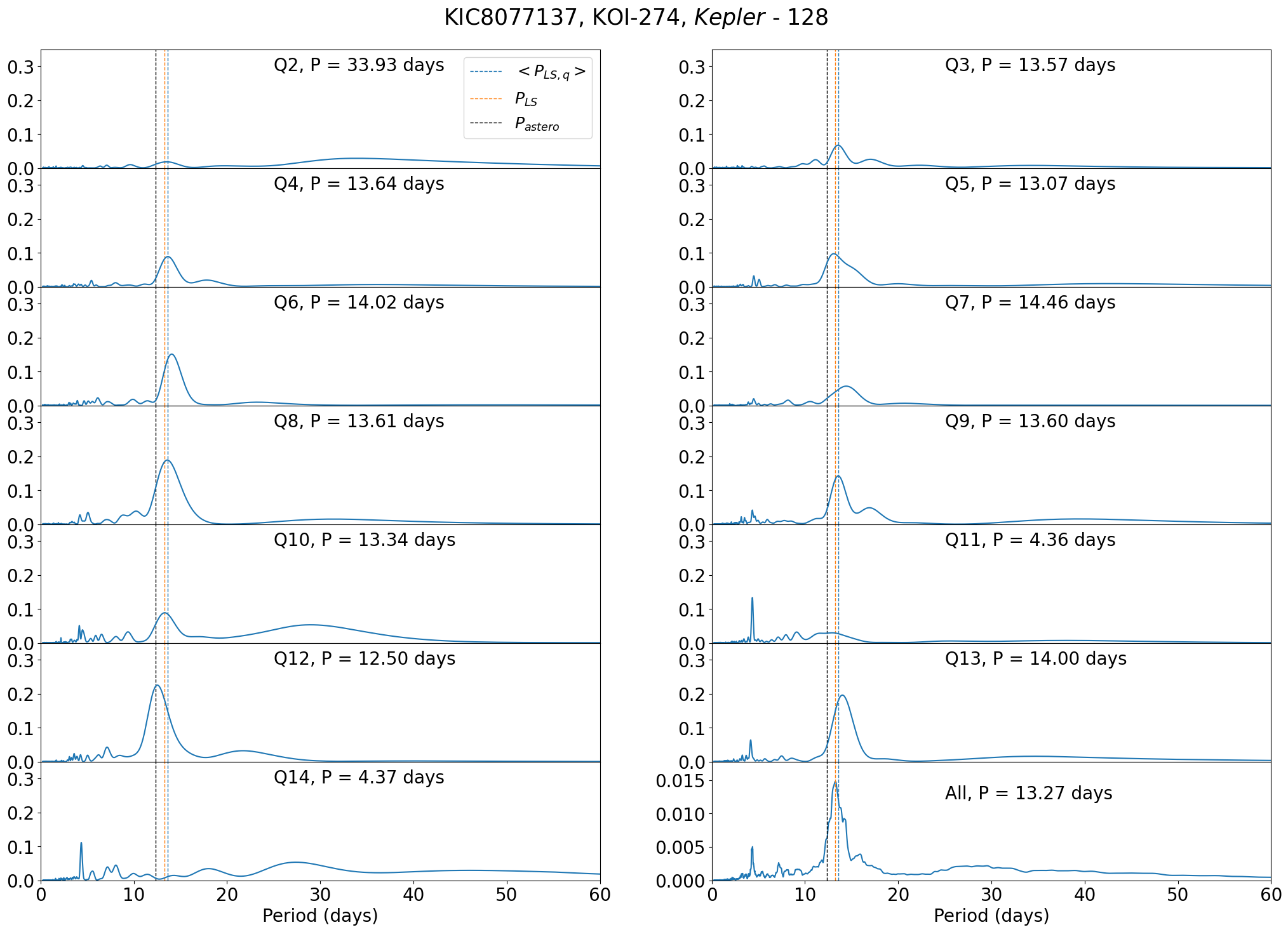}   
\caption{The LS periodograms of \textit{Kepler}-128  (Group Ba)
    for different quarters.}  
	\label{fig:kepler128-LSQ}
\end{figure}

\begin{figure}[ht]
  \centering
  \includegraphics[width=8cm]{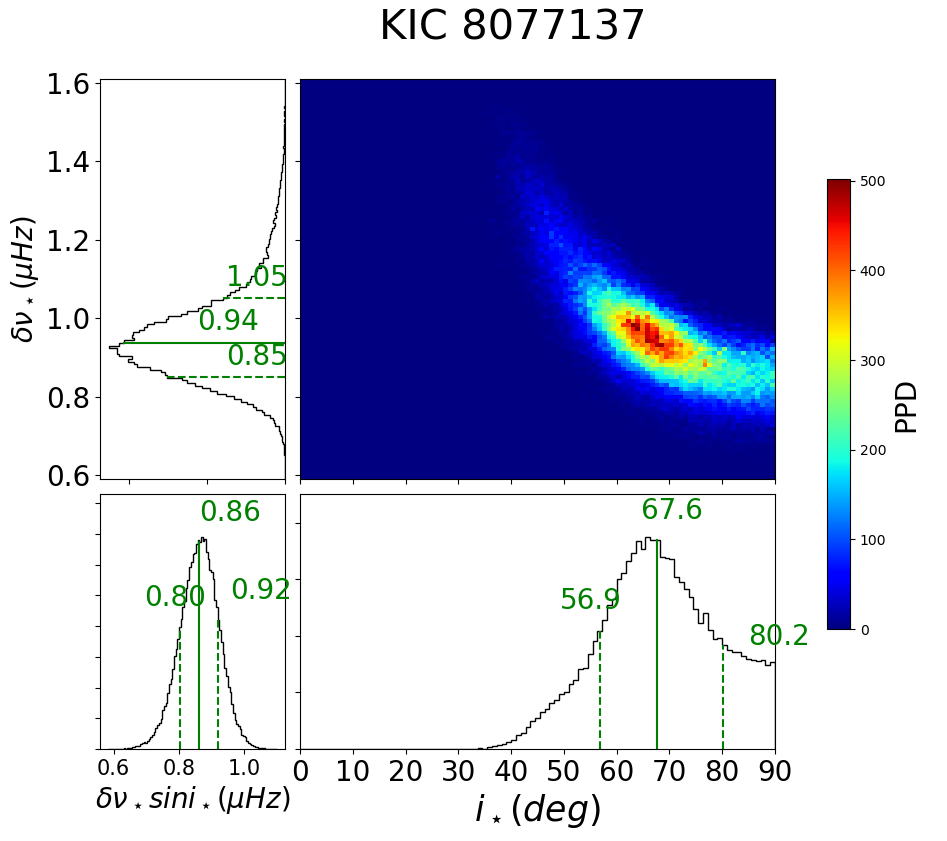}
  \caption{Asteroseismic constraints on $\delta\nu_*$ and $i_*$
        for \textit{Kepler}-128  (Group Ba).}
\label{fig:seismickepler128}
\end{figure}

\clearpage
\subsection{\textit{Kepler}-409  (KIC 9955598, KOI-1925) \label{app:kepler409}}

\begin{figure}[ht]
\centering
\includegraphics[width=14cm]{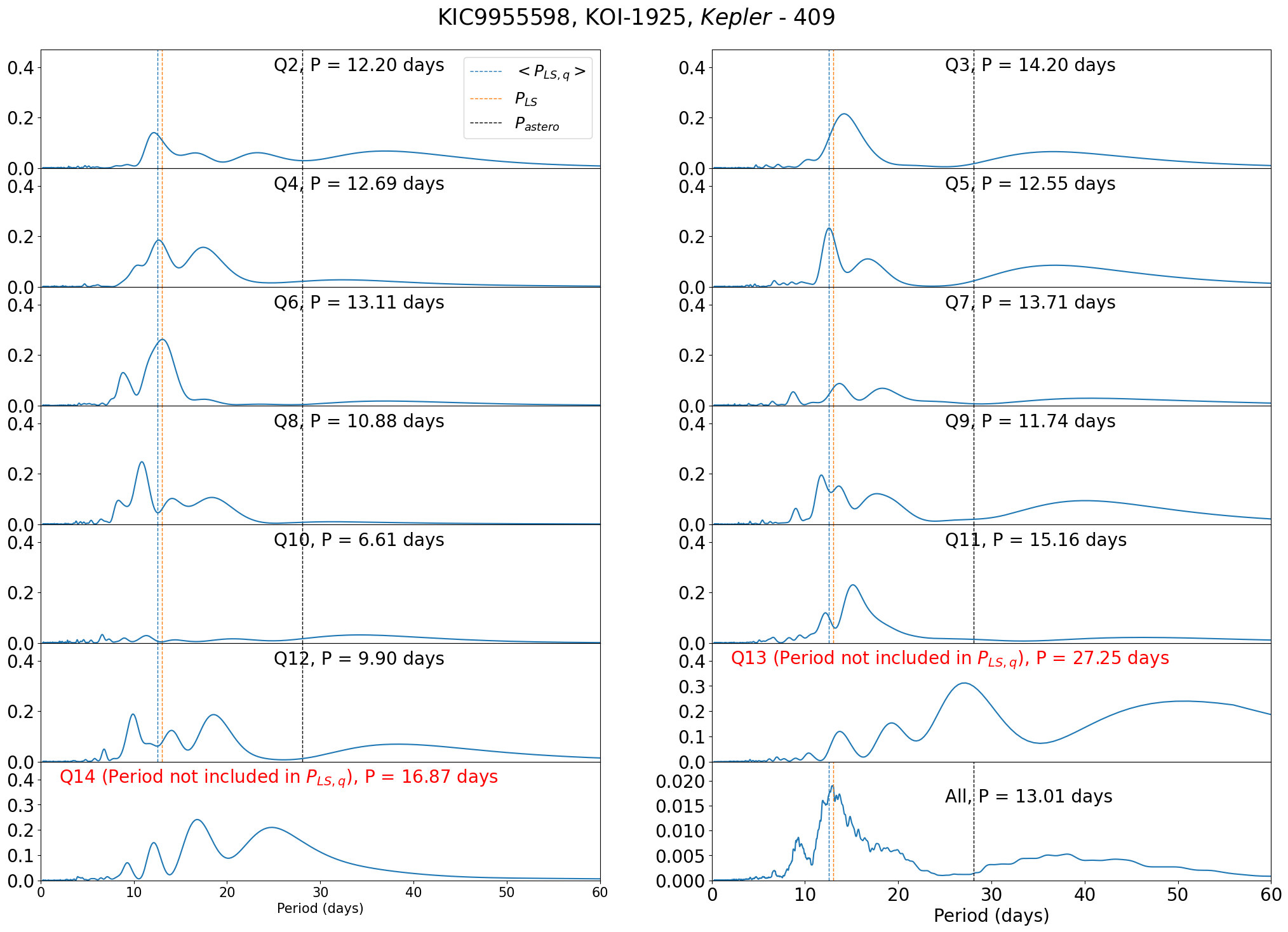}   
\caption{The LS periodograms of \textit{Kepler}-409  (Group Ab)
    for different quarters.}  
	\label{fig:kepler409-LSQ}
\end{figure}
\begin{figure}[ht]
  \centering
  \includegraphics[width=8cm]{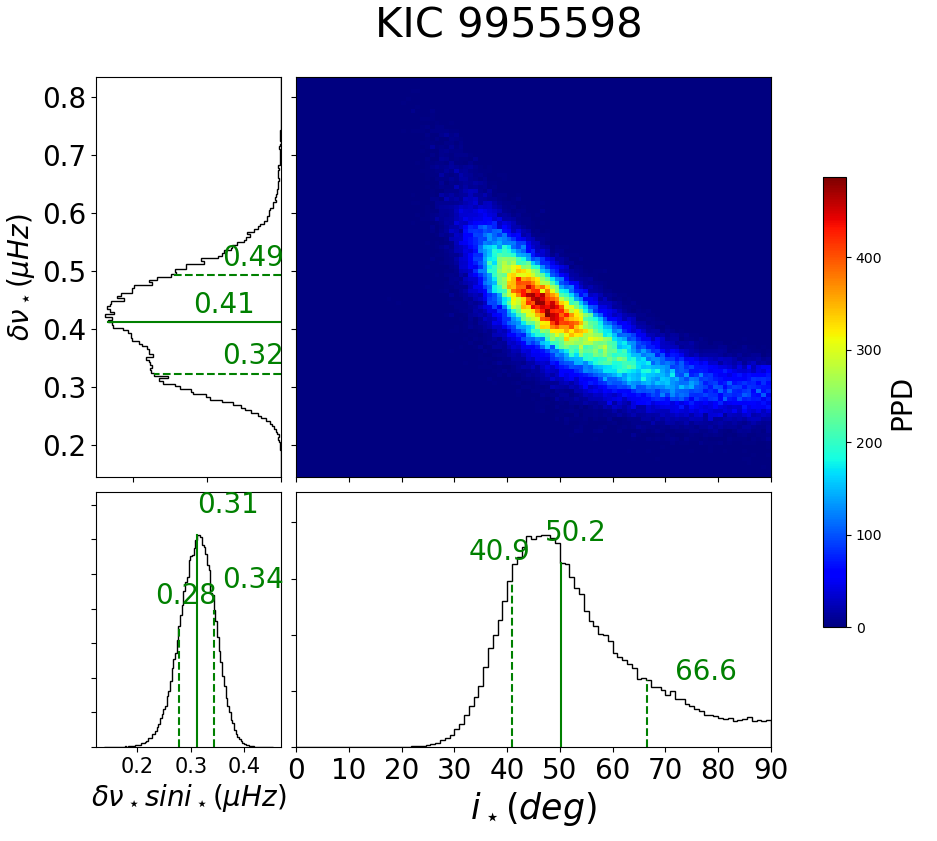}
  \caption{Asteroseismic constraints on $\delta\nu_*$ and $i_*$
        for \textit{Kepler}-409  (Group Ab).}
\label{fig:seismickepler409}
\end{figure}

\clearpage
\subsection{KIC 7970740 \label{app:kic7970740}}

\begin{figure}[ht]
\centering
\includegraphics[width=14cm]{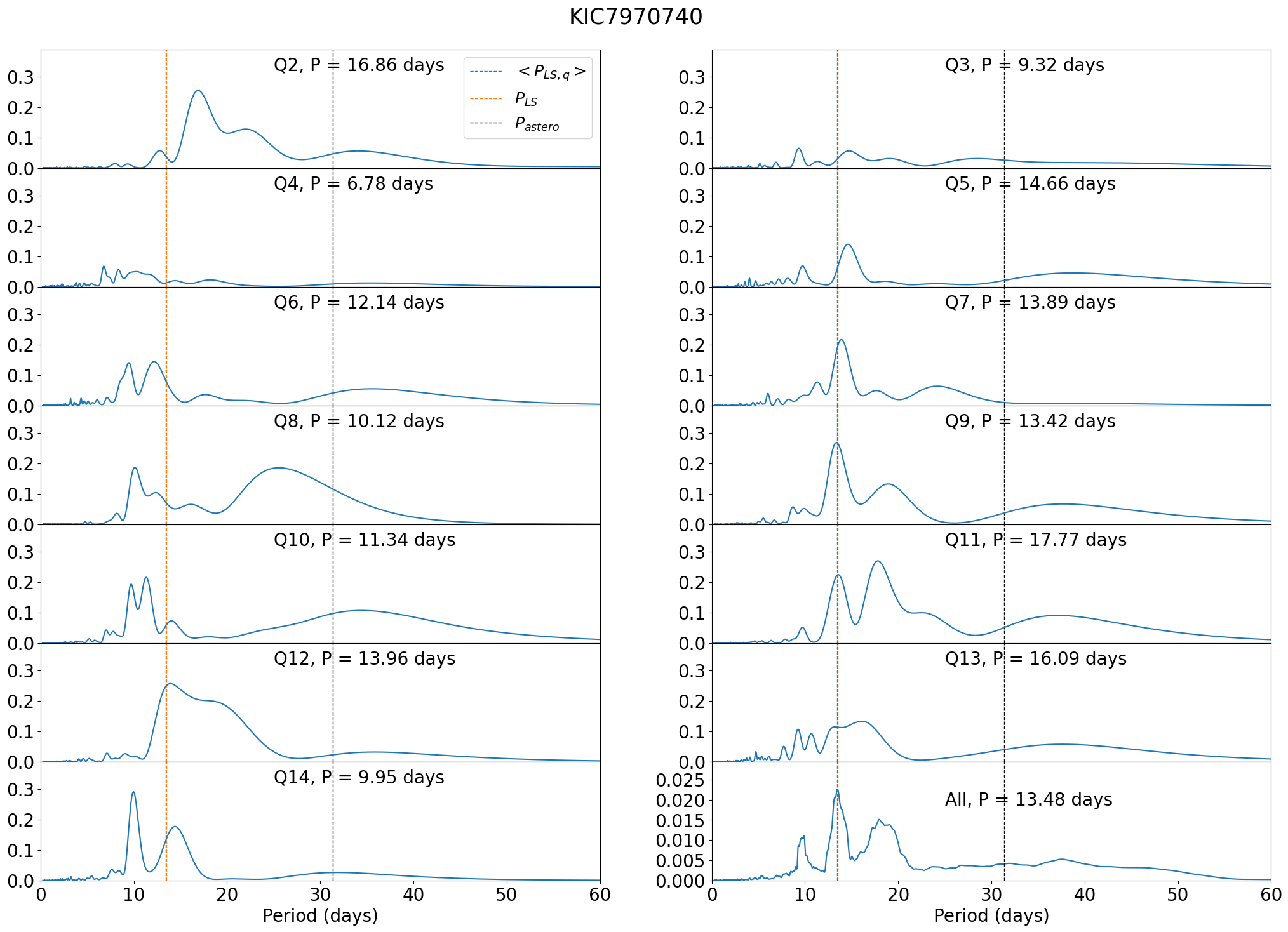}   
\caption{The LS periodograms of KIC 7970740  (Group Ab)
    for different quarters.}  
	\label{fig:7970740-LSQ}
\end{figure}
\begin{figure}[ht]
  \centering
  \includegraphics[width=8cm]{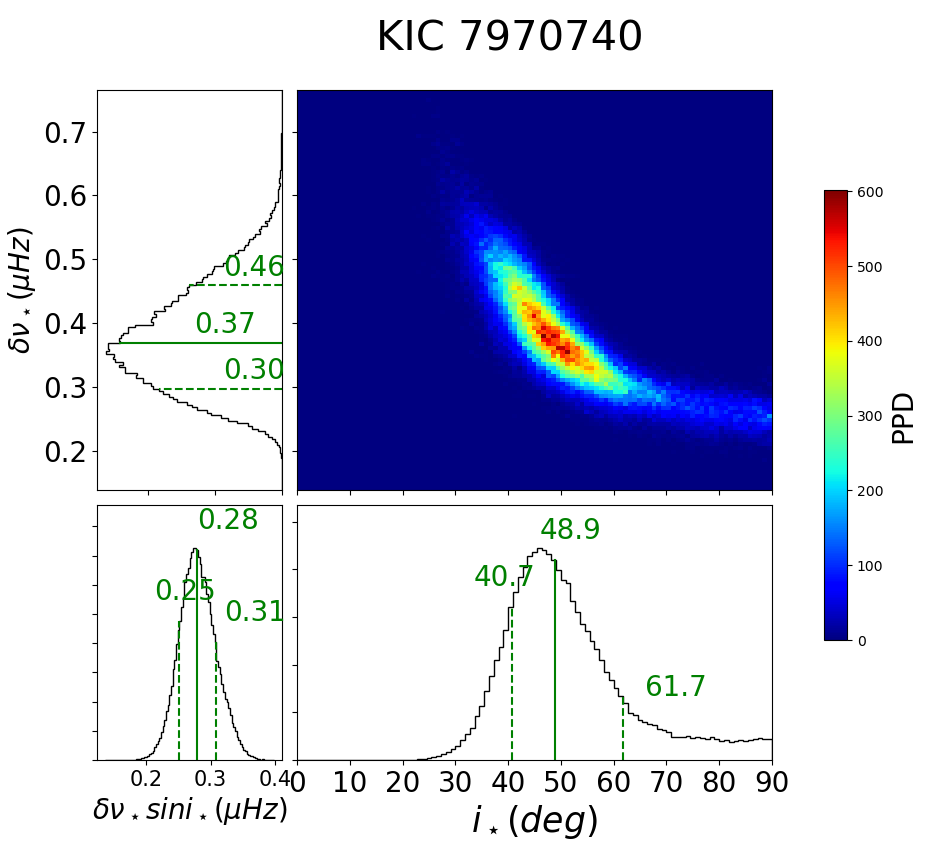}
  \caption{Asteroseismic constraints on $\delta\nu_*$ and $i_*$
        for KIC 7970740  (Group Ab).}
\label{fig:seismic7970740}
\end{figure}

\clearpage
\subsection{KOI-268  (KIC 3425851) \label{app:koi268}}

\begin{figure}[ht]
\centering
\includegraphics[width=14cm]{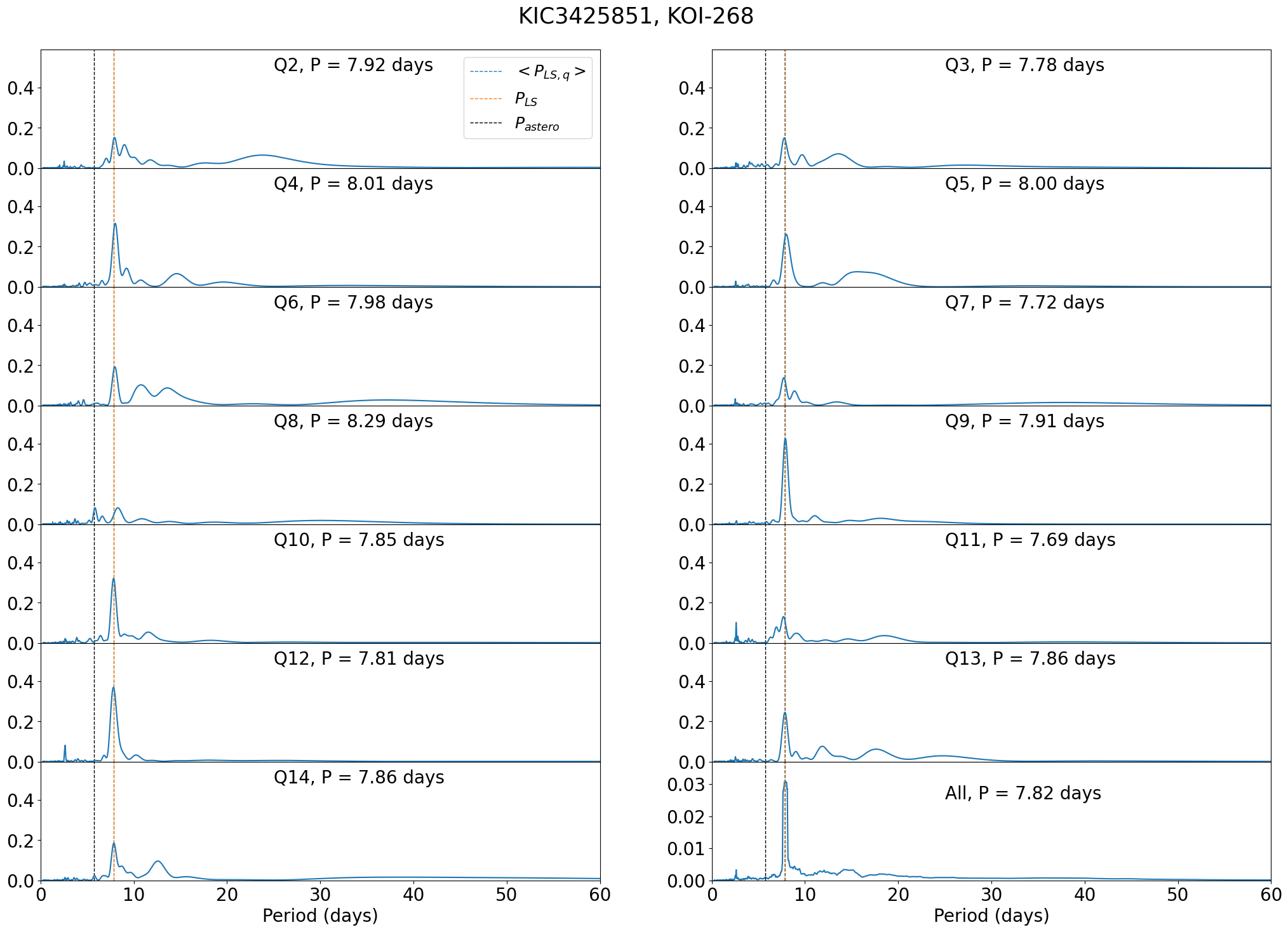}   
\caption{The LS periodograms of KOI-268  (Group Ab)
    for different quarters.}  
	\label{fig:koi268-LSQ}
\end{figure}
\begin{figure}[ht]
  \centering
  \includegraphics[width=8cm]{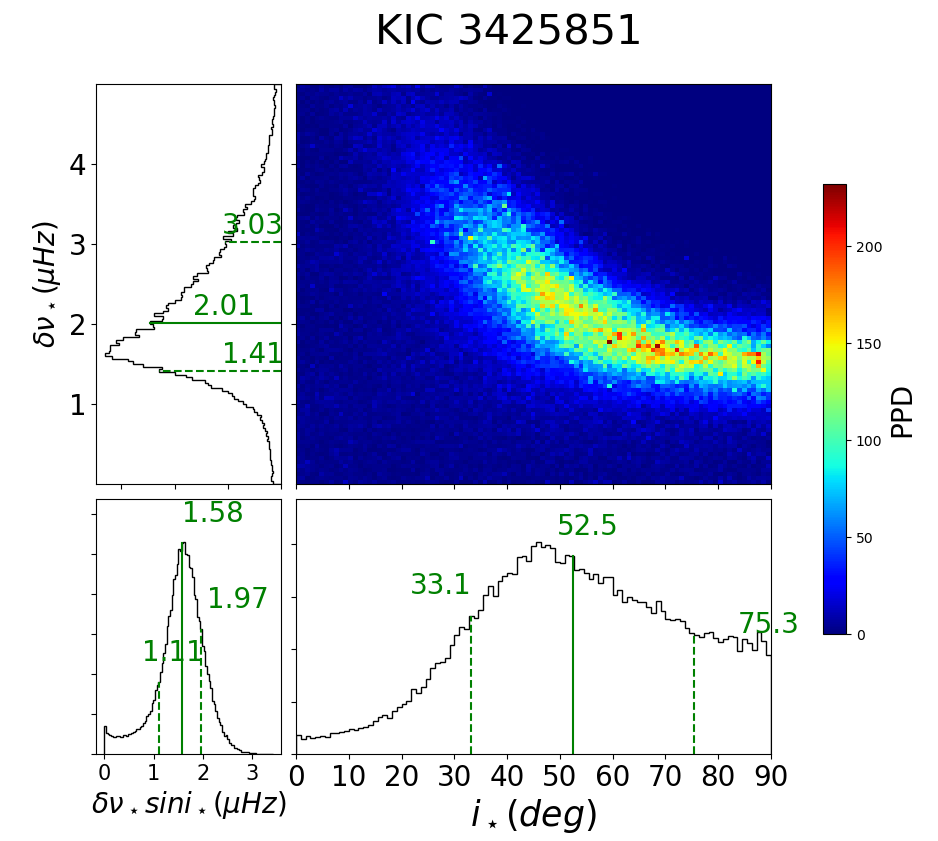}
  \caption{Asteroseismic constraints on $\delta\nu_*$ and $i_*$
        for KOI-268  (Group Ab).}
\label{fig:seismickoi268}
\end{figure}

\clearpage
\subsection{KIC 6225718 \label{app:6225718}}

\begin{figure}[ht]
\centering
\includegraphics[width=14cm]{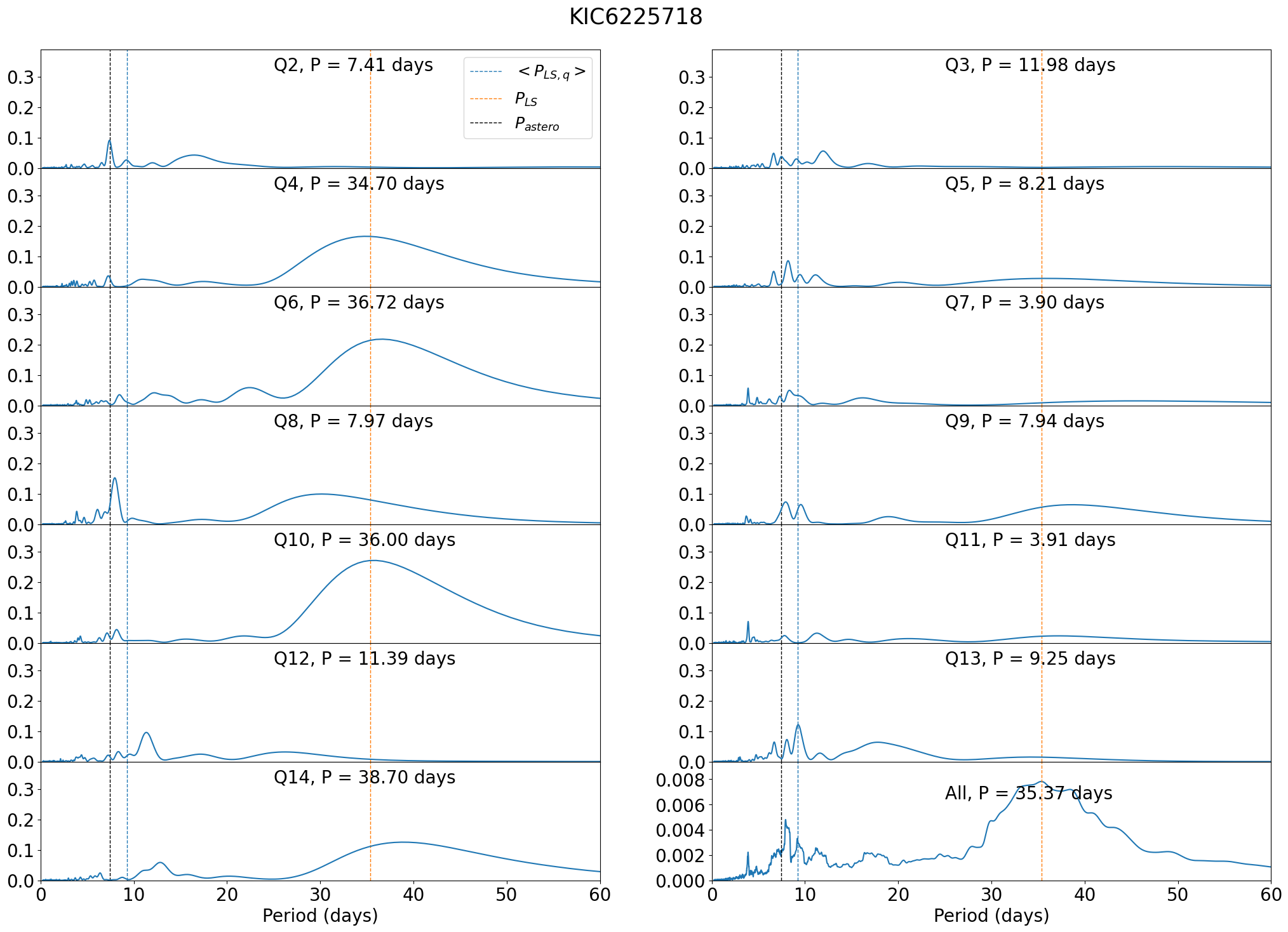}   
\caption{The LS periodograms of KIC 6225718 (Group Ba)
    for different quarters.}  
	\label{fig:6225718-LSQ}
\end{figure}
\begin{figure}[ht]
  \centering
  \includegraphics[width=8cm]{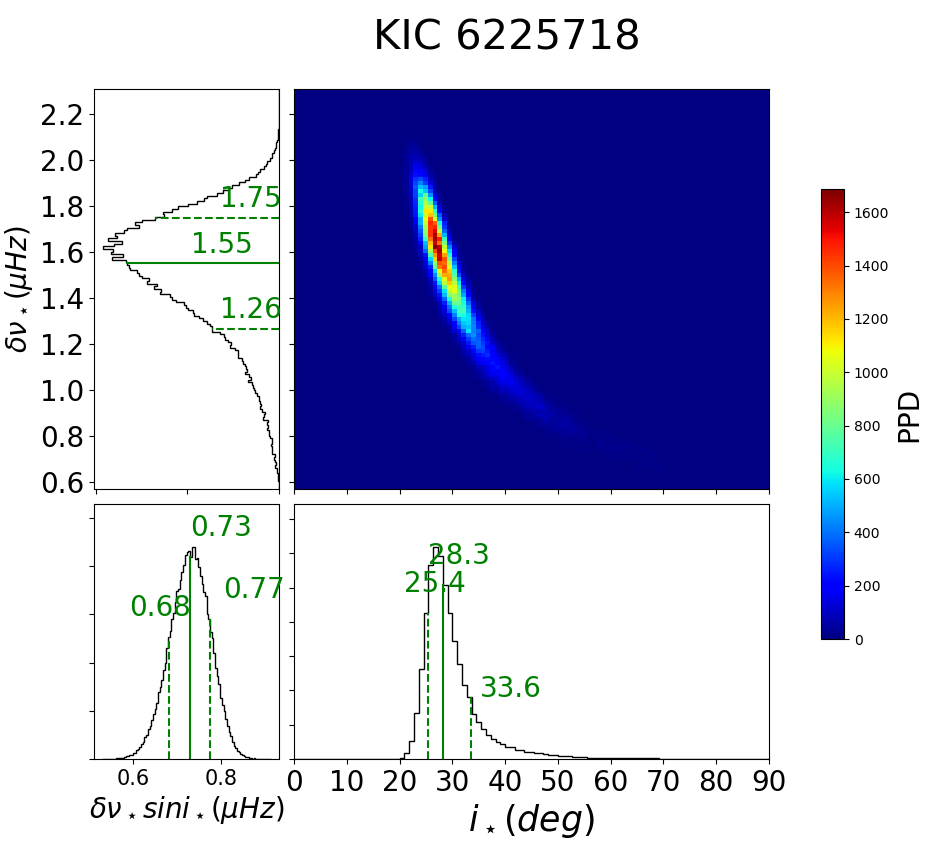}
  \caption{Asteroseismic constraints on $\delta\nu_*$ and $i_*$
        for KIC 6225718  (Group Ba).}
\label{fig:seismic6225718}
\end{figure}

\clearpage
\subsection{KIC 9965715 \label{app:9965715}}

\begin{figure}[ht]
\centering
\includegraphics[width=14cm]{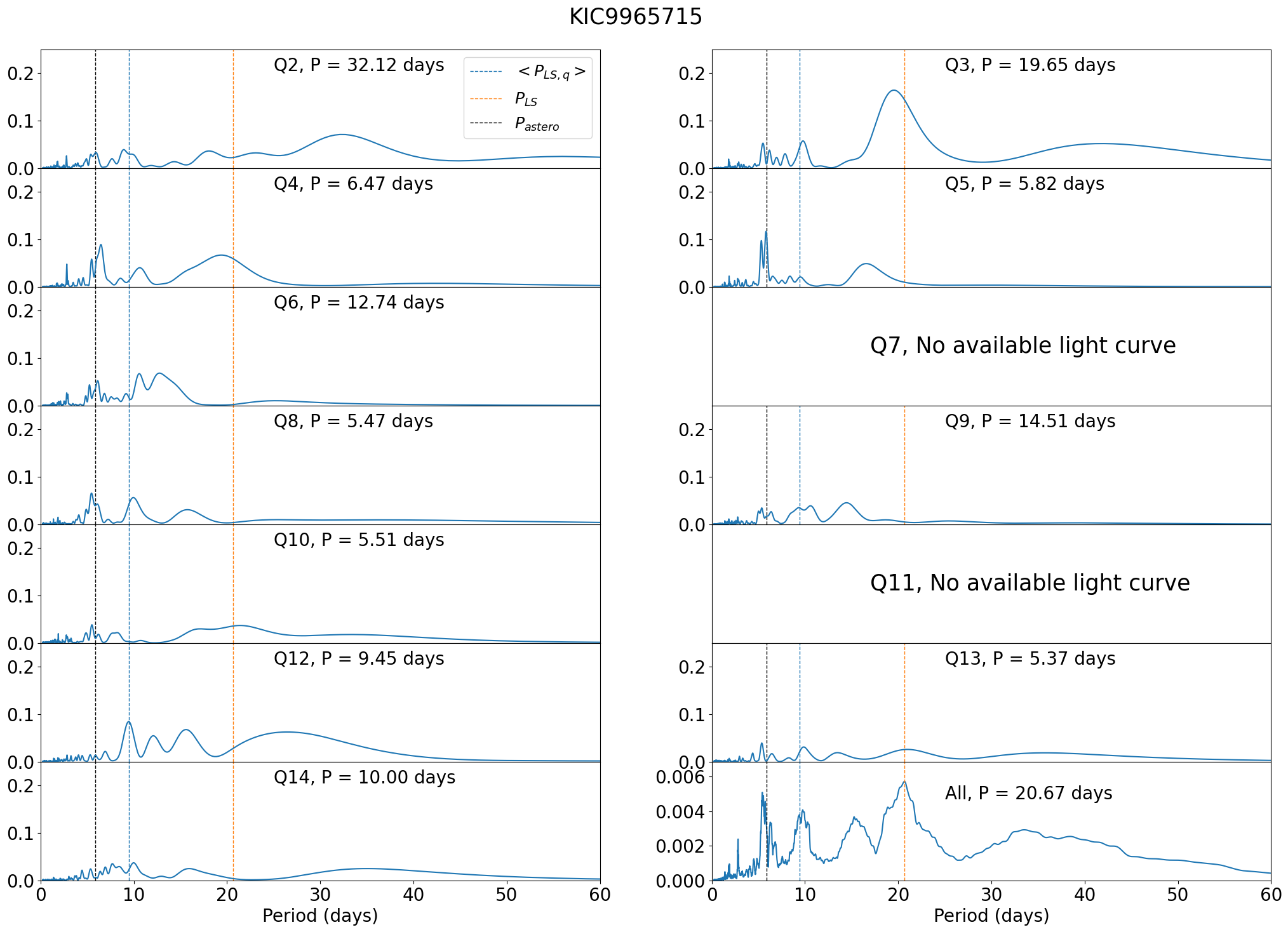}   
\caption{The LS periodograms of KIC 9965715  (Group Ba)
    for different quarters.}  
	\label{fig:9965715-LSQ}
\end{figure}
\begin{figure}[ht]
  \centering
  \includegraphics[width=8cm]{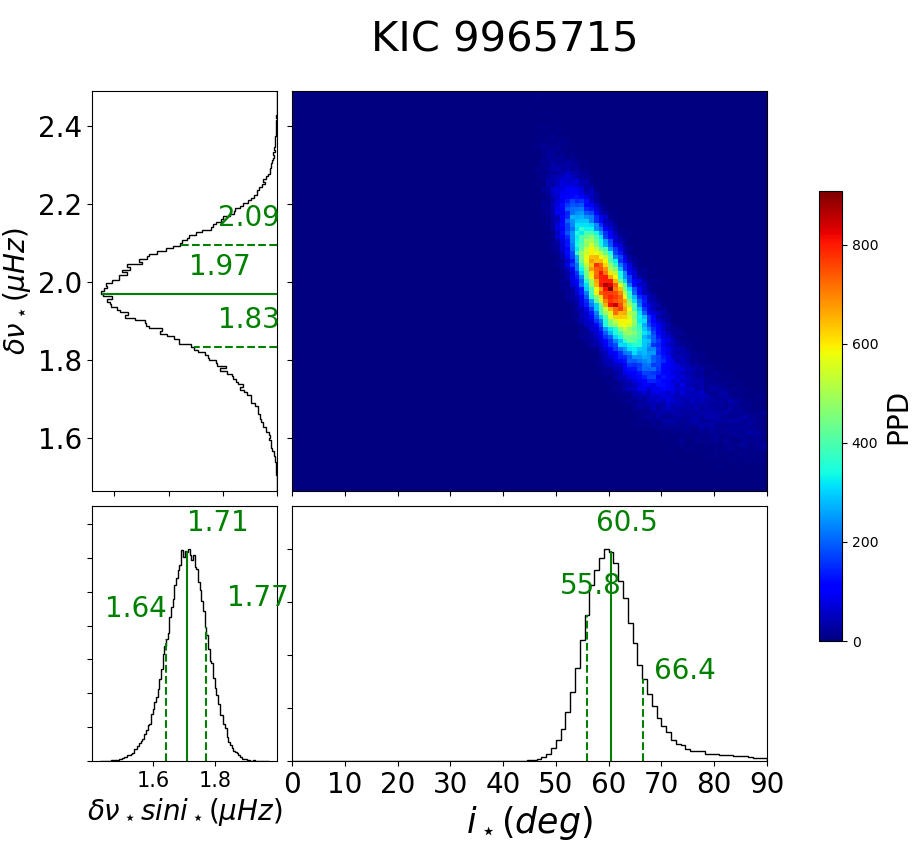}
  \caption{Asteroseismic constraints on $\delta\nu_*$ and $i_*$
        for KIC 9965715  (Group Ba).}
\label{fig:seismic9965715}
\end{figure}

\clearpage
\subsection{\textit{Kepler}-2 (HAT-P-7, KOI-2, KIC 10666592)  \label{app:kepler2}}

\begin{figure}[ht]
\centering
\includegraphics[width=14cm]{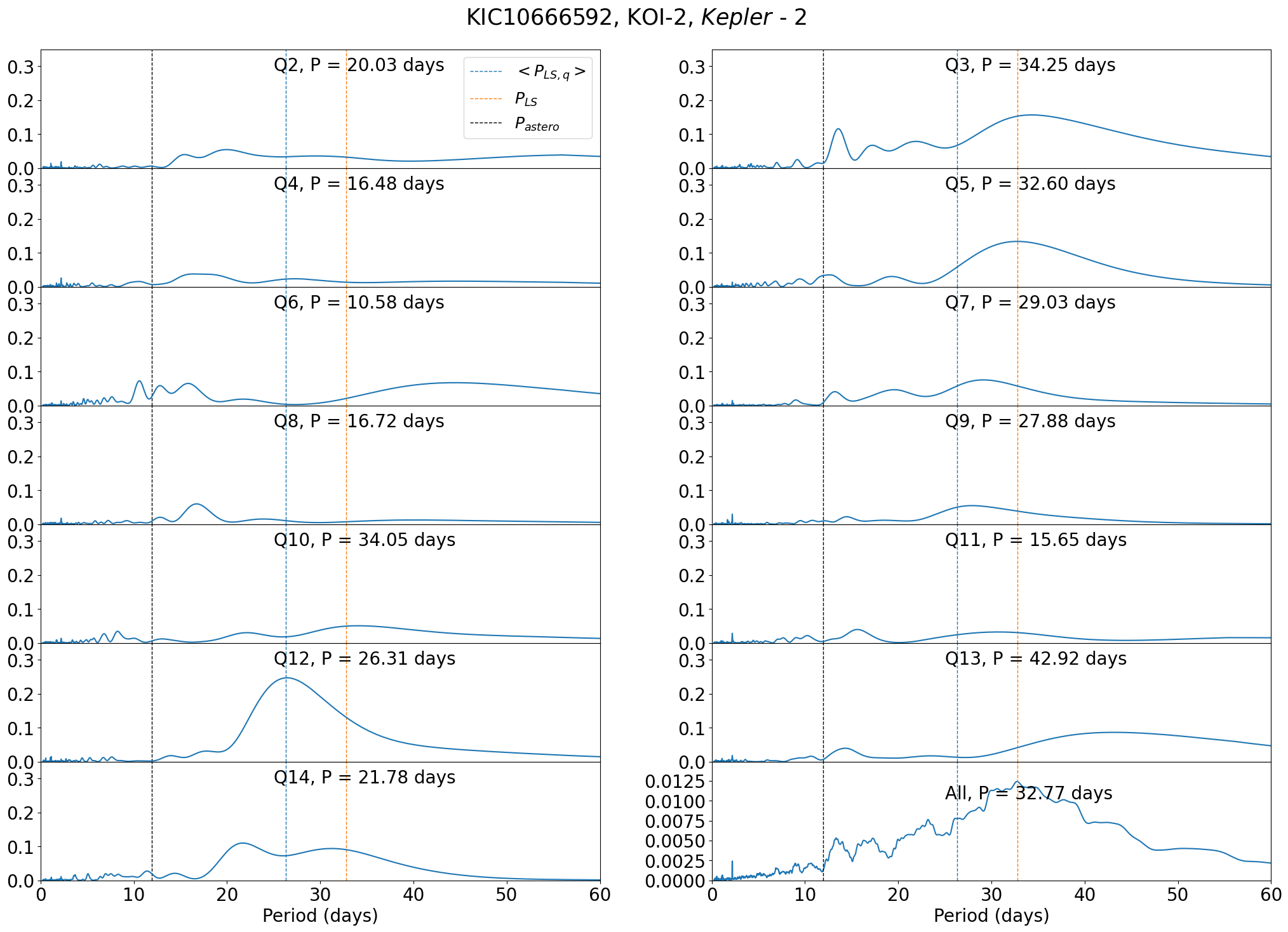}   
\caption{The LS periodograms of \textit{Kepler}-2
    for different quarters.}  
	\label{fig:kepler2-LSQ}
\end{figure}
\begin{figure}[ht]
  \centering
\includegraphics[width=8cm]{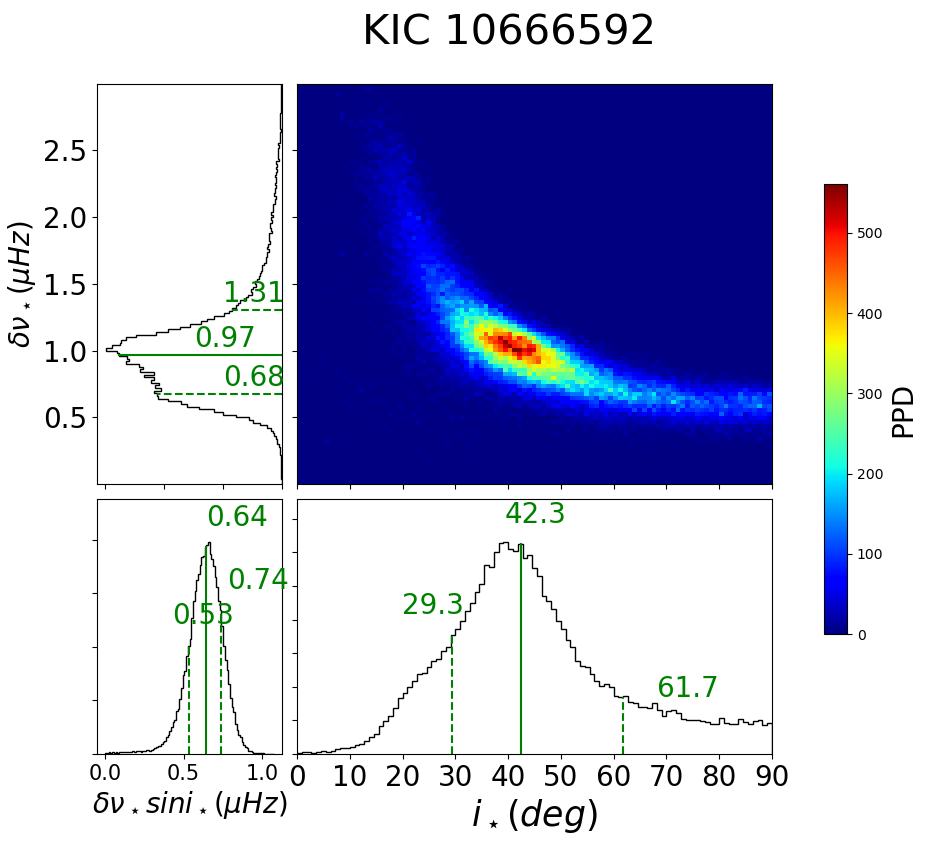}
  \caption{Asteroseismic constraints on $\delta\nu_*$ and $i_*$
        for \textit{Kepler}-2.}
\label{fig:seismickepler2}
\end{figure}

\clearpage
\subsection{\textit{Kepler}-410 (KOI-42, KIC 8866102)  \label{app:kepler410}}

\begin{figure}[ht]
\centering
\includegraphics[width=14cm]{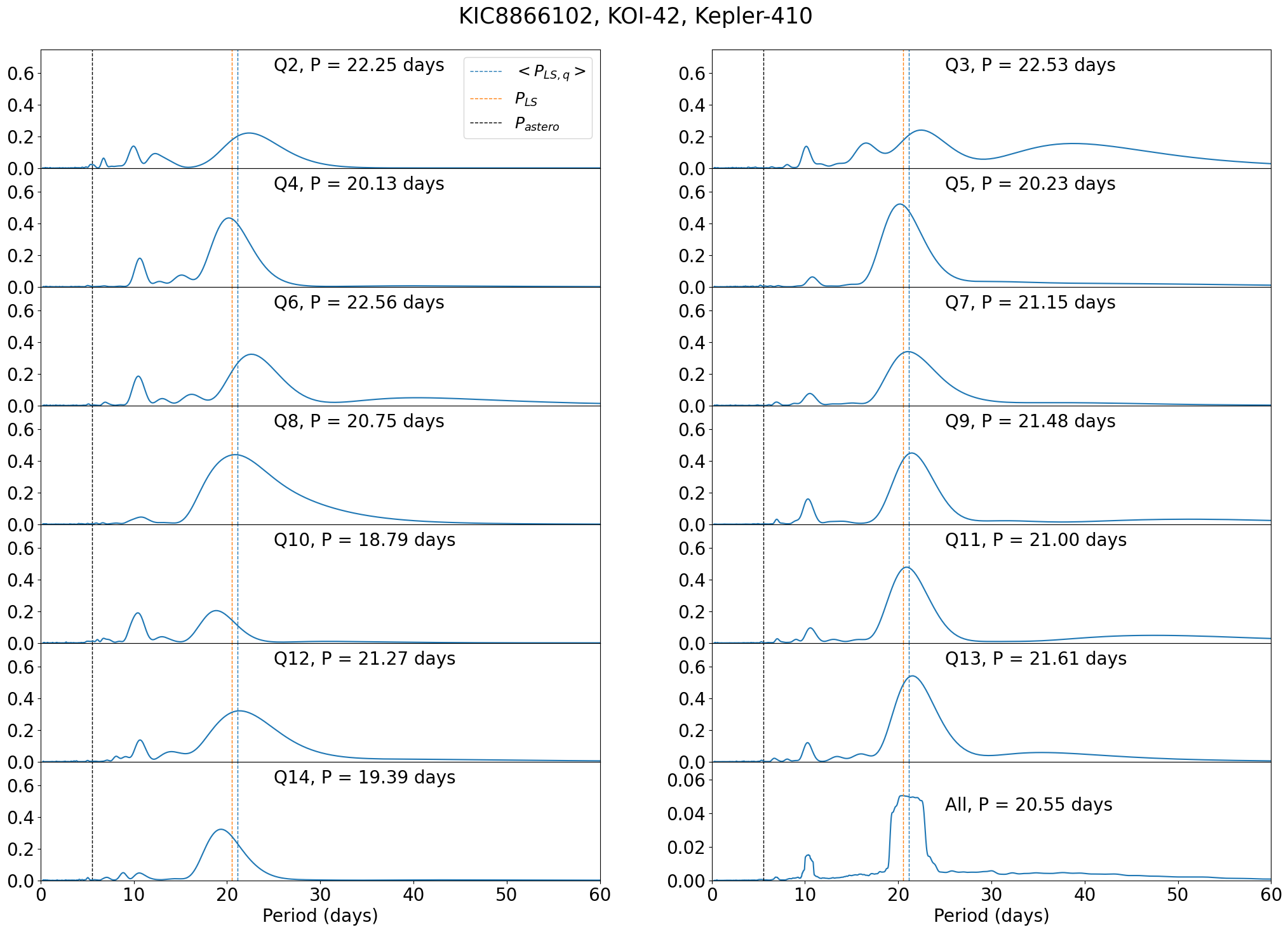}   
\caption{The LS periodograms of \textit{Kepler}-410
    for different quarters.}  
	\label{fig:kepler410-LSQ}
\end{figure}
\begin{figure}[ht]
  \centering
\includegraphics[width=8cm]{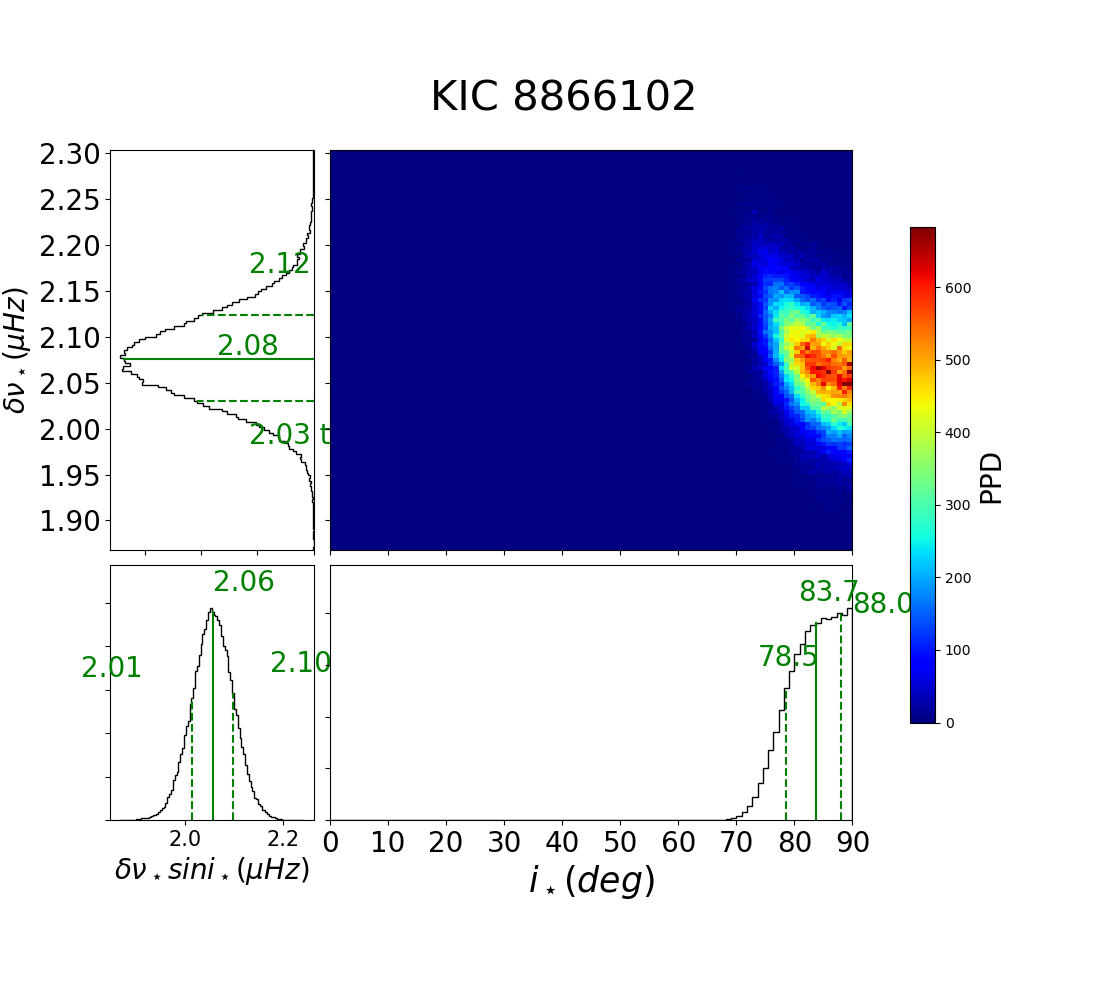}
  \caption{Asteroseismic constraints on $\delta\nu_*$ and $i_*$
        for \textit{Kepler}-410.}
\label{fig:seismickepler410}
\end{figure}

\clearpage
\subsection{KOI-288 (KIC 9592705)  \label{app:koi288}}

\begin{figure}[ht]
\centering
\includegraphics[width=14cm]{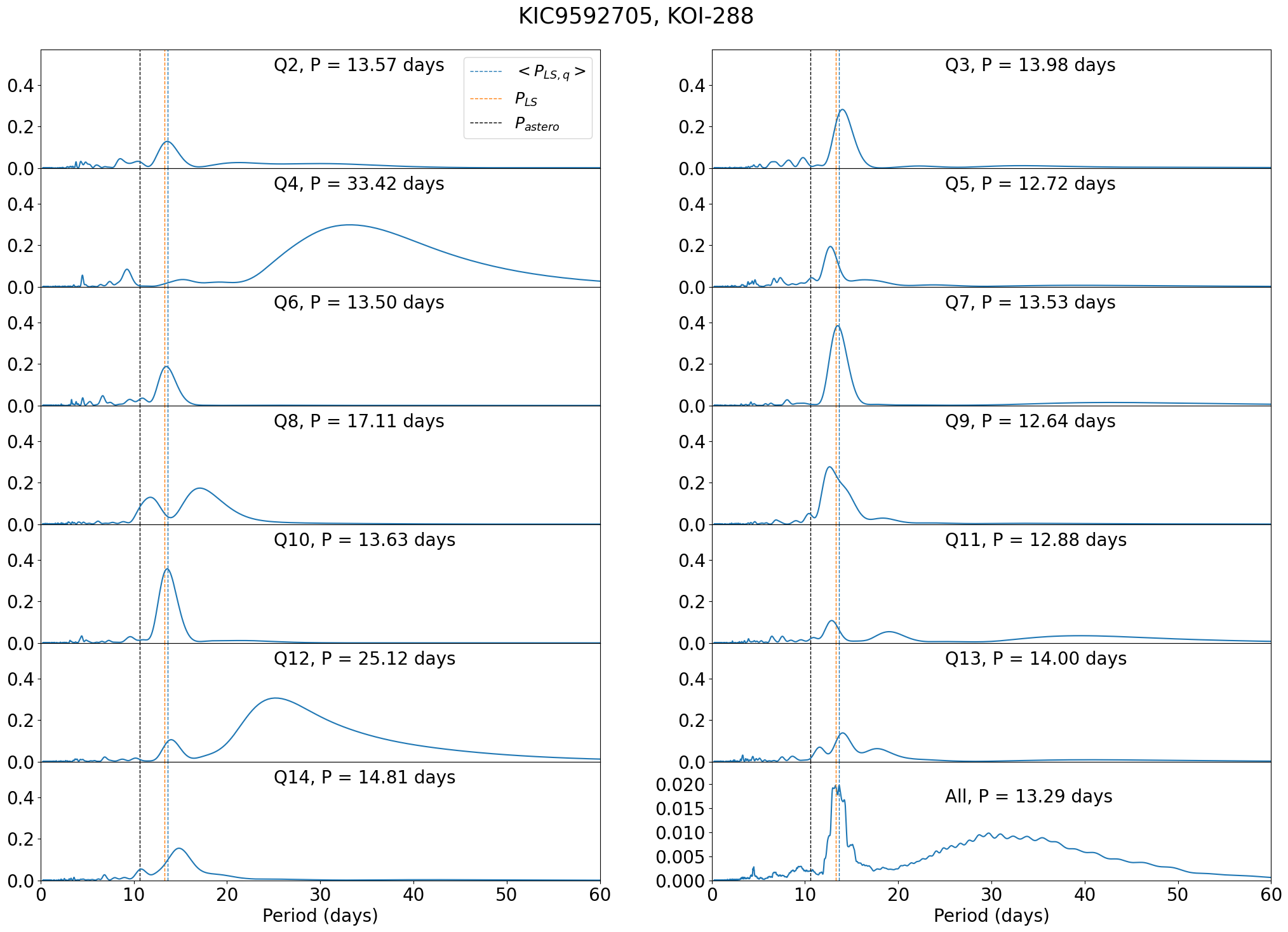}   
\caption{The LS periodograms of KOI-288
    for different quarters.}  
	\label{fig:koi288-LSQ}
\end{figure}
\begin{figure}[ht]
  \centering
\includegraphics[width=8cm]{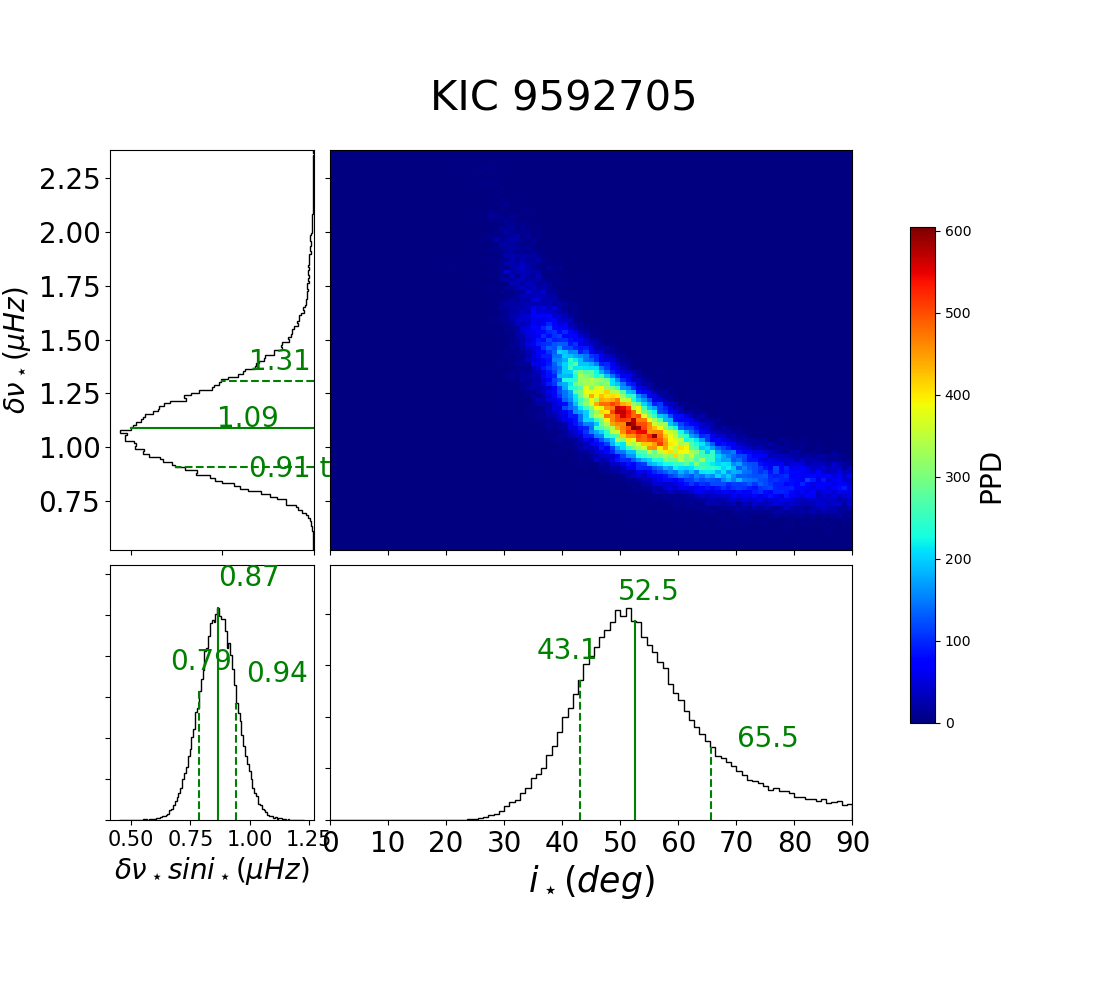}
  \caption{Asteroseismic constraints on $\delta\nu_*$ and $i_*$
        for KOI-288.}
\label{fig:seismickoi288}
\end{figure}

\clearpage

\begin{table*}[!ht]
\centering
\caption{Estimated rotation periods
  for 22 possible binary/multiple-star systems.}
\resizebox{1.0\linewidth}{!}{
\begin{threeparttable}
\begin{tabular}{lllclllllc} 
\toprule
    KIC & KOI\tnote{b} & \textit{Kepler} ID & RUWE &$\left<P_{\rm LS, q}\right>$ & $P_{\rm LS}$ & $P_{\rm ACF}$ & $P_{\rm wavelet}$ & $P_{\rm astero}\tnote{a}$ & Ref. \\ 
     &  &  &  & (Day) & (Day) & (Day) & (Day) & (Day) \\\hline
11295426 & KOI-246 & $\textit{Kepler}$-68 & 0.92 & $16.07_{-5.74}^{+4.39}$ & $32.90_{-5.81}^{+12.02}$ & $33.26_{-5.62}^{+11.45}$ & $32.85_{-7.28}^{+13.89}$ & $37.33_{-12.30}^{+15.92}$ & 1 \\
8866102 & KOI-42 & $\textit{Kepler}$-410 & 0.97 & $21.15_{-0.97}^{+0.78}$ & $20.54_{-1.42}^{+2.40}$ & $20.78_{-1.67}^{+2.12}$ & $20.97_{-3.15}^{+3.25}$ & $5.59_{-0.13}^{+0.11}$ & 6 \\
10666592 & KOI-2 & HAT-P-7 & 0.98 & $26.31_{-9.71}^{+7.02}$ & $32.77_{-7.09}^{+11.74}$ & $32.11_{-6.08}^{+12.77}$ & $32.75_{-11.52}^{+11.62}$ & $11.95_{-3.08}^{+5.20}$ & 4 \\
6278762 & KOI-3158 & $\textit{Kepler}$-444 & 1 & $31.68_{-5.72}^{+5.89}$ & $31.81_{-5.74}^{+19.00}$ & $30.71_{-4.35}^{+10.34}$ & $31.61_{-7.39}^{+18.34}$ & $29.68_{-6.06}^{+3.39}$ & 3 \\
3632418 & KOI-975 & $\textit{Kepler}$-21 & 1.05 & $12.63_{-4.71}^{+1.09}$ & $12.68_{-0.72}^{+1.07}$ & $12.74_{-0.80}^{+0.63}$ & $12.81_{-2.22}^{+2.53}$ & $12.29_{-1.14}^{+0.75}$ & 4 \\
9592705 & KOI-288 &  & 1.67 & $13.63_{-0.45}^{+2.33}$ & $13.29_{-0.57}^{+1.18}$ & $13.10_{-0.30}^{+1.33}$ & $13.64_{-2.32}^{+2.53}$ & $10.61_{-1.78}^{+2.11}$ & 1 \\
8554498 & KOI-5 &  & 4.89 & $17.00_{-6.80}^{+13.72}$ & $13.25_{-1.16}^{+3.99}$ & $35.52_{-7.03}^{+11.28}$ & $35.74_{-11.00}^{+11.21}$ & $24.31_{-17.44}^{+82.77}$ & 1 \\
9139151 &  &  & 0.8 & $11.82_{-3.46}^{+1.79}$ & $6.30_{-0.28}^{+0.22}$ & $14.08_{-1.40}^{+0.42}$ & $12.09_{-3.46}^{+7.18}$ & $11.59_{-1.04}^{+0.81}$ & 2 \\
9139163 &  &  & 0.93 & $6.08_{-5.47}^{+8.21}$ & $5.91_{-0.14}^{+0.24}$ & $16.41_{-3.49}^{+2.76}$ & $15.81_{-5.94}^{+23.09}$ & $3.30_{-0.30}^{+0.44}$ & 2 \\
3427720 &  &  & 1 & $13.59_{-1.02}^{+0.53}$ & $13.17_{-0.69}^{+1.51}$ & $13.24_{-0.36}^{+1.45}$ & $13.43_{-2.53}^{+2.63}$ & $25.74_{-12.76}^{+28.39}$ & 2 \\
9098294 &  &  & 1.64 & $19.20_{-0.53}^{+1.04}$ & $19.31_{-0.84}^{+2.74}$ & $20.12_{-1.63}^{+1.86}$ & $20.14_{-3.05}^{+3.25}$ & $26.57_{-9.22}^{+7.72}$ & 5 \\
8179536 &  &  & 1.8 & $5.38_{-0.24}^{+0.27}$ & $5.35_{-0.26}^{+0.20}$ & $13.17_{-1.18}^{+0.23}$ & $5.37_{-0.98}^{+1.29}$ & $6.69_{-0.64}^{+0.75}$ & 5 \\
8424992 &  &  & 2.08 & $16.73_{-6.05}^{+10.11}$ & $33.78_{-6.28}^{+11.06}$ & $35.21_{-3.32}^{+9.34}$ & $36.05_{-10.28}^{+9.14}$ & $34.44_{-23.43}^{+70.79}$ & 5 \\
12317678 &  &  & 3.93 & $9.42_{-4.36}^{+4.56}$ & $13.36_{-1.32}^{+5.05}$ & $13.48_{-1.17}^{+3.49}$ & $14.15_{-4.18}^{+33.42}$ & $10.82_{-2.21}^{+1.58}$ & 5 \\
7510397 &  &  & 4.28 & $19.53_{-3.05}^{+12.67}$ & $33.50_{-4.22}^{+12.02}$ & $39.63_{-7.82}^{+5.59}$ & $35.23_{-20.40}^{+12.65}$ & $8.37_{-1.43}^{+3.15}$ & 2 \\
4914923 &  &  & 5.14 & $10.80_{-2.72}^{+1.95}$ & $7.93_{-0.20}^{+0.46}$ & $7.89_{-0.18}^{+0.39}$ & $14.57_{-7.28}^{+6.35}$ & $19.69_{-2.96}^{+3.57}$ & 5 \\
9025370 &  &  & 5.16 & $14.55_{-2.58}^{+8.18}$ & $12.35_{-0.79}^{+3.57}$ & $29.80_{-4.65}^{+8.50}$ & $30.89_{-11.73}^{+8.11}$ & $24.53_{-3.45}^{+3.36}$ & 2 \\
1435467 &  &  & 5.73 & $7.12_{-0.57}^{+0.91}$ & $7.07_{-0.72}^{+0.52}$ & $6.98_{-0.54}^{+0.23}$ & $6.71_{-1.39}^{+2.74}$ & $7.01_{-0.65}^{+0.86}$ & 5 \\
10454113 &  &  & 5.96 & $14.56_{-0.55}^{+0.79}$ & $14.55_{-1.04}^{+1.18}$ & $14.33_{-0.83}^{+0.98}$ & $14.67_{-2.22}^{+2.53}$ & $9.89_{-2.39}^{+3.74}$ & 5 \\
7871531 &  &  & 9.7 & $16.99_{-1.44}^{+16.78}$ & $34.71_{-4.25}^{+7.68}$ & $35.01_{-4.02}^{+7.49}$ & $35.12_{-6.46}^{+6.25}$ & $28.93_{-4.48}^{+4.85}$ & 5 \\
6933899 &  &  & 13.14 & $16.25_{-0.92}^{+0.55}$ & $16.18_{-1.58}^{+1.06}$ & $15.93_{-1.38}^{+0.74}$ & $15.81_{-3.46}^{+3.25}$ & $30.84_{-5.37}^{+3.76}$ & 5 \\
8379927 &  &  & 32.79 & $17.32_{-1.32}^{+0.71}$ & $17.06_{-0.97}^{+1.71}$ & $17.33_{-1.22}^{+1.39}$ & $17.25_{-2.53}^{+2.63}$ & $10.06_{-0.40}^{+0.41}$ & 2\\ \hline
\end{tabular}
\begin{tablenotes}[para,flushleft]
    \item[a] Data adopted from \cite{Kamiaka2018}\\  
    \item[b] KOI number only available for systems with reported planet detection.\\
    \textbf{References} (1)EB catalogue; (2)\cite{Garcia2014} (3)\cite{LilloBox2014} (4)NASA Exoplanet Archive (5)RUWE (6)\cite{VanEylen2014}.
\end{tablenotes}
\label{tab:multiplestar}
\end{threeparttable}
}
\end{table*}


\begin{table*}[!ht]
\centering
\caption{Estimated rotation periods for 14 single stars
     in Group Aa.}
\resizebox{0.8\linewidth}{!}{
\begin{threeparttable}
\begin{tabular}{lllllllll} 
\toprule
    KIC & KOI\tnote{b} & \textit{Kepler} ID & $\left<P_{\rm LS,q}\right>$  &$P_{\rm LS}$  & $P_{\rm ACF}$ & $P_{\rm wavelet}$  & $P_{\rm rot,\;\rm astero}\tnote{a}$ & Remark\\ 
     &  &  & (Day) &  (Day) & (Day) & (Day) & (Day) & \\\hline
11807274 & KOI-262 & $\textit{Kepler}$ - 50 & $7.82_{-0.58}^{+0.34}$ & $8.09_{-0.89}^{+0.35}$ & $7.95_{-0.24}^{+0.34}$ & $7.95_{-1.60}^{+2.01}$ & $7.60_{-0.80}^{+0.55}$ &  \\
5866724 & KOI-85 & $\textit{Kepler}$ - 65 & $8.23_{-0.39}^{+0.46}$ & $8.19_{-0.36}^{+0.56}$ & $8.20_{-0.35}^{+0.57}$ & $8.16_{-1.39}^{+1.50}$ & $8.19_{-0.58}^{+0.58}$ &  \\
6521045 & KOI-41 & $\textit{Kepler}$ - 100 & $12.92_{-4.08}^{+0.20}$ & $13.78_{-2.37}^{+0.76}$ & $13.73_{-1.26}^{+0.75}$ & $13.12_{-3.36}^{+3.25}$ & $24.98_{-2.28}^{+1.96}$ &  \\
8292840 & KOI-260 & $\textit{Kepler}$ - 126 & $7.66_{-0.70}^{+0.25}$ & $7.09_{-0.35}^{+1.23}$ & $7.09_{-0.27}^{+0.31}$ & $7.44_{-1.70}^{+2.74}$ & $7.86_{-0.61}^{+0.58}$ &  \\
10963065 & KOI-1612 & $\textit{Kepler}$ - 408 & $12.32_{-0.49}^{+0.41}$ & $12.59_{-0.99}^{+1.08}$ & $12.63_{-1.03}^{+0.78}$ & $12.50_{-2.32}^{+2.22}$ & $11.69_{-0.95}^{+1.40}$ &  \\
7670943 & KOI-269 &  & $5.25_{-0.08}^{+0.11}$ & $5.35_{-0.20}^{+0.15}$ & $5.30_{-0.17}^{+0.13}$ & $5.27_{-0.88}^{+0.88}$ & $6.08_{-0.46}^{+0.37}$ &  \\
9414417 & KOI-974 &  & $10.87_{-0.07}^{+0.11}$ & $10.75_{-0.39}^{+0.63}$ & $10.85_{-0.48}^{+0.54}$ & $10.85_{-1.50}^{+1.60}$ & $10.93_{-1.75}^{+1.56}$ &  \\
5773345 &  &  & $11.54_{-0.30}^{+2.03}$ & $11.20_{-0.60}^{+0.93}$ & $11.24_{-0.51}^{+0.79}$ & $11.57_{-2.22}^{+5.42}$ & $6.49_{-1.06}^{+1.19}$ &  \\
7103006 &  &  & $4.71_{-0.08}^{+0.20}$ & $4.74_{-0.17}^{+0.17}$ & $4.68_{-0.12}^{+0.12}$ & $4.75_{-0.88}^{+1.08}$ & $5.88_{-0.53}^{+0.67}$ &  \\
7206837 &  &  & $4.07_{-0.03}^{+0.03}$ & $4.11_{-0.13}^{+0.01}$ & $4.04_{-0.06}^{+0.08}$ & $4.13_{-0.67}^{+0.57}$ & $4.21_{-0.40}^{+0.48}$ &  \\
7771282 &  &  & $12.01_{-0.44}^{+0.21}$ & $12.27_{-1.01}^{+0.25}$ & $12.05_{-0.82}^{+0.39}$ & $11.88_{-2.01}^{+2.12}$ & $9.64_{-1.38}^{+1.27}$ &  \\
7940546 &  &  & $11.72_{-0.89}^{+0.46}$ & $11.63_{-0.81}^{+1.02}$ & $11.61_{-0.58}^{+1.07}$ & $11.67_{-2.22}^{+2.74}$ & $10.70_{-1.30}^{+1.13}$ &  \\
9812850 &  &  & $6.00_{-1.09}^{+1.13}$ & $6.76_{-0.59}^{+0.31}$ & $6.73_{-0.34}^{+0.33}$ & $5.48_{-1.39}^{+3.25}$ & $7.32_{-1.14}^{+0.90}$ &  \\
12009504 &  &  & $9.53_{-0.85}^{+0.49}$ & $9.62_{-0.56}^{+0.72}$ & $9.61_{-0.49}^{+0.33}$ & $9.71_{-1.81}^{+2.22}$ & $9.66_{-0.53}^{+0.60}$ &  \\\hline
\end{tabular}
\begin{tablenotes}[para,flushleft]
    \item[a] Data adopted from \cite{Kamiaka2018}\\  
    \item[b] KOI number only available for systems with reported planet detection.\\
    \item[M] Suggested candidate of rotation period, $\left<P_{\rm LS, q}\right>$ (days). Target with multiple periodic component.\\
    \item[A] Suggested candidate of rotation period, $\left<P_{\rm LS, q}\right>$ (days). Target with abnormal quarter.\\ 
    \item[W] Targets with weak or no periodic signals.\\
\end{tablenotes}
\label{tab:singlestarAa}
\end{threeparttable}
}
\end{table*}

\begin{table*}[!ht]
\centering
\caption{Same as Figure \ref{tab:singlestarAa} but
    for 9 single stars in Group Ab.}
\resizebox{0.8\linewidth}{!}{
\begin{threeparttable}
\begin{tabular}{lllllllll} 
\toprule
    KIC & KOI\tnote{b} & \textit{Kepler} ID & $\left<P_{\rm LS,q}\right>$  &$P_{\rm LS}$  & $P_{\rm ACF}$ & $P_{\rm wavelet}$  & $P_{\rm rot,\;\rm astero}\tnote{a}$ & Remark\\ 
     &  &  & (Day) &  (Day) & (Day) & (Day) & (Day) & \\\hline
6196457 & KOI-285 & $\textit{Kepler}$ - 92 & $15.64_{-4.05}^{+1.54}$ & $16.25_{-1.13}^{+1.88}$ & $16.71_{-1.57}^{+1.08}$ & $16.12_{-3.56}^{+3.15}$ & $5.92_{-3.02}^{+29.78}$ &  \\
8494142 & KOI-370 & $\textit{Kepler}$ - 145 & $13.53_{-1.71}^{+0.94}$ & $14.24_{-1.49}^{+0.97}$ & $14.10_{-1.31}^{+0.38}$ & $13.84_{-3.87}^{+4.60}$ & $10.49_{-3.82}^{+2.20}$ &  \\
9955598 & KOI-1925 & $\textit{Kepler}$ - 409 & $12.55_{-1.68}^{+1.15}$ & $13.01_{-1.64}^{+2.36}$ & $12.98_{-1.31}^{+2.44}$ & $13.12_{-3.25}^{+4.49}$ & $28.10_{-4.60}^{+7.73}$ &  \\
3425851 & KOI-268 &  & $7.86_{-0.06}^{+0.13}$ & $7.82_{-0.20}^{+0.34}$ & $7.87_{-0.26}^{+0.28}$ & $7.95_{-1.19}^{+1.39}$ & $5.76_{-1.93}^{+2.45}$ &  \\
7970740 &  &  & $13.42_{-3.38}^{+1.96}$ & $13.48_{-0.72}^{+1.07}$ & $13.48_{-0.56}^{+1.12}$ & $13.95_{-3.98}^{+7.28}$ & $31.35_{-6.14}^{+7.68}$ &  \\
9353712 &  &  & $11.20_{-1.21}^{+2.89}$ & $10.74_{-0.63}^{+4.75}$ & $14.98_{-1.49}^{+0.39}$ & $11.88_{-3.56}^{+6.46}$ & $6.18_{-1.86}^{+6.66}$ &  \\
9410862 &  &  & $12.02_{-1.22}^{+1.04}$ & $12.31_{-1.73}^{+1.59}$ & $11.32_{-0.76}^{+1.67}$ & $11.98_{-2.53}^{+3.15}$ & $9.79_{-4.08}^{+13.03}$ &  \\
10162436 &  &  & $11.67_{-3.05}^{+1.39}$ & $11.92_{-1.14}^{+2.67}$ & $12.30_{-1.12}^{+1.01}$ & $12.29_{-2.84}^{+6.15}$ & $17.27_{-6.04}^{+6.10}$ &  \\
12258514 &  &  & $14.67_{-2.26}^{+0.38}$ & $15.24_{-1.15}^{+1.06}$ & $14.92_{-0.80}^{+1.23}$ & $14.88_{-2.94}^{+2.74}$ & $24.54_{-10.35}^{+15.43}$ & \\ \hline
\end{tabular}
\label{tab:singlestarAb}
\end{threeparttable}
}
\end{table*}

\begin{table*}[!ht]
\centering
\caption{Same as Figure \ref{tab:singlestarAa} but
    for 19 single stars in Group Ba.}
\resizebox{0.8\linewidth}{!}{
\begin{threeparttable} 
\begin{tabular}{lllllllll} 
\toprule
    KIC & KOI\tnote{b} & \textit{Kepler} ID & $\left<P_{\rm LS,q}\right>$  &$P_{\rm LS}$  & $P_{\rm ACF}$ & $P_{\rm wavelet}$  & $P_{\rm rot,\;\rm astero}\tnote{a}$ & Remark\\ 
     &  &  & (Day) &  (Day) & (Day) & (Day) & (Day) & \\\hline
4349452 & KOI-244 & $\textit{Kepler}$ - 25 & $11.90_{-2.91}^{+6.27}$ & $22.24_{-1.22}^{+3.63}$ & $22.49_{-1.36}^{+3.36}$ & $22.93_{-12.34}^{+4.18}$ & $7.75_{-0.48}^{+0.46}$ &  \\
3544595 & KOI-69 & $\textit{Kepler}$ - 93 & $14.23_{-2.56}^{+12.87}$ & $15.09_{-1.74}^{+1.13}$ & $14.39_{-0.90}^{+1.10}$ & $14.57_{-2.94}^{+3.87}$ & $23.33_{-2.92}^{+3.86}$ &  \\
8077137 & KOI-274 & $\textit{Kepler}$ - 128 & $13.60_{-0.82}^{+0.40}$ & $13.27_{-0.81}^{+1.18}$ & $13.29_{-0.49}^{+1.14}$ & $13.43_{-2.43}^{+2.63}$ & $12.34_{-1.32}^{+1.28}$ &  \\
2837475 &  &  & $3.72_{-0.06}^{+4.47}$ & $3.68_{-0.05}^{+0.07}$ & $3.66_{-0.04}^{+0.08}$ & $3.82_{-0.88}^{+2.01}$ & $3.83_{-0.14}^{+0.18}$ &  \\
5184732 &  &  & $9.75_{-3.02}^{+5.43}$ & $6.70_{-0.32}^{+0.31}$ & $6.70_{-0.35}^{+0.21}$ & $18.08_{-12.34}^{+4.80}$ & $20.27_{-2.05}^{+1.32}$ &  \\
6106415 &  &  & $22.85_{-6.95}^{+10.53}$ & $28.39_{-13.60}^{+21.37}$ & $15.70_{-0.97}^{+1.49}$ & $28.20_{-15.65}^{+21.23}$ & $16.35_{-0.96}^{+0.71}$ &  \\
6116048 &  &  & $9.91_{-3.99}^{+4.19}$ & $16.81_{-1.55}^{+2.26}$ & $16.80_{-1.31}^{+2.34}$ & $16.63_{-5.22}^{+3.87}$ & $17.93_{-1.42}^{+0.98}$ &  \\
6225718 &  &  & $9.25_{-1.58}^{+26.10}$ & $35.37_{-5.88}^{+8.65}$ & $35.52_{-6.27}^{+5.95}$ & $35.23_{-6.87}^{+7.90}$ & $7.46_{-0.85}^{+1.69}$ & $9.25\pm2.73\tnote{M}$ \\
6508366 &  &  & $3.89_{-0.37}^{+0.84}$ & $3.75_{-0.10}^{+0.19}$ & $6.84_{-0.08}^{+0.37}$ & $3.82_{-0.77}^{+4.08}$ & $5.28_{-0.17}^{+0.18}$ &  \\
6679371 &  &  & $10.39_{-4.69}^{+5.07}$ & $5.80_{-0.25}^{+0.05}$ & $5.71_{-0.17}^{+0.11}$ & $14.98_{-10.28}^{+3.77}$ & $6.21_{-0.34}^{+0.29}$ &  \\
8006161 &  &  & $28.70_{-13.56}^{+5.10}$ & $33.04_{-4.22}^{+9.43}$ & $32.37_{-2.83}^{+9.69}$ & $33.78_{-6.56}^{+8.11}$ & $21.51_{-3.38}^{+3.88}$ &  \\
8394589 &  &  & $11.71_{-0.66}^{+0.27}$ & $11.50_{-0.93}^{+1.22}$ & $11.53_{-0.38}^{+0.87}$ & $11.67_{-3.15}^{+5.42}$ & $10.95_{-0.56}^{+0.60}$ &  \\
8694723 &  &  & $8.48_{-0.71}^{+6.55}$ & $7.86_{-0.41}^{+0.16}$ & $7.83_{-0.40}^{+0.14}$ & $7.95_{-1.81}^{+11.11}$ & $9.18_{-1.29}^{+1.30}$ &  \\
9965715 &  &  & $9.45_{-3.94}^{+5.06}$ & $20.67_{-2.40}^{+2.35}$ & $20.39_{-2.29}^{+1.35}$ & $20.14_{-6.97}^{+5.73}$ & $5.88_{-0.35}^{+0.44}$ &  $10.00\pm4.19\tnote{M}$ \\
10079226 &  &  & $10.20_{-2.57}^{+5.01}$ & $14.87_{-0.50}^{+1.89}$ & $15.74_{-1.35}^{+0.77}$ & $15.29_{-3.36}^{+2.94}$ & $15.35_{-2.94}^{+2.47}$ &  \\
10516096 &  &  & $11.04_{-3.93}^{+7.11}$ & $6.90_{-0.35}^{+0.34}$ & $6.75_{-0.20}^{+0.37}$ & $19.21_{-13.07}^{+4.91}$ & $23.44_{-3.53}^{+2.16}$ &  \\
11081729 &  &  & $2.73_{-0.04}^{+0.14}$ & $2.70_{-0.05}^{+0.04}$ & $18.14_{-1.68}^{+1.12}$ & $2.79_{-0.57}^{+2.22}$ & $3.40_{-0.10}^{+0.19}$ &  \\
11253226 &  &  & $3.82_{-0.08}^{+6.30}$ & $3.80_{-0.09}^{+0.04}$ & $13.32_{-1.21}^{+0.28}$ & $13.12_{-10.18}^{+8.83}$ & $3.56_{-0.15}^{+0.16}$ &  \\
11772920 &  &  & $15.99_{-6.39}^{+1.39}$ & $16.09_{-1.67}^{+1.68}$ & $15.72_{-0.58}^{+2.01}$ & $16.43_{-4.49}^{+23.71}$ & $34.72_{-5.70}^{+5.47}$ & \\\hline
\end{tabular}
\label{tab:singlestarBa}
\end{threeparttable}
}
\end{table*}

\begin{table*}[!ht]
\centering
\caption{Same as Figure \ref{tab:singlestarAa} but
    for 28 single stars in Group Bb.}
\resizebox{0.8\linewidth}{!}{
\begin{threeparttable}
\begin{tabular}{lllllllll} 
\toprule
    KIC & KOI\tnote{b} & \textit{Kepler} ID & $\left<P_{\rm LS,q}\right>$  &$P_{\rm LS}$  & $P_{\rm ACF}$ & $P_{\rm wavelet}$  & $P_{\rm rot,\;\rm astero}\tnote{a}$ & Remark\\ 
     &  &  & (Day) &  (Day) & (Day) & (Day) & (Day) & \\\hline
11853905 & KOI-7 & $\textit{Kepler}$ - 4 & $19.53_{-4.37}^{+13.86}$ & $35.68_{-16.05}^{+9.07}$ & $36.28_{-6.37}^{+9.46}$ & $34.09_{-14.31}^{+9.97}$ & $28.16_{-15.88}^{+23.45}$ &  \\
11904151 & KOI-72 & $\textit{Kepler}$ - 10 & $16.46_{-2.76}^{+8.09}$ & $15.40_{-2.36}^{+4.22}$ & $14.20_{-0.63}^{+2.68}$ & $15.60_{-3.56}^{+5.22}$ & $33.53_{-19.20}^{+31.84}$ &  \\
11401755 & KOI-277 & $\textit{Kepler}$ - 36 & $14.02_{-3.35}^{+3.35}$ & $16.82_{-1.10}^{+2.14}$ & $17.06_{-1.25}^{+1.32}$ & $16.63_{-4.29}^{+3.67}$ & $17.60_{-3.92}^{+3.75}$ &  \\
8478994 & KOI-245 & $\textit{Kepler}$ - 37 & $14.36_{-1.77}^{+9.38}$ & $32.81_{-6.31}^{+9.88}$ & $33.64_{-2.96}^{+9.79}$ & $33.78_{-9.25}^{+8.21}$ & $15.93_{-6.06}^{+9.80}$ &  $14.40\pm2.12\tnote{M}$\\
8349582 & KOI-122 & $\textit{Kepler}$ - 95 & $16.49_{-1.48}^{+0.90}$ & $16.97_{-1.02}^{+1.66}$ & $17.36_{-1.42}^{+1.16}$ & $16.94_{-3.46}^{+3.25}$ & $40.18_{-19.59}^{+44.77}$ &  \\
4914423 & KOI-108 & $\textit{Kepler}$ - 103 & $16.48_{-7.39}^{+3.41}$ & $19.39_{-3.44}^{+4.69}$ & $21.72_{-3.14}^{+2.13}$ & $19.42_{-10.59}^{+5.94}$ & $18.03_{-10.08}^{+13.47}$ &  \\
5094751 & KOI-123 & $\textit{Kepler}$ - 109 & $13.48_{-3.68}^{+2.54}$ & $7.29_{-0.25}^{+0.36}$ & $16.50_{-3.10}^{+0.85}$ & $15.08_{-8.52}^{+6.46}$ & $17.92_{-12.49}^{+50.67}$ & $13.58\pm2.08\tnote{M}$ \\
10586004 & KOI-275 & $\textit{Kepler}$ - 129 & $14.16_{-4.11}^{+7.78}$ & $10.10_{-0.62}^{+0.42}$ & $10.15_{-0.72}^{+0.19}$ & $14.67_{-6.56}^{+23.50}$ & $16.05_{-5.80}^{+12.92}$ &  \\
11133306 & KOI-276 & $\textit{Kepler}$ - 509 & $16.53_{-2.10}^{+1.76}$ & $15.75_{-2.28}^{+2.59}$ & $15.85_{-2.57}^{+2.15}$ & $15.70_{-3.46}^{+5.42}$ & $17.95_{-10.48}^{+28.71}$ & \\
4143755 & KOI-281 & $\textit{Kepler}$ - 510 & $16.94_{-4.79}^{+5.03}$ & $15.18_{-2.17}^{+12.04}$ & $15.26_{-2.27}^{+2.08}$ & $15.70_{-4.60}^{+19.89}$ & $33.21_{-20.75}^{+62.62}$ &  \\
4141376 & KOI-280 & $\textit{Kepler}$ - 1655 & $13.61_{-6.62}^{+12.22}$ & $13.54_{-0.99}^{+1.29}$ & $13.70_{-1.28}^{+0.85}$ & $13.53_{-3.36}^{+3.46}$ & $11.74_{-3.35}^{+2.53}$ &  \\
7296438 & KOI-364 &  & $13.17_{-2.95}^{+6.19}$ & $26.42_{-3.49}^{+3.03}$ & $25.83_{-3.31}^{+3.22}$ & $25.93_{-4.49}^{+5.01}$ & $22.00_{-11.74}^{+43.25}$ & $13.17\pm4.39\tnote{A}$ \\
3456181 &  &  & $8.76_{-4.38}^{+7.72}$ & $18.30_{-3.57}^{+3.31}$ & $11.57_{-0.46}^{+0.77}$ & $27.69_{-16.99}^{+7.90}$ & $11.51_{-3.05}^{+3.57}$ &  \\
3656476 &  &  & $17.24_{-0.66}^{+12.26}$ & $16.91_{-1.37}^{+2.12}$ & $17.09_{-0.55}^{+2.37}$ & $16.94_{-3.46}^{+3.36}$ & $36.64_{-9.34}^{+10.72}$ &  \\
3735871 &  &  & $9.51_{-3.85}^{+3.66}$ & $12.77_{-1.93}^{+1.21}$ & $13.04_{-0.83}^{+0.87}$ & $12.40_{-3.56}^{+3.05}$ & $15.96_{-4.21}^{+2.39}$ &  \\
5950854 &  &  & $20.99_{-6.23}^{+11.19}$ & $31.44_{-10.91}^{+16.83}$ & $36.50_{-17.77}^{+14.91}$ & $30.78_{-11.73}^{+16.06}$ & $9.76_{-4.85}^{+96.92}$ &  \\
6603624 &  &  & $13.85_{-3.28}^{+13.89}$ & $19.22_{-4.18}^{+3.26}$ & $33.54_{-5.38}^{+6.02}$ & $32.95_{-6.87}^{+9.76}$ & $34.86_{-26.93}^{+258.42}$ &  \\
7106245 &  &  & $10.53_{-0.99}^{+8.99}$ & $36.89_{-11.88}^{+15.82}$ & $12.85_{-0.92}^{+0.49}$ & $35.33_{-26.39}^{+16.79}$ & $19.69_{-11.26}^{+19.14}$ & $10.53\pm3.58\tnote{A}$  \\
7680114 &  &  & $24.62_{-11.84}^{+5.07}$ & $30.87_{-6.37}^{+13.31}$ & $30.02_{-5.49}^{+1.54}$ & $28.62_{-7.18}^{+18.65}$ & $19.37_{-7.08}^{+22.09}$ &  \\
8150065 &  &  & $10.73_{-1.17}^{+4.58}$ & $45.39_{-7.90}^{+12.81}$ & $45.30_{-7.72}^{+12.82}$ & $45.76_{-8.63}^{+10.59}$ & $22.24_{-13.95}^{+39.49}$ &  $10.73\pm1.25\tnote{A}$\\
8228742 &  &  & $10.96_{-3.21}^{+7.94}$ & $19.90_{-2.75}^{+2.83}$ & $19.22_{-2.02}^{+2.36}$ & $19.52_{-10.28}^{+4.80}$ & $13.88_{-3.46}^{+6.35}$ &  \\
8760414 &  &  & $16.00_{-3.46}^{+21.16}$ & $34.91_{-3.49}^{+9.19}$ & $34.91_{-3.10}^{+9.31}$ & $36.57_{-6.56}^{+6.87}$ & $22.58_{-15.98}^{+112.30}$ &  \\
8938364 &  &  & $12.89_{-3.62}^{+4.27}$ & $16.58_{-3.62}^{+1.52}$ & $16.52_{-1.61}^{+0.80}$ & $15.60_{-4.29}^{+4.49}$ & $45.61_{-28.32}^{+52.07}$ &  \\
9206432 &  &  & $10.12_{-2.05}^{+1.92}$ & $9.96_{-0.66}^{+0.68}$ & $9.94_{-0.40}^{+0.77}$ & $9.92_{-3.36}^{+3.05}$ & $6.56_{-2.42}^{+3.91}$ &  \\
10068307 &  &  & $9.72_{-0.41}^{+8.18}$ & $9.35_{-0.39}^{+0.57}$ & $9.33_{-0.15}^{+0.65}$ & $19.73_{-5.84}^{+9.87}$ & $14.88_{-2.89}^{+4.67}$ &  \\
10644253 &  &  & $11.46_{-0.98}^{+2.13}$ & $33.31_{-3.68}^{+11.45}$ & $34.02_{-4.33}^{+8.63}$ & $34.61_{-6.46}^{+15.55}$ & $31.29_{-22.90}^{+85.88}$ & $11.46\pm0.72\tnote{M}$ \\
10730618 &  &  & $8.67_{-0.50}^{+1.44}$ & $8.90_{-0.77}^{+0.60}$ & $9.24_{-1.12}^{+0.18}$ & $8.57_{-2.53}^{+2.53}$ & $8.46_{-4.15}^{+11.43}$ &  \\
12069127 &  &  & $8.71_{-7.77}^{+9.10}$ & $17.38_{-3.15}^{+1.61}$ & $17.36_{-1.60}^{+1.42}$ & $16.84_{-3.87}^{+5.42}$ & $17.27_{-7.43}^{+13.21}$ & \\\hline
\end{tabular}
\label{tab:singlestarBb}
\end{threeparttable}
}
\end{table*}

\clearpage

\bibliographystyle{aasjournal}
\bibliography{ref-lbks.bib}

\end{document}